        \documentclass[useAMS,usenatbib]{mn2e}
        \usepackage{natbib}
        \usepackage{times}
        \usepackage{graphicx}

\newcommand{\aj}{AJ}
\newcommand{\aap}{A\&A}
\newcommand{\aaps}{A\&ASS}
\newcommand{\apj}{ApJ}
\newcommand{\apjl}{ApJ (Letters)}
\newcommand{\apjs}{ApJS}
\newcommand{\araa}{ARA\&A}
\newcommand{\pasp}{PASP}
\newcommand{\mnras}{MNRAS}

\usepackage{here}                       
\usepackage{dcolumn}                    

\newcolumntype{d}[1]{D{.}{.}{#1}}       

\newcommand{\x}{$\times$\ }
\newcommand{\ha}{H$\alpha$}
\newcommand{\ox}{[O\,{\sc iii}]}

    \title[Angular diameters, fluxes and extinction of compact planetary nebulae]
          {Angular diameters, fluxes and extinction of compact planetary nebulae:
           further evidence for steeper extinction towards the Bulge}

    \author[P. M. E. Ruffle et al.]
           {P. M. E. Ruffle$^1$\thanks{E-mail: paul.ruffle@postgrad.umist.ac.uk},
            A. A. Zijlstra$^1$\thanks{E-mail: a.zijlstra@umist.ac.uk},
            J. R. Walsh$^2$\thanks{E-mail: jwalsh@eso.org}, 
            M. D. Gray$^1$\thanks{E-mail: malcolm.gray@umist.ac.uk}
    \newauthor
            K. Gesicki$^3$, 
            D. Minniti$^4$ and 
            F. Comeron$^5$ \\
            $^1$ Department of Physics, University of Manchester Institute of Science and Technology, 
                 PO Box 88, Manchester M60 1QD, United Kingdom \\
            $^2$ ESO/ECF, Karl Schwarzschildstrasse 2 D-85748 Garching, Germany \\
            $^3$ Centrum Astronomii UMK, ul.Gagarina 11, PL-87-100 Torun, Poland \\
            $^4$ Department of Astronomy, P. Universidad Cat{\'o}lica, Casilla 306, Santiago 22, Chile \\
            $^5$ ESO, Karl Schwarzschildstrasse 2 D-85748 Garching, Germany}

\date{Accepted 2004 June 8. Received 2004 June 5; in original form 2003 November 13}

\pagerange{\pageref{firstpage}--\pageref{lastpage}} 

\pubyear{2004}

\begin{document}

\label{firstpage}

\maketitle

\begin{abstract}
We present values for angular diameter, flux and extinction for 70 
Galactic planetary nebulae observed using narrow band filters. 
Angular diameters are derived using constant emissivity shell and 
photoionization line emission models. The mean of the results 
from these two models are presented as our best estimate. 
Contour plots of 36 fully resolved objects are included and the low 
intensity contours often reveal an elliptical structure that is not 
always apparent from FWHM measurements.  
Flux densities are determined, and for both H$\alpha$ and [O\,{\sc iii}] 
there is little evidence of any systematic differences between observed 
and catalogued values. 
Observed H$\alpha$ extinction values are determined using observed H$\alpha$ 
and catalogued radio fluxes. H$\alpha$ extinction values are also derived
from catalogued H$\alpha$ and H$\beta$ flux values by means of an
$R_\mathrm{V}$ dependent extinction law. $R_\mathrm{V}$ is then calculated
in terms of observed extinction values and catalogued H$\alpha$ and
H$\beta$ flux values.  Comparing observed and catalogue extinction
values for a subset of Bulge objects, observed values tend to be 
lower than catalogue values calculated with $R_\mathrm{V} = 3.1$. 
For the same subset we calculate $\langle R_\mathrm{V} \rangle = 2.0$, confirming 
that toward the Bulge interstellar extinction is steeper than $R_\mathrm{V} = 3.1$.
For the inner Galaxy a relation with the higher supernova rate is suggested, and 
that the low-density warm ionized medium is the site of the anomalous extinction. 
Low values of extinction are also derived using dust models with a turnover 
radius of 0.08 microns.
\end{abstract}

\begin{keywords}
planetary nebulae: diameters, flux -- ISM: dust, extinction -- Galaxy: centre, bulge
\end{keywords}


\section{Introduction}\label{intro}

The angular diameter and flux of a planetary nebula (PNe) are much
needed parameters when modelling the optical spectrum and evolution of
the nebula and its central star. From these parameters the density and
mass of the PNe can be calculated, as well as the extinction along the line 
of sight. PNe also provide an independent tracer of stellar populations
and extinction in regions where visual tracers, such as hot luminous
stars, are lacking.

At the distance of the Bulge of $R = 8$ kpc \citep{reid} many PNe are only 
a few arc seconds across. Therefore determination of angular diameters, 
by deconvolving FWHM measurements, need to use methods that take into 
account ionisation structure and the particular emission line observed. 
\citet[][hereafter BZ]{bedding} and \citet[][hereafter VH]{vanhoof} 
provide two suitable methods of deconvolution.

Spectroscopic flux measurements can lead to a possible underestimation
if the slit is not properly aligned with the source. Therefore accurate 
determination of fluxes can be achieved using calibrated narrow-band 
filters, provided that correction for the underlying continuum and any 
unwanted lines is applied. Extinction can then be calculated from observed 
H$\alpha$ and catalogue radio flux ratios, and compared with values 
calculated from catalogue H$\alpha$/H$\beta$ flux ratios. A method for 
determining $R_\mathrm{V}$, the ratio of total to selective extinction, 
can also be derived.

Correction for interstellar extinction is necessary for determining
emission line ratios, absolute fluxes, and some distance
calculations. Usually, a single Galactic extinction law is used for
dereddening. However, the emission mechanism of ionized media is well
understood, and PNe are intrinsically very bright. This allows one to
use PNe to determine extinction constants for Galactic dust distribution 
studies, for line of sights where few other good tracers are available. 

In order to improve the accuracy of available PNe data, we have
obtained narrow-band H$\alpha$ and [O\,{\sc iii}] CCD images of a 
sample of Galactic PNe and calculated values for angular diameter, 
flux and extinction.


\begin{table}
    \caption[]{Filter characteristics (\AA).}
    \label{filterchars}
    \setlength{\tabcolsep}{7.0pt}
    \begin{tabular}{lccccc}
        \hline
        \vspace{-8pt} \\
        Filter                         & ESO\# & Peak  & Centre  & FWHM  & $\int T \mathrm{d}\lambda$  \\
        \vspace{-8pt} \\
        \hline
        \vspace{-8pt} \\
        H$\alpha$ on-band*             & 654   & 6546  & 6549    & 32    & 11  \\
        H$\alpha$ off-band             & 598   & 6675  & 6673    & 65    & 37  \\
        $[$O\,{\sc iii}$]$ on-band*    & 589   & 5016  & 5014    & 55    & 36  \\
        $[$O\,{\sc iii}$]$ off-band    & 591   & 5100  & 5110    & 60    & 43  \\
        \vspace{-8pt} \\
        \hline
        \vspace{-8pt} \\
    \end{tabular}
*Calculated from observed spectra.
\end{table}

\section{Observations and data processing}\label{obsdat}

\begin{figure}
    \hbox{\hspace{3mm}\includegraphics[width=77mm]{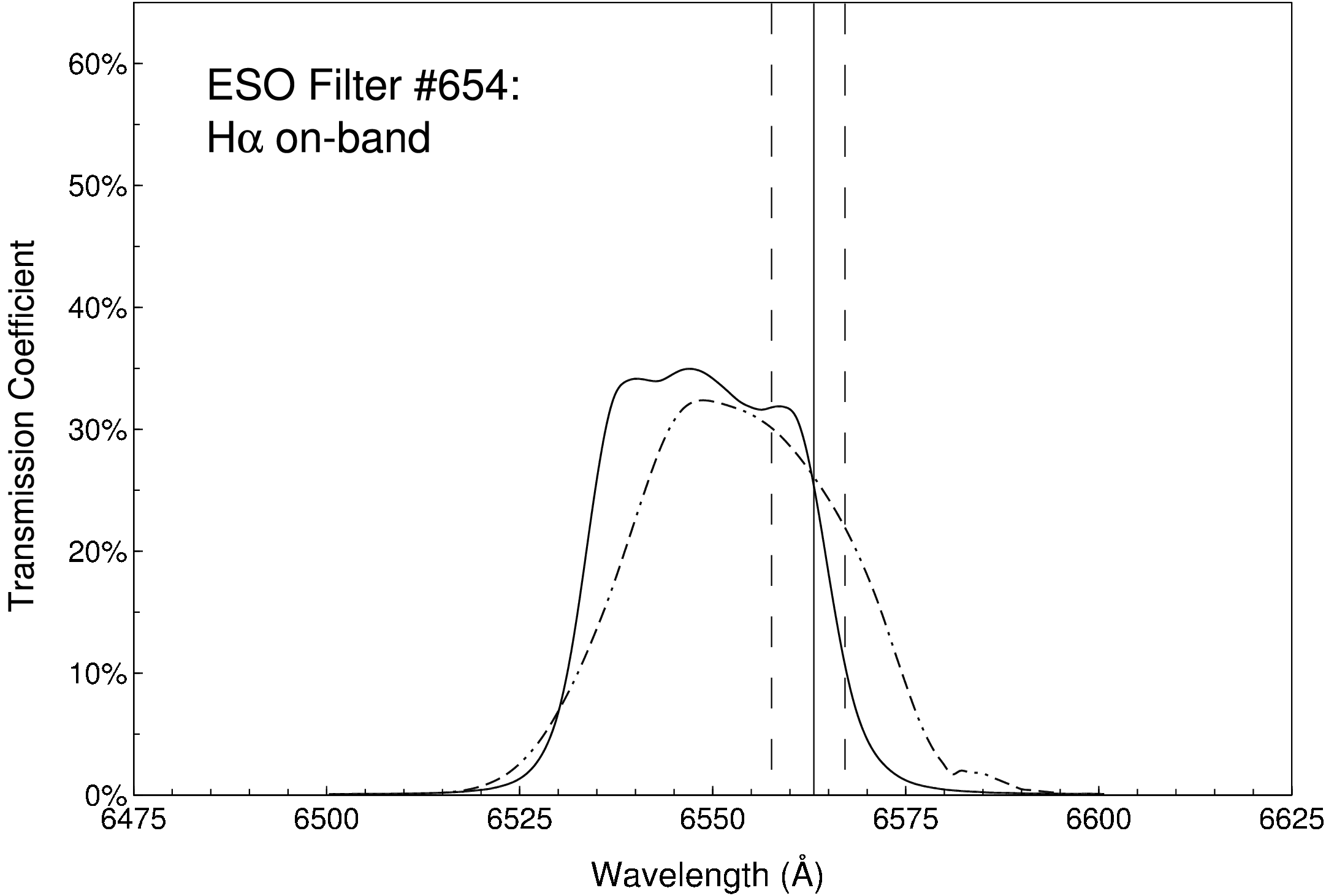}}
    \vspace{5mm}
    \hbox{\hspace{3mm}\includegraphics[width=77mm]{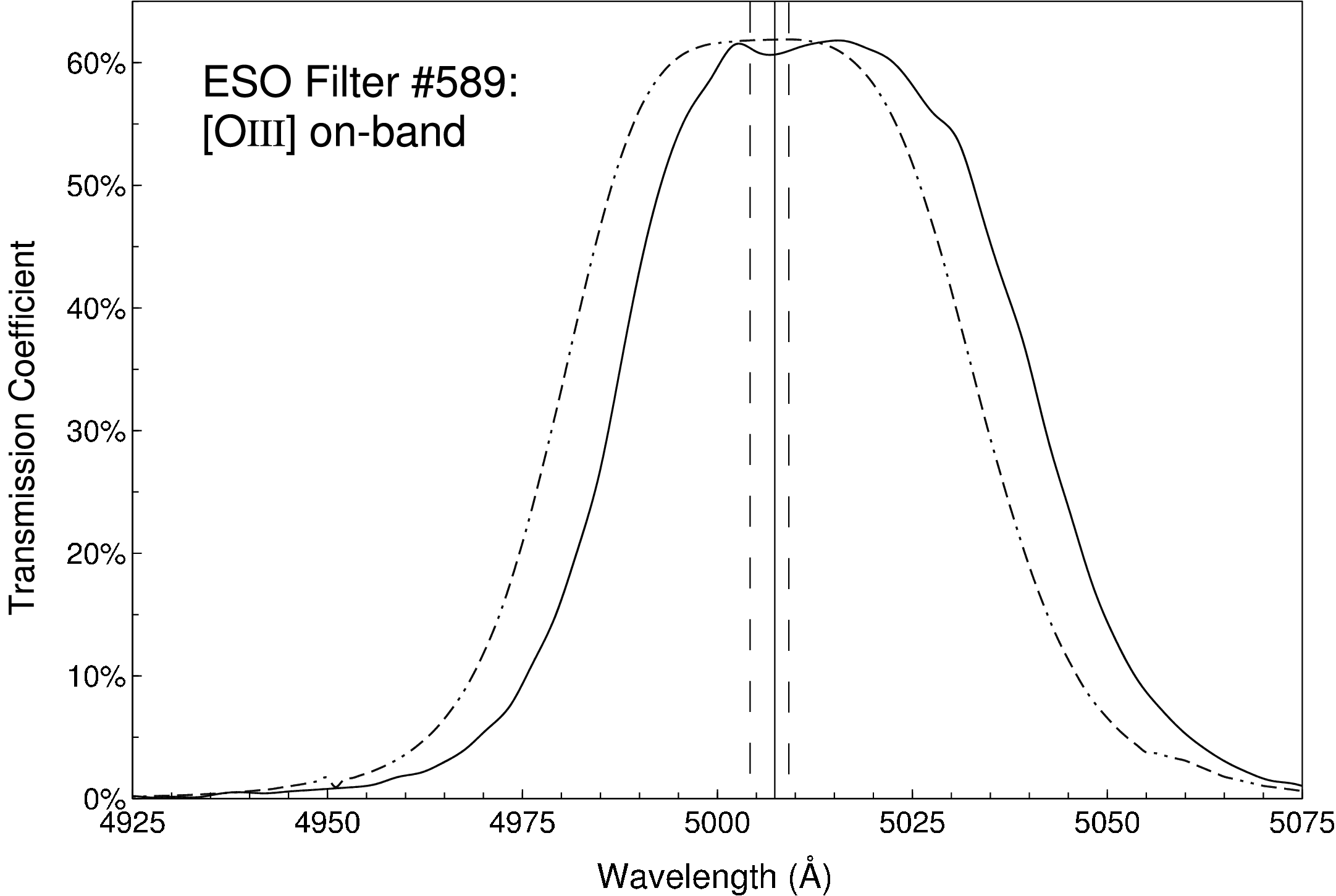}}
    \caption{Transmission curves for on-band H$\alpha$ and [O\,{\sc iii}] filters.
             Dashed curves are filter response values provided by ESO and 
             solid curves are those based on observed calibration spectra. 
             Solid vertical lines indicate La Silla-air wavelengths, with dashed vertical 
             lines indicating the spread of observed wavelengths due to Doppler shifting.} 
    \label{filterplots}
\end{figure}

We observed 70 Galactic PNe 
during the two nights of 2002 June 24 \& 25, using the ESO 3.5-m New 
Technology Telescope (NTT) in Chile.  We used the EMMI camera (ESO 
Multi-Mode Instrument), which at the time had a 2046~$\times$~2046~pixel
Tektronix CCD with an image scale of 0\farcs27/pixel. For each object a 
pair of exposures were taken using on- and off-band H$\alpha$ filters. 
16 of the objects were also observed with on- and off-band [O\,{\sc iii}] filters. 
The FWHM of the on-band filters are adapted to the high velocity range of the observed 
PNe (200 to 300 km s$^{-1}$ in the Galactic bulge, corresponding to $\sim$ 5\AA). 
The characteristics of the filters, in \AA, are listed in 
Table~\ref{filterchars} and plotted in Fig.~\ref{filterplots}. 
It should be noted that, to avoid contamination by the [N\,{\sc ii}] 6584\AA\ line, 
the H$\alpha$ on-band filter is centred on 6549\AA\ and not 6563\AA. 

In Fig.~\ref{filterplots} the dashed curves are the filter response values 
provided by ESO, and the solid curves are response values calculated from 
calibration spectra taken at the NTT (H$\alpha$ in 2004 February using Grism \#3 
at a resolution of 1.5\AA, interpolated to 0.1\AA; [O\,{\sc iii}] in 2003 June 
using Grism \#3 at a resolution of 3\AA, interpolated to 1\AA).
The solid vertical lines denote La Silla-air wavelengths and the dashed 
vertical lines the spread of observed wavelengths 
due to Doppler shifting. Assuming the same equivalent widths, the H$\alpha$ 
on-band filter \#654 has blue-shifted by $\sim5$\AA\ and the [O\,{\sc iii}] 
on-band filter \#589 has red-shifted by $\sim7$\AA. It can be seen that for 
[O\,{\sc iii}] this shift makes little difference to the calculation of actual 
flux values. However, for H$\alpha$ the shift is significant, as actual 
observed wavelengths are in the steep redward side of the curve (especially 
for positive radial velocities relative to the telescope). 

Because the filters in the red arm of EMMI are used in the parallel beam, 
there is a large change in central wavelength with position on the detector. 
The solid curve is given for the optical centre of the field. Because objects 
were not always positioned at the optical centre of the CCD array (up to 
$\pm20$ pixels on the first night and up to $\pm50$ pixels on the second), 
wavelength corrections were calculated to take into account the filter angle 
of 7.5\degr relative to the CCD array. This only affected offsets in the 
$y$-axis, with a 20 pixel offset generating a wavelength shift of 0.6\AA. 
For these reasons, in particular, we chose to calculate precise Doppler 
wavelength shifts when determining actual observed flux values. We have 
not been able to determine the reason for the shift in filter transmission 
curves, so would therefore recommend always taking calibration spectra from 
the optical centre of the field in order to check filter response data. 

To reduce the CCD readout time and hence improve the observing duty
cycle, only the central 400~$\times$~400 pixels
(108\arcsec~$\times$~108\arcsec) of the detector were read
out. Calibration of the CCD images followed standard procedures:
subtraction of a constant value to represent the bias, and removal of
the background continuum, achieved by subtracting the off-band image,
scaled by the integrated transmission ratio of the two filters ($\int
T_\mathrm{on} \mathrm{d}\lambda\ / \int T_\mathrm{off}
\mathrm{d}\lambda$), from the on-band image. The ratios were 0.31 for
the H$\alpha$ filter and 0.84 for the [O\,{\sc iii}] filter. Where
necessary and prior to this subtraction, pairs of images were brought
into alignment by integer pixel shifts of the off-band image in the
$x$ and/or $y$ direction.  For the first night's observations this
involved shifts of only one or two pixels.  However, due to taking
spectra between on- and off-band exposures, the second night's
observations required shifts of between 20 and 30 pixels. One off-band
image (356.1$+$02.7 Th3-13) had to be rotated as well, to bring it into
alignment. This procedure removed the background field stars, except
for objects with spectral lines within the filter band. 
Correction with dome flats was not considered necessary due
to the sources being very small and pixel to pixel variations in the
centre of the CCD contributing a possible error in flux values of only 1 per cent.

All exposures were for 30 seconds with the autoguider deactivated, as the NTT can track 
without guiding for several minutes. The resulting images show no sign of elongation of 
the field stars. Ten second exposures of two tertiary standard stars were also taken 
for flux calibration. Seeing conditions varied between one and two arc seconds.
Subsequent image analysis was done using the ESO Munich Image Data Analysis System ({\sc midas}).

Bulge PNe were concentrated on, but non-Bulge objects were also observed 
to make best use of telescope time. Using a Galactic Bulge criteria for 
the galactic coordinates (PN G) of $|l| < 15\degr $ and $|b| < 15\degr $ 
gives 56 Bulge PNe and 14 non-Bulge PNe. However, adopting the criteria of 
\citet{bensby}: $|l| < 10\degr $ and $|b| < 10\degr $; $0\farcs9 < \theta 
< 20\arcsec $; 4.2 mJy $< S_{5\,\mathrm{GHz}} <$ 45 mJy, gives 39 Bulge 
PNe and 31 non-Bulge PNe. The non-Bulge PNe include two Sagittarius dwarf 
galaxy objects \citep{dudziak}.


\section{Diameter determinations}\label{diams}

\subsection{Method}

The images presented here are broadened by the seeing. 
This must be taken into account when deriving angular diameters and we now 
describe the methods used to deconvolve the point spread function (beam size).

Using the method of BZ, the point spread function (PSF) was
approximated with a Gaussian FWHM of $\theta_\mathrm{PSF}$, by
fitting Gaussian profiles to four distinct field stars in each of the
on-band images, taking the mean value and expressing the error as the
standard deviation of the four readings. The results are listed in
column four of Table~\ref{diamsa}, with an overall mean value for
$\theta_\mathrm{PSF}$ of 1\farcs55. The FWHM of the nebulae
$\theta_\mathrm{FWHM}$, was measured by fitting two dimensional
Gaussian profiles to the corrected on-band images. As BZ point out, 
if the objects were Gaussian, deconvolution would be trivial, and the
true FWHM of the nebulae would simply be

\begin{equation}
    \theta_\mathrm{gauss} = \sqrt{(\theta_\mathrm{FWHM})^2 - (\theta_\mathrm{PSF})^2}.
\end{equation}

Although $\theta_\mathrm{gauss}$ is not a realistic model for PNe, a
correction factor $\gamma$, can be applied in order to determine the
true diameter, $\theta_\mathrm{true} = \gamma \theta_\mathrm{gauss}$.
Various methods of calculating this factor have been investigated by
VH and the conversion factor $\gamma$ is found to be a function of
$\beta$, the resolution of the observation, i.e. the ratio of the
Gaussian diameter and the point spread function, 
$\beta = \theta_\mathrm{gauss} / \theta_\mathrm{PSF}$.
According to VH, $\gamma(\beta)$ can be approximated extremely well 
by the following analytic function (VH method A):

\begin{equation}
    \gamma(\beta) = \frac{a_1}{1 + a_2\beta^2} + a_3,
    \label{gamma}
\end{equation}

\noindent
which will be used throughout this paper (appropriate parameters for
equation \ref{gamma} are given in Table~\ref{params}). 

\begin{table}
    \caption[]{The parameters \citep[tables 1 \& 2]{vanhoof} for calculating 
               conversion factors using equation \ref{gamma}.}
    \label{params}
    \setlength{\tabcolsep}{10pt}
    \begin{tabular}{lccc}
        \hline
        \vspace{-9pt} \\
        Model                                   &  a$_1$        &  a$_2$        &  a$_3$        \\
        \vspace{-9pt} \\
        \hline
        \vspace{-9pt} \\
        Constant emissivity shell 0.8           &  0.3893       &  0.7940       &  1.2281       \\
        H$\alpha$ emission line shell           &  0.3605       &  0.7822       &  1.4937       \\
        $[$O\,{\sc iii}$]$ emission line shell  &  0.4083       &  0.7892       &  1.3176       \\
        \vspace{-9pt} \\
        \hline
    \end{tabular}
\end{table}

VH finds that the method used by BZ for the calculation of $\gamma$ for disk, 
sphere and shell geometries is essentially identical to his own method B, 
with results that are also in good agreement with his method A (see VH fig.~3), 
apart from a slight overestimation of $\gamma$ for larger values of $\beta$. 
In the case of a constant emissivity shell, VH fig. 2 shows that the shape of the 
curve for the conversion factor $\gamma$ does not change very much as a function of 
$r_\mathrm{i} / r_\mathrm{s}$, but that the height of the curve does. Therefore, the 
correct value for $\gamma$ depends critically on the assumed value for the inner radius, 
$r_\mathrm{i}$, with differences of up to 40 per cent, depending on the choice $r_\mathrm{i}$. 
VH also examines what he considers to be more realistic geometries based on a 
photoionization model of PNe. This model takes into account emissivity and absorption 
of specific lines as a function of distance to the star. As VH point out, the surface 
brightness profile is very sensitive to optical depth effects and thus different 
emission lines will yield different diameters. Measured diameters will also be different, 
depending on which line is used, because of ionisation stratification effects. 
VH also present an alternative algorithm for determination of the intrinsic FWHM of a 
source, using the observed surface brightness distribution and the FWHM of the beam (PSF). 
This was not investigated as knowledge of the intrinsic surface brightness distribution is 
still needed to convert the intrinsic FWHM of the source to the true diameter. 
The reader is referred to VH for a full discussion of the various methods that can be used 
to measure angular diameters of extended sources. 

Based on the above we chose to calculate diameters using two models: 
a constant emissivity shell, $\theta_\mathrm{shell}$ (where an inner diameter of 
$0.8 \theta_\mathrm{shell}$ was chosen because it represents a filling factor of 0.5); 
and photoionization line emission, $\theta_\mathrm{line}$. In Table~\ref{diamsa} we list 
deconvolved Gaussian FWHM values, $\theta_\mathrm{gauss}$ and PSF (beam size) values, 
$\theta_\mathrm{PSF}$, so that calculations based on alternative intrinsic surface 
brightness distributions can be made.

\begin{table*}
\caption[]{H$\alpha$ angular diameters derived from 18 restored on-band images, 
           compared with diameters derived directly from on-band--minus--off-band images (Table~\ref{diamsa}).
           Images restored with the Richardson-Lucy algorithm \citep{rich72,lucy7406}. 
           Diameters are in arc seconds and are defined in section~\ref{diams}.}
\label{restdiams}
\setlength{\tabcolsep}{7.6pt}
\begin{tabular}{llcccccccc}
\vspace{-10pt} \\
\hline
\vspace{-6pt} \\
PN G & Object & 
\multicolumn{1}{c}{$\theta_\mathrm{rest\,PSF}$}   & 
\multicolumn{1}{c}{$\theta_\mathrm{rest\,gauss}$} & 
\multicolumn{1}{c}{$\theta_\mathrm{rest}$}       & 
\multicolumn{1}{c}{$\theta_\mathrm{rest\,10\%}$}  & 
\multicolumn{1}{c}{$\theta_\mathrm{PSF}$}   & 
\multicolumn{1}{c}{$\theta_\mathrm{gauss}$} & 
\multicolumn{1}{c}{$\theta_\mathrm{mean}$}  & 
\multicolumn{1}{c}{$\theta_\mathrm{10\%}$}  \\ 
\vspace{-6pt} \\
\hline
\vspace{-6pt} \\
000.3$-$04.6 & M2-28    & 1.04 & 4.27 \x 3.60 &  5.9 \x  5.0 &  7.5 \x  7.0 & 1.57 & 4.10         & 5.8        &  9.0 \x  8.0 \\
000.3$-$02.8 & M3-47    & 0.82 & 5.63 \x 3.60 &  7.7 \x  5.0 &  6.0 \x  2.5 & 1.10 & 5.39 \x 3.93 & 7.4 \x 5.5 &  9.0 \x  8.0 \\
000.6$-$01.3 & Bl3-15   & 0.79 & 1.87 \x 1.55 &  2.7 \x  2.3 &  5.0 \x  4.0 & 1.50 & 2.27 \x 2.03 & 3.4 \x 3.1 &  6.0 \x  4.5 \\
002.3$-$03.4 & H2-37    & 0.72 & 3.88 \x 1.50 &  5.3 \x  2.2 &  5.5 \x  3.0 & 1.88 & 3.83 \x 2.04 & 5.5 \x 3.2 &  7.0 \x  5.5 \\
002.8$-$02.2 & Pe2-12   & 0.85 & 6.73 \x 1.89 &  9.2 \x  2.7 &  7.0 \x  3.0 & 2.00 & 7.01 \x 2.14 & 9.8 \x 3.3 & 10.5 \x  5.0 \\
003.0$-$02.6 & KFL4     & 0.63 & 1.19 \x 1.04 &  1.7 \x  1.5 &  2.5         & 1.81 & 1.46         & 2.2        &              \\
003.7$-$04.6 & M2-30    & 0.68 & 2.81 \x 2.81 &  3.9 \x  3.9 &  4.0         & 1.57 & 3.00 \x 2.58 & 4.4 \x 3.8 &              \\
004.8$-$05.0 & M3-26    & 0.70 & 6.11 \x 5.84 &  8.4 \x  8.0 & 10.0 \x  8.5 & 1.54 & 5.93         & 7.7        & 11.0 \x  9.5 \\
005.2$-$18.6 & StWr2-21 & 0.85 & 0.74 \x 0.76 &  1.2 \x  1.2 &  2.5 \x  2.0 & 1.80 & 1.58 \x 1.21 & 2.5 \x 2.0 &              \\
008.6$-$07.0 & He2-406  & 1.16 & 3.73 \x 2.57 &  5.2 \x  3.7 &  6.5 \x  5.5 & 1.56 & 4.06 \x 3.37 & 5.8 \x 4.9 &  8.0 \x  7.5 \\
283.8$+$02.2 & My60     & 0.80 & 6.04 \x 5.60 &  8.3 \x  7.7 & 10.0         & 1.65 & 5.65         & 7.9        & 11.0 \x 10.5 \\
346.3$-$06.8 & Fg2      & 0.46 & 3.39 \x 2.93 &  4.7 \x  4.0 &  5.5 \x  4.5 & 1.17 & 3.44 \x 2.96 & 4.9 \x 4.2 &  6.5 \x  5.5 \\
352.6$+$00.1 & H1-12    & 1.18 & 6.29 \x 5.19 &  8.7 \x  7.2 & 10.0 \x  8.5 & 1.13 & 5.76         & 7.9        & 12.0 \x 10.0 \\
352.6$+$03.0 & H1-8     & 0.51 & 1.80 \x 1.50 &  2.5 \x  2.1 &  4.0 \x  3.0 & 1.36 & 2.10 \x 1.73 & 3.1 \x 2.6 &              \\
352.8$-$00.2 & H1-13    & 0.57 & 9.88 \x 9.03 & 13.5 \x 12.3 & 13.5 \x 12.0 & 1.57 & 9.27         & 12.7       & 13.5 \x 12.0 \\
354.9$+$03.5 & Th3-6    & 1.28 & 1.74 \x 1.78 &  2.6 \x  2.7 &  3.0         & 2.38 & 2.64 \x 2.34 & 4.1 \x 3.7 &              \\
357.1$-$06.1 & M3-50    & 0.49 & 4.81 \x 0.62 &  6.6 \x  0.9 &  7.0 \x  2.0 & 1.25 & 4.82 \x 1.07 & 6.7 \x 1.7 &  8.5 \x  3.5 \\
357.1$-$04.7 & H1-43    & 0.72 & 0.54 \x 0.34 &  0.9 \x  0.6 &  2.0         & 1.71 & 1.05 \x 0.75 & 1.7 \x 1.3 &              \\
\vspace{-6pt} \\
\hline
\end{tabular}
\vspace{-3pt} \\
\end{table*}

\subsection{Image Restoration}

Since the seeing was not exceptional during the observations it
was considered desirable to improve the morphological information of
the images by restoring them. The critical component of restoration
is fidelity of the PSF, and for that reason only those images were
selected which had a bright well separated star image on the frames. 
Therefore images with many stars, or with no bright stars, were not 
deemed suitable for restoration. 18 pairs of H$\alpha$ on- and off-band
images were restored using the Richardson-Lucy algorithm \citep{rich72,lucy7406} 
as implemented in the IRAF\footnote{IRAF is distributed by 
the National Optical Astronomy Observatories, operated by the 
Association of Universities for Research in Astronomy, Inc., under 
contract to the National Science Foundation of the United States.} 
stsdas.analysis.restore package. The images were background
subtracted by a constant and the PSF was cut from the image. Each
image was restored either for 50 iterations or until convergence
($\chi^2 \sim 1$). Using these on-band restored images, diameters 
for $\theta_\mathrm{rest\,PSF}$, $\theta_\mathrm{rest\,gauss}$, 
$\theta_\mathrm{rest\,shell}$ and $\theta_\mathrm{rest\,line}$ were 
then calculated using the same methods as above. The results are 
listed in Table~\ref{restdiams} with $\theta_\mathrm{rest}$ the mean 
of $\theta_\mathrm{rest\,shell}$ and $\theta_\mathrm{rest\,line}$. 
Diameters derived directly from the on-band--minus--off-band images 
(Table~\ref{diamsa}) are also included for ease of comparison. 
An attempt to generate restored continuum-free images was not made 
by subtracting the restored off-band image from the restored on-band 
image. Such difference images would likely suffer from high frequency 
noise, since the restored PSFs would not be identical in both images 
and therefore could not be used for flux determination.

\subsection{Results}

\begin{figure}
    \hbox{\hspace{4mm}\includegraphics[width=72mm]{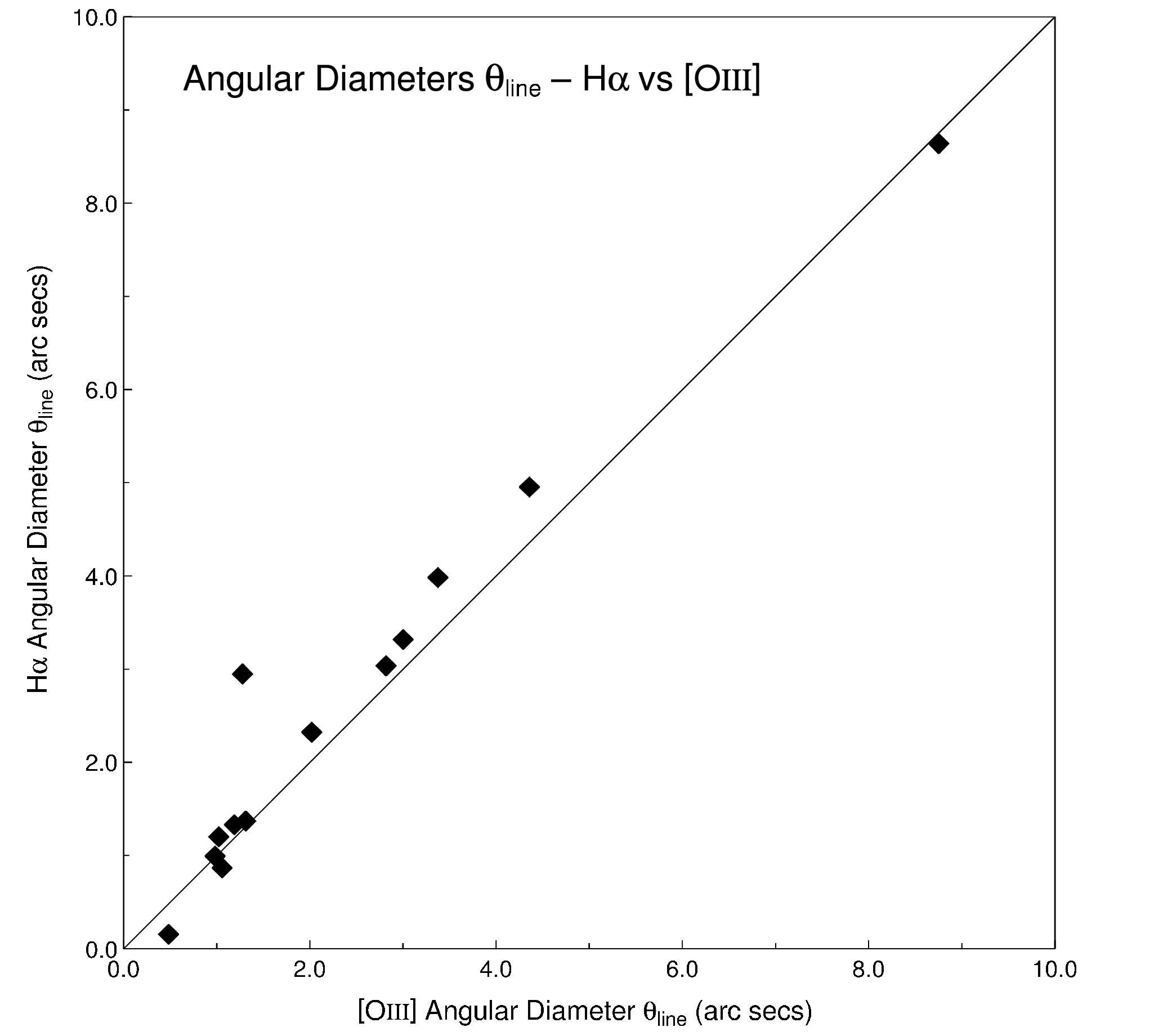}}
    \caption{Comparison of H$\alpha$ and [O\,{\sc iii}] angular diameters $\theta_\mathrm{line}$.}
    \label{hao3angdias}
\end{figure}

From the results listed in Table~\ref{diamsa} it can be seen that the ionization 
stratification model, $\theta_\mathrm{line}$, consistently yields larger diameter 
values than the constant emissivity shell 0.8 model, $\theta_\mathrm{shell}$. 
Calculating $\theta_\mathrm{shell\,0.5}$ ($r_\mathrm{i} = 0.5$, filling factor 0.9) 
yields values within a few per cent of $\theta_\mathrm{line}$. 
However, we believe $\theta_\mathrm{shell}$ with a filling factor of 0.5 
($r_\mathrm{i} = 0.8$) to be more reasonable. 
Therefore we present the mean value, $\theta_\mathrm{mean}$, of the two models 
as our best estimate, with $\theta_\mathrm{line}$ and $\theta_\mathrm{shell}$ the 
upper and lower limits. For $\theta_\mathrm{mean}$, both axis diameters are shown 
where the axis ratio $> 1.1$ and the axis difference $> 0.3$ arc seconds.
Fig.~\ref{hao3angdias} shows that calculated [O\,{\sc iii}] 
$\theta_\mathrm{line}$ diameters are in most cases smaller 
than those for H$\alpha$, as would be expected. 
As Table~\ref{diamsa} shows, 36 of the objects are fully resolved
($\theta_\mathrm{line} \geq 2 \theta_\mathrm{PSF}$), and Fig.~\ref{plots} 
plots 26\arcsec~$\times$~26\arcsec\ images of these objects, 
with contours expressed as percentages of the peak nebula flux. 
The morphology of the nebulae are apparent and the structures seen in the 
images (elliptical, disc, bipolar, etc.) can be compared with the 
evolutionary sequences of \citet{balick}.

25 objects were sufficiently resolved so that a direct diameter 
measurement could be made. For PNe with a well defined outer radius (i.e., 
with a steep drop-off to zero surface brightness at a certain radius) 
it is easy to measure directly the radius for a specific surface brightness. 
Published data and HST images show that flux usually drops very fast between 
contour levels in the interval 15 to 5 per cent \citep{bensby}. Therefore 
10 per cent of the peak surface brightness contour diameter measurements were 
made with an estimated error of $\pm 0 \farcs 5$, and these are listed in 
column nine of Table~\ref{diamsa}. 
It is worth noting that in these cases the low intensity contours seen
in Fig.~\ref{plots} often reveal a larger elliptical structure than
that apparent from the 2D FWHM axis ratio measurement of the bright
nebula core (column twelve of Table~\ref{diamsa}). For comparison
columns ten and eleven of Table~\ref{diamsa} list the catalogued radio
and optical diameters from the compilation available in the Strasbourg
- ESO Catalogue \citep{acker}. \citet{tylenda0307} have recently
redetermined a large set of PNe diameters: we find good agreement with 
their results for all but the most compact ($<2''$) objects.

Table~\ref{restdiams} lists H$\alpha$ angular diameters derived from 
the 18 restored on-band images, together with diameters derived directly 
from the on-band--minus--off-band images (Table~\ref{diamsa}). 
For 12 objects column eight ($\theta_\mathrm{gauss}$) also shows both 
axis values, which are only averaged in Table~\ref{diamsa}.
$\theta_\mathrm{rest}$ is presented as the mean value of 
$\theta_\mathrm{rest\,shell}$ and $\theta_\mathrm{rest\,line}$. 
Fig.~\ref{meanvsrestdias} compares angular diameters 
$\theta_\mathrm{mean}$ and $\theta_\mathrm{rest}$, 
and it can be seen that, regardless of PNe axis, restored diameters 
tend to be around one arc second smaller below 4 arc seconds. 
The data suggests the simple linear relationship 
$\theta_\mathrm{rest} = 1.1 \theta_\mathrm{mean} - 1$.
Direct diameter measurements, $\theta_\mathrm{rest\,10\%}$, were made at 
10 per cent of the peak surface brightness contour for 13 of the restored 
on-band images, and as Table~\ref{restdiams} shows, these are consistently 
lower by 1--2 arc seconds than the non-restored 10 per cent measurements, 
$\theta_\mathrm{10\%}$ (except for 352.8$-$00.2, H1-13). 
It could be that the restoration algorithm removes the detail that can be 
seen in the very low intensity contours of some PNe in Fig.~\ref{plots}.

\begin{figure*}
    \hbox{\includegraphics[width=55mm]{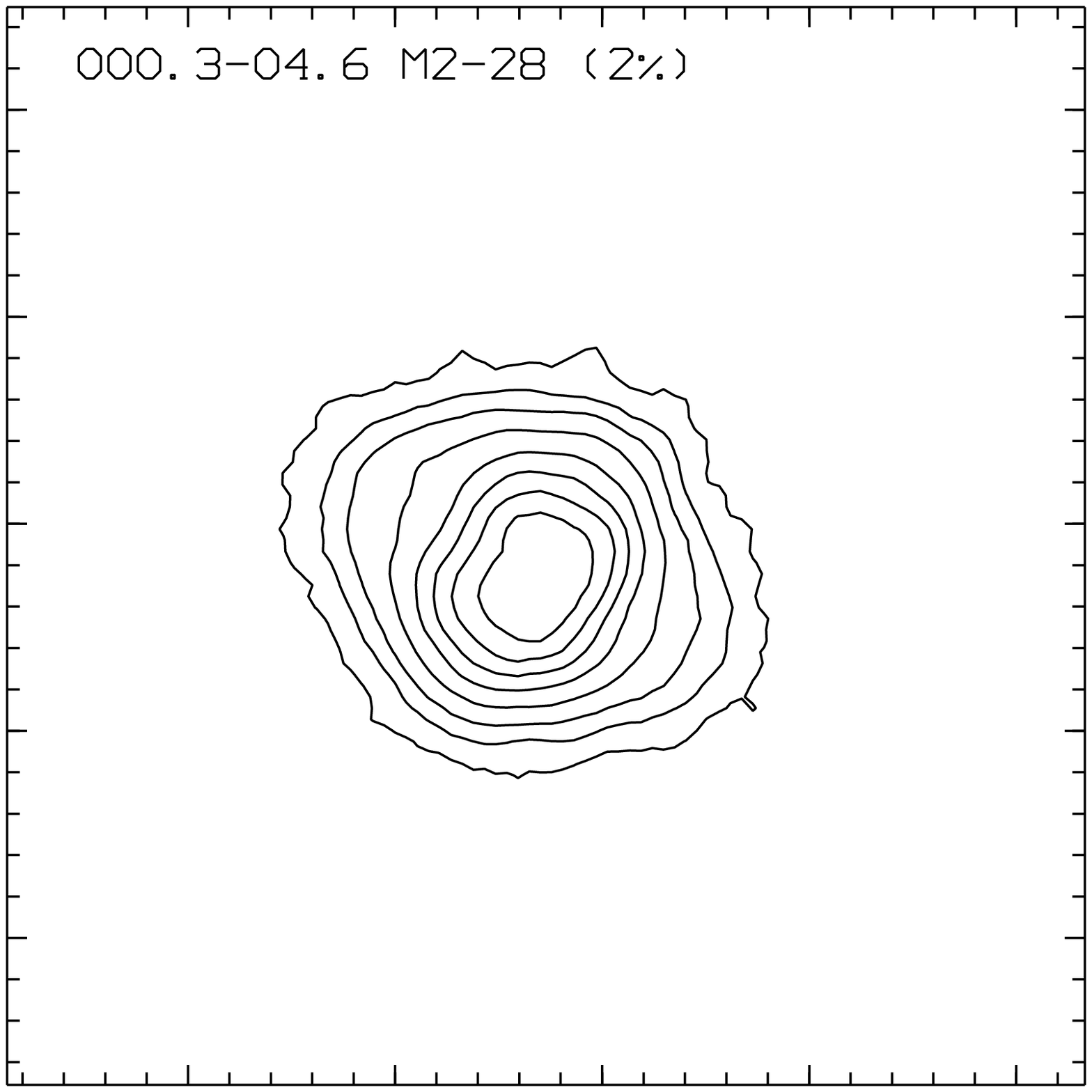}
          \hspace{4.5mm}
          \includegraphics[width=55mm]{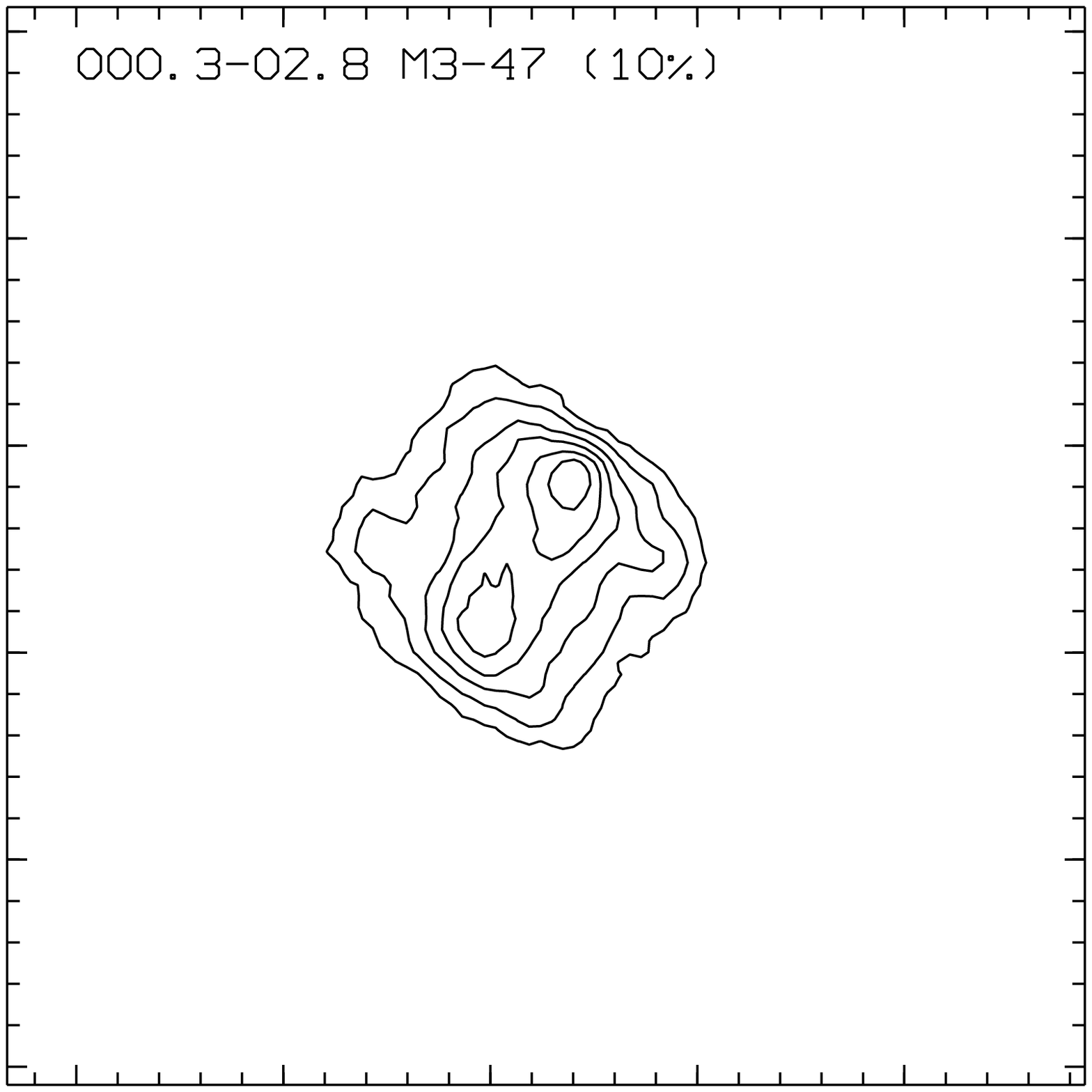}
          \hspace{4.5mm}
          \includegraphics[width=55mm]{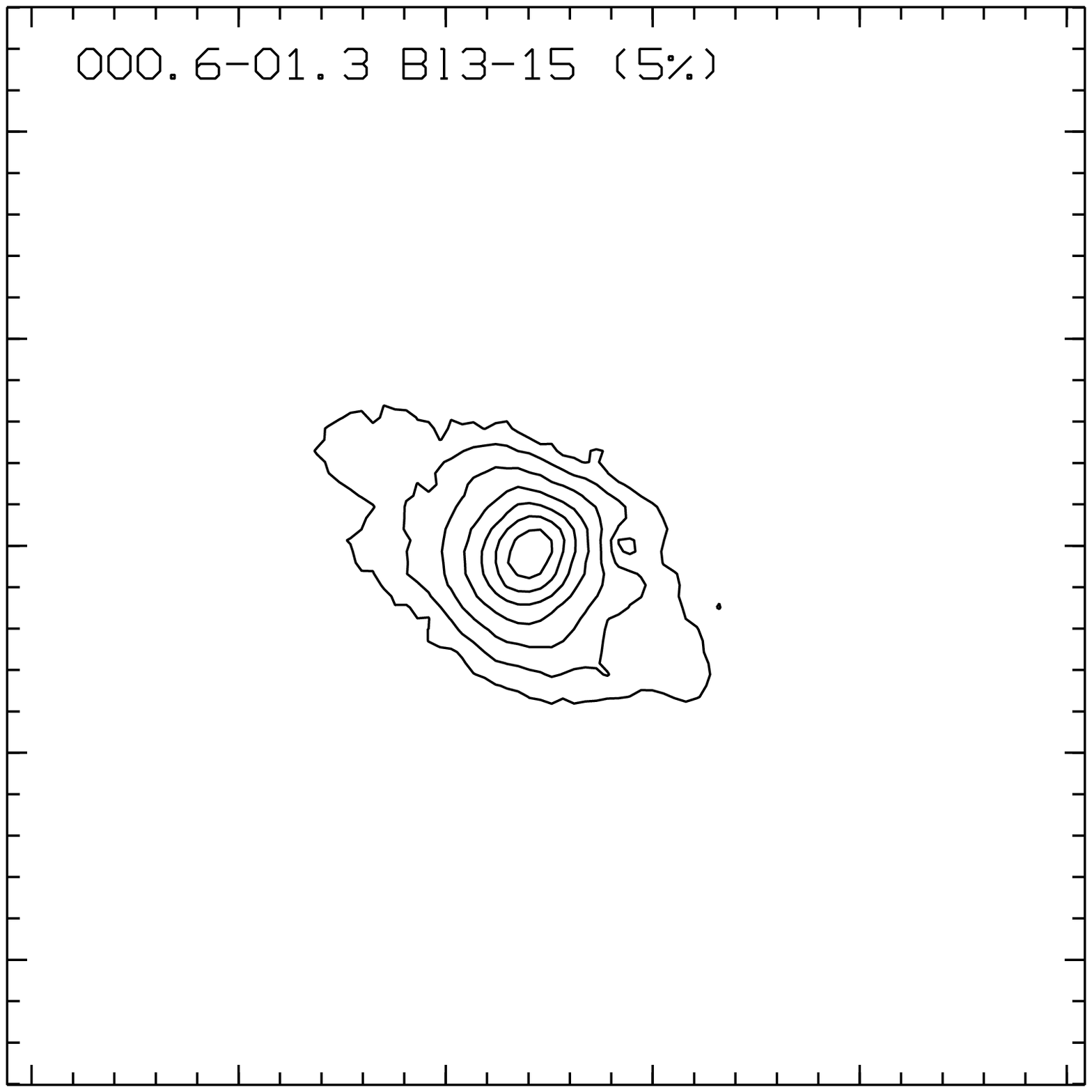}}
    \vspace{3mm}
    \hbox{\includegraphics[width=55mm]{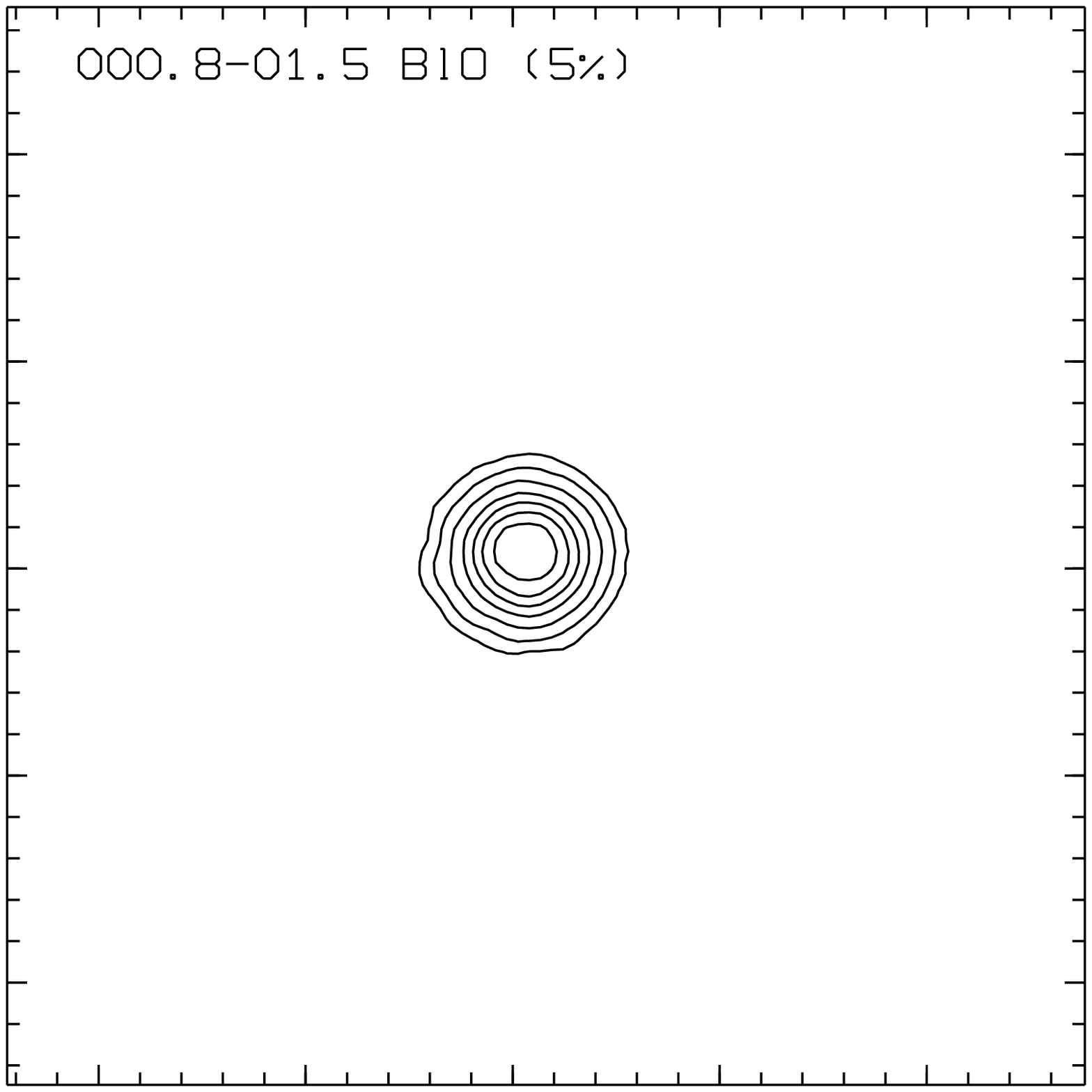}
          \hspace{4.5mm}
          \includegraphics[width=55mm]{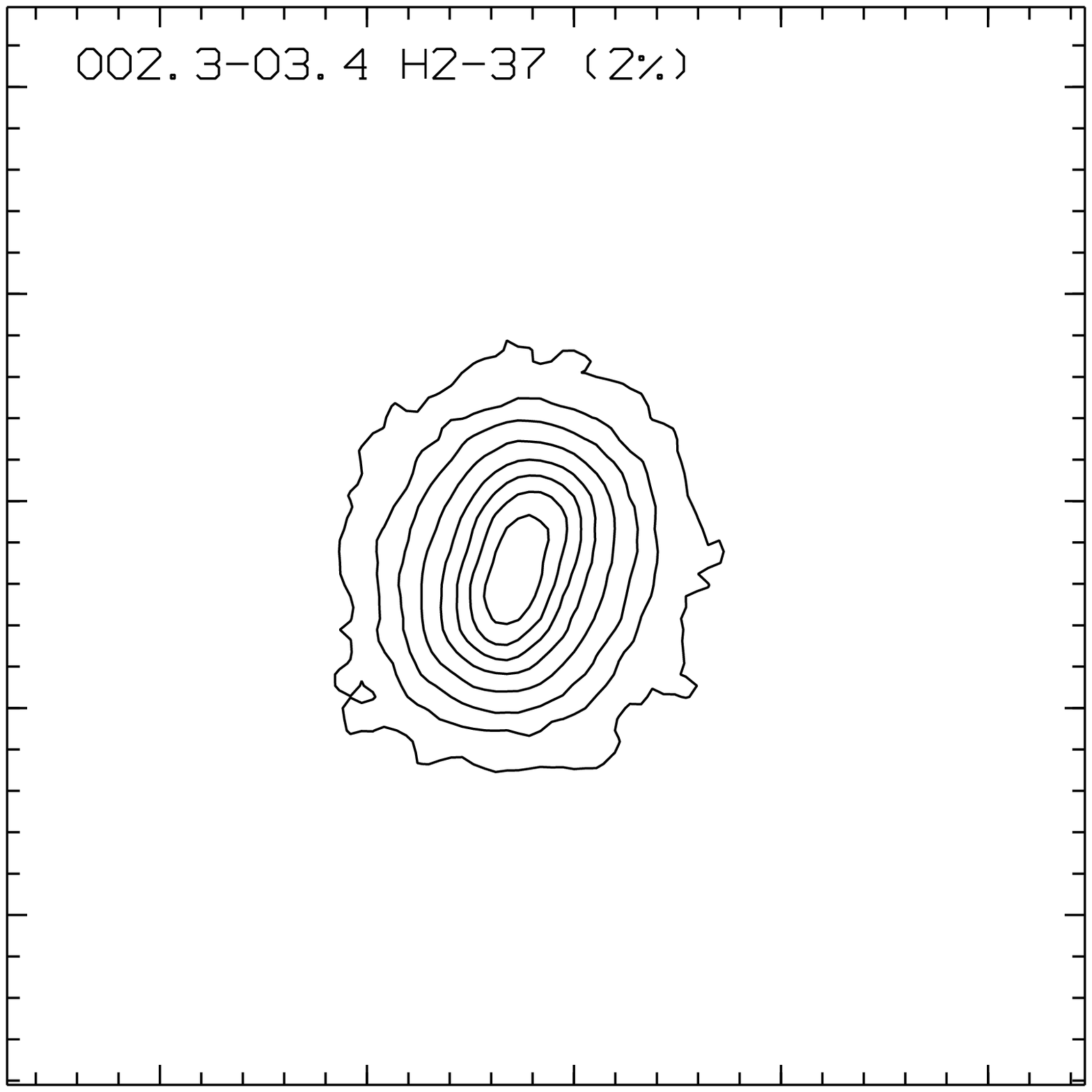}
          \hspace{4.5mm}
          \includegraphics[width=55mm]{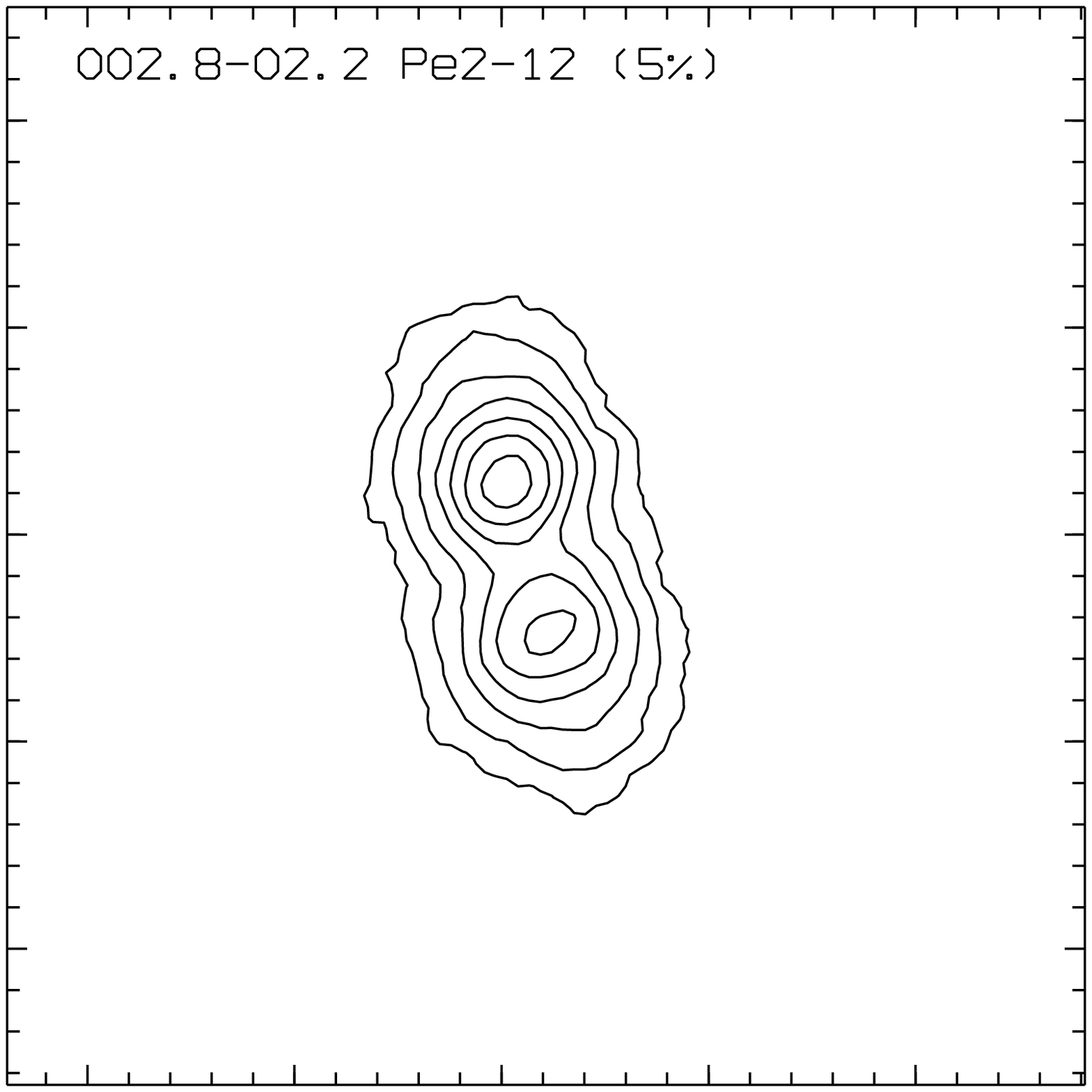}}
    \vspace{3mm}
    \hbox{\includegraphics[width=55mm]{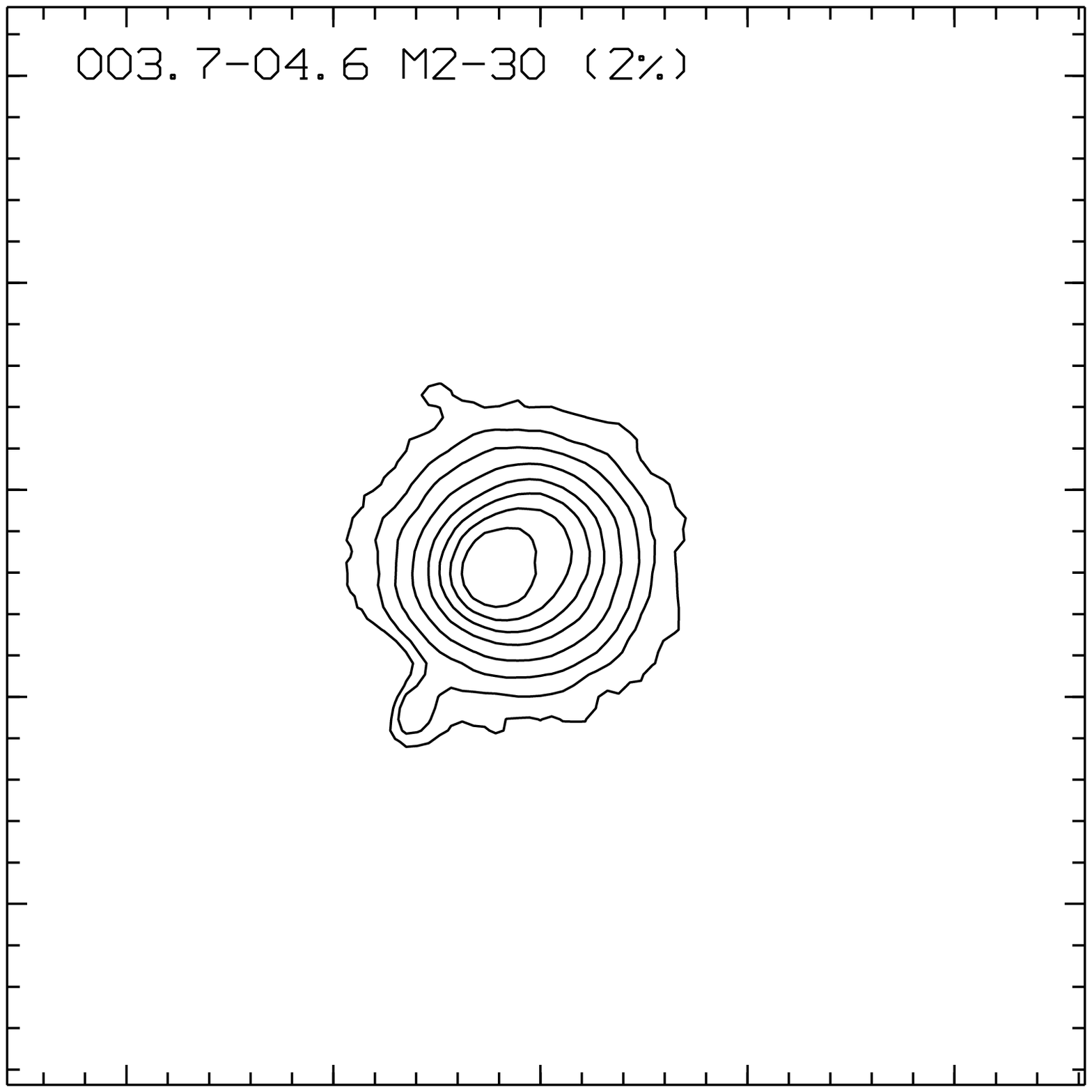}
          \hspace{4.5mm}
          \includegraphics[width=55mm]{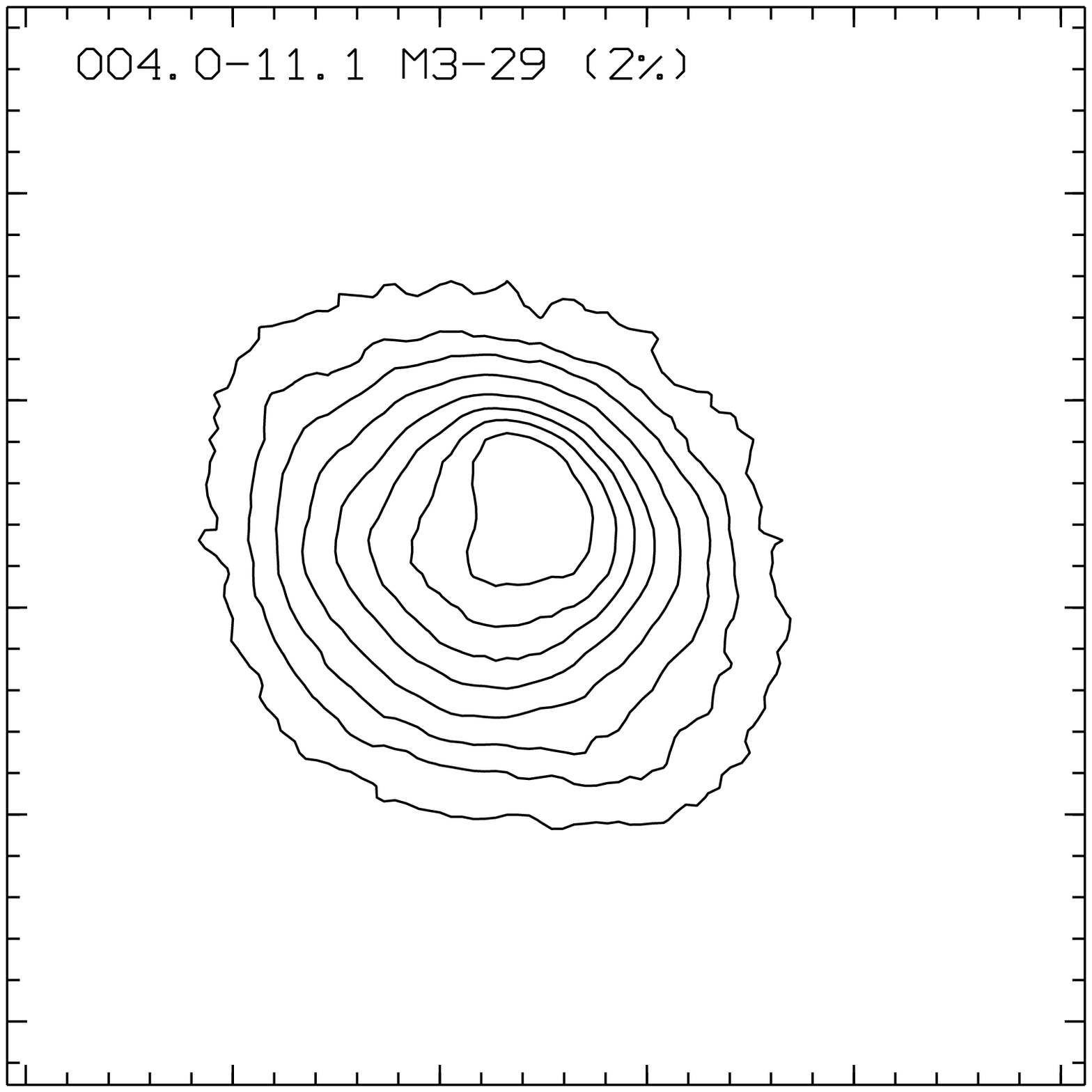}
          \hspace{4.5mm}
          \includegraphics[width=55mm]{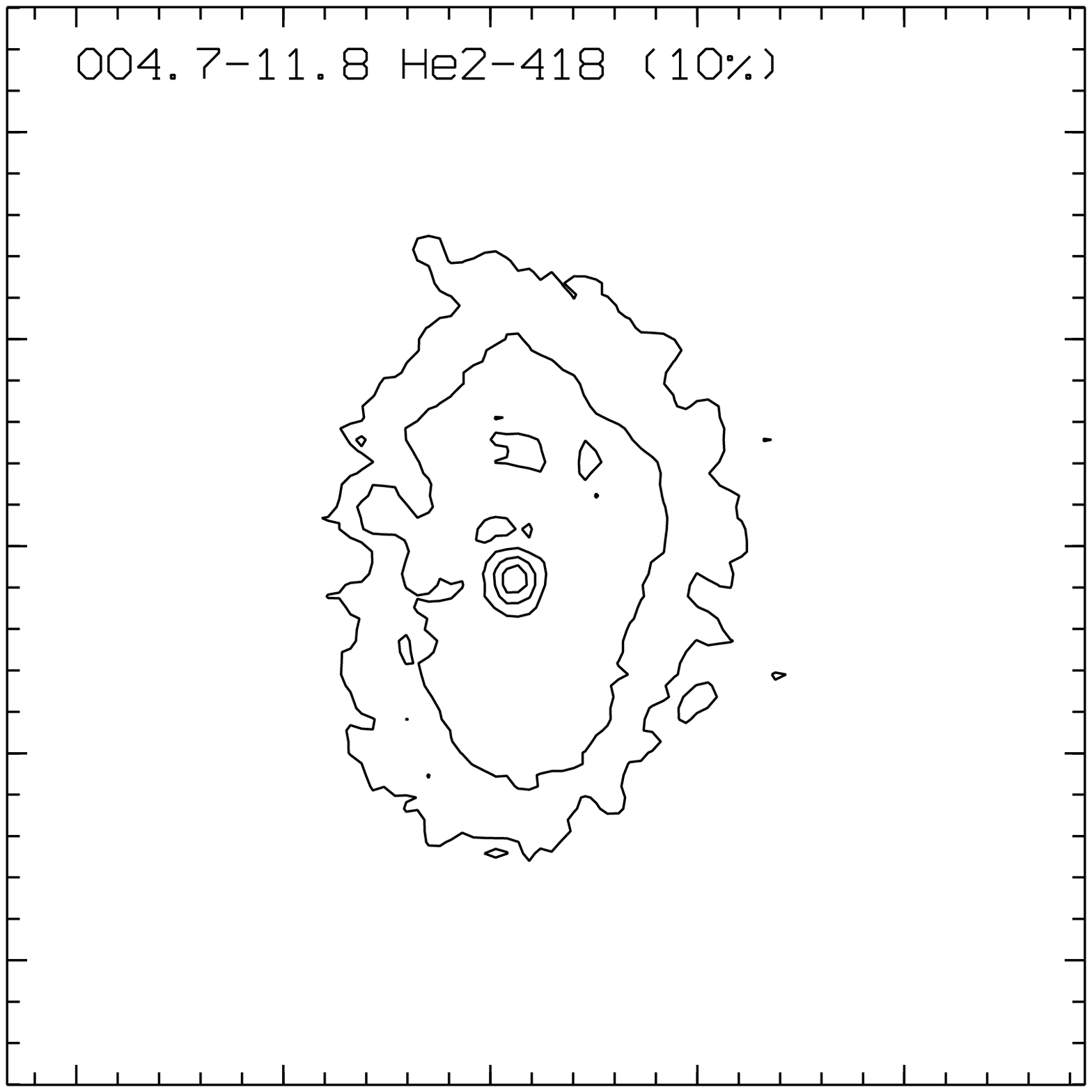}}
    \vspace{3mm}
    \hbox{\includegraphics[width=55mm]{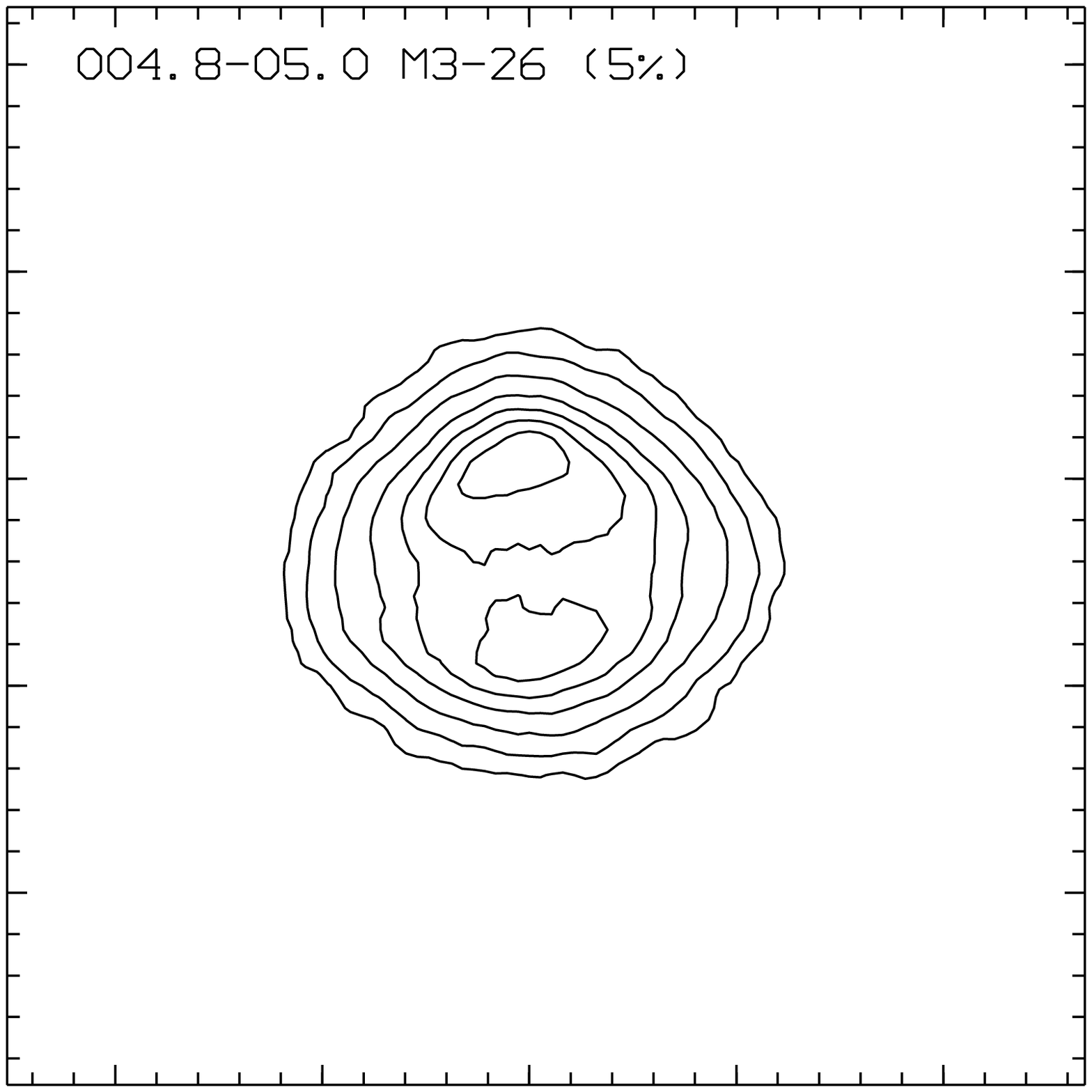}
          \hspace{4.5mm}
          \includegraphics[width=55mm]{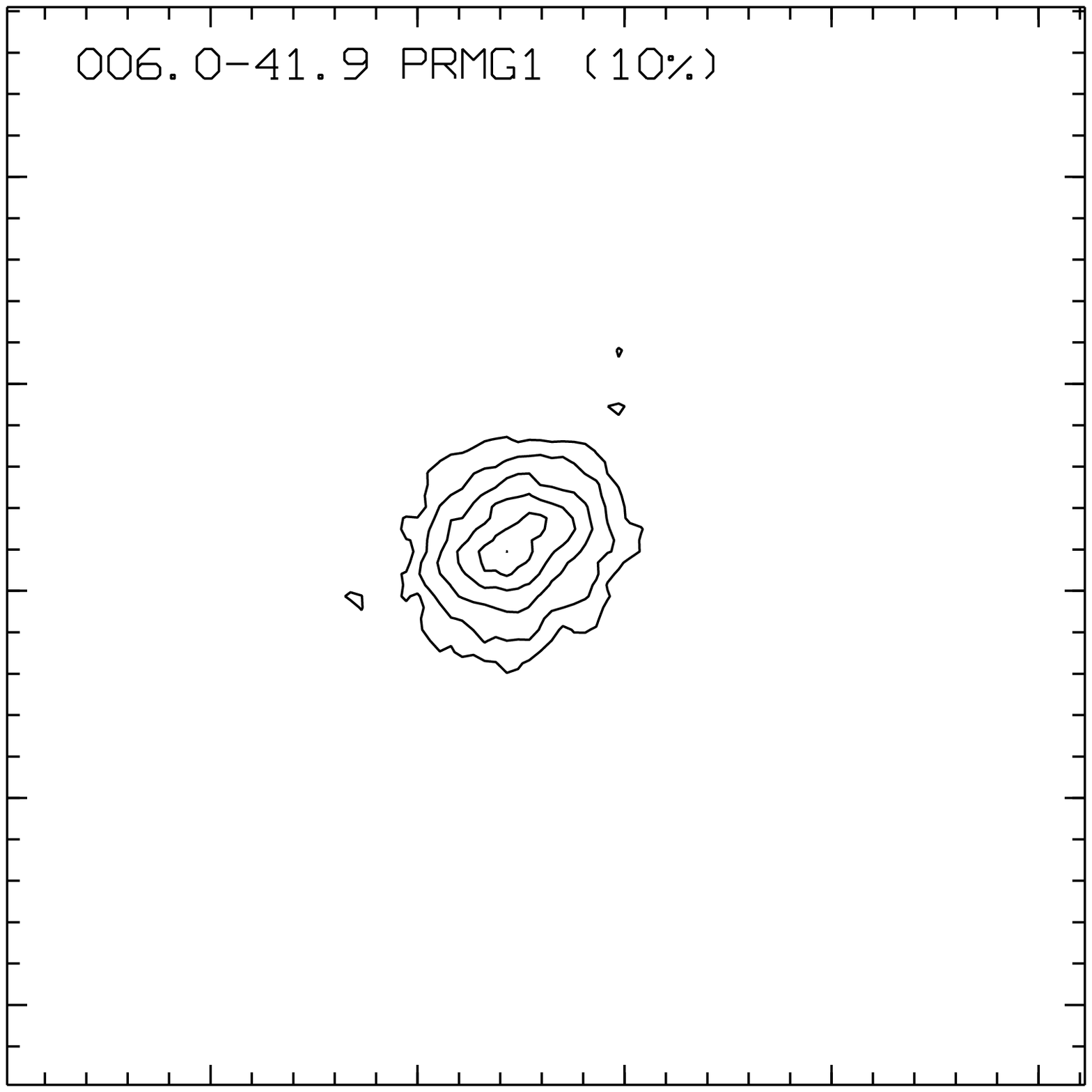}
          \hspace{4.5mm}
          \includegraphics[width=55mm]{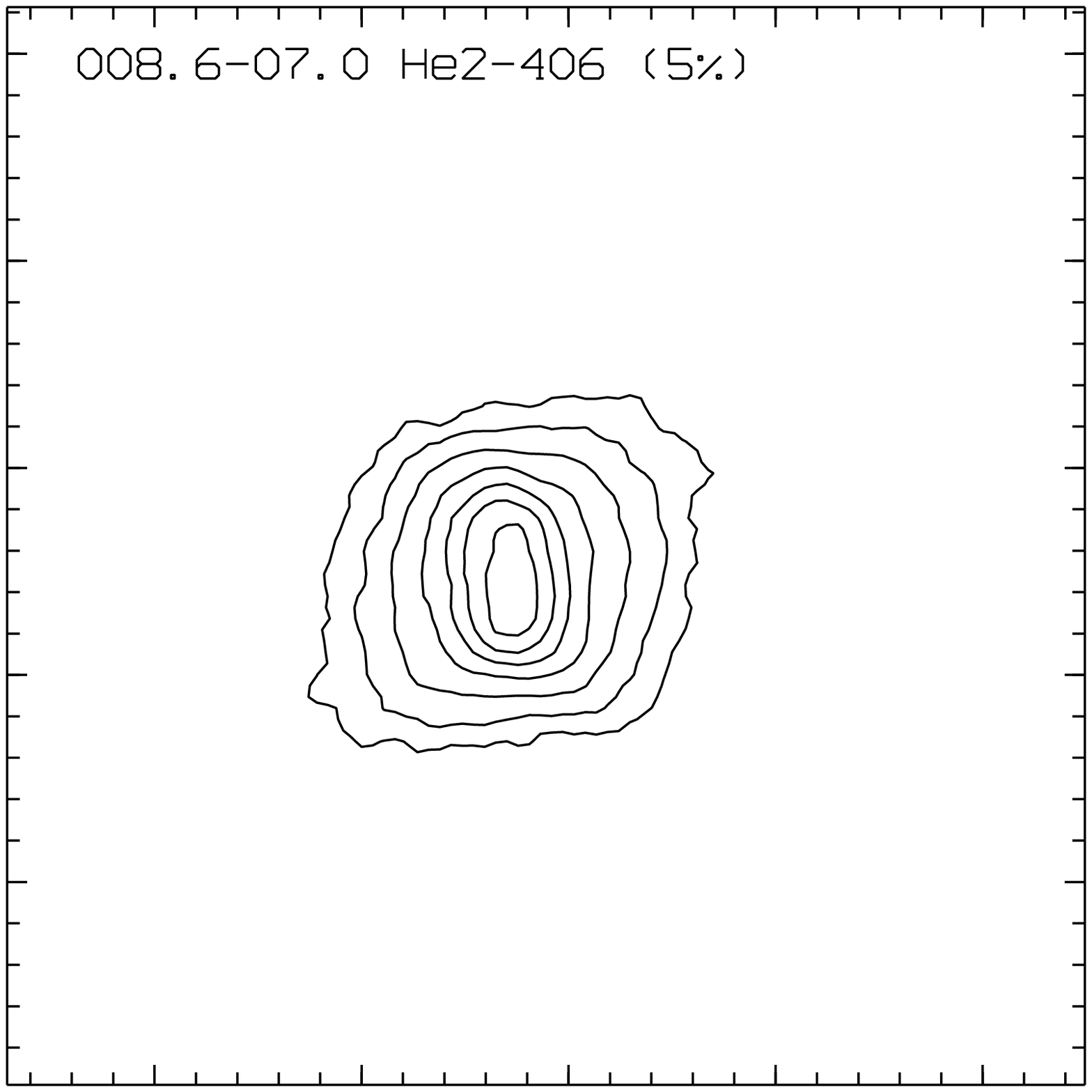}}
    \caption{H$\alpha$ images with contour levels at 80, 65, 50, 35, 20, 10, (5, 2, 1, 0.5, 0.2, 0.1) per cent of the nebula peak 
             (lowest indicated after PNe name). All images are 26\arcsec~$\times$~26\arcsec\ with north up and east to the left. 
             To filter the noise, the images have been smoothed with a 3 pixel (0\farcs81) box.}
    \label{plots}
\end{figure*}

\addtocounter{figure}{-1}

\begin{figure*}
    \hbox{\includegraphics[width=55mm]{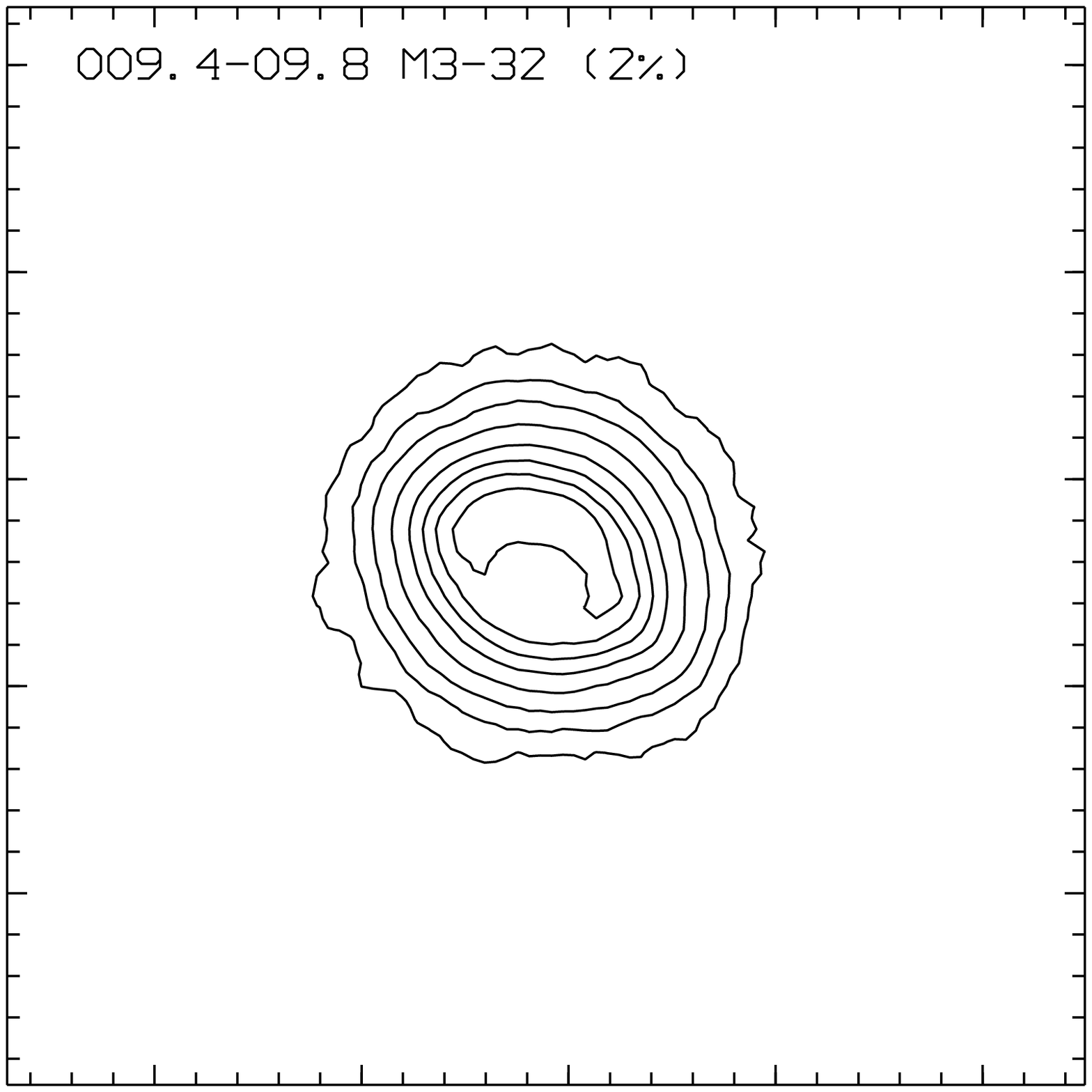}
          \hspace{4.5mm}
          \includegraphics[width=55mm]{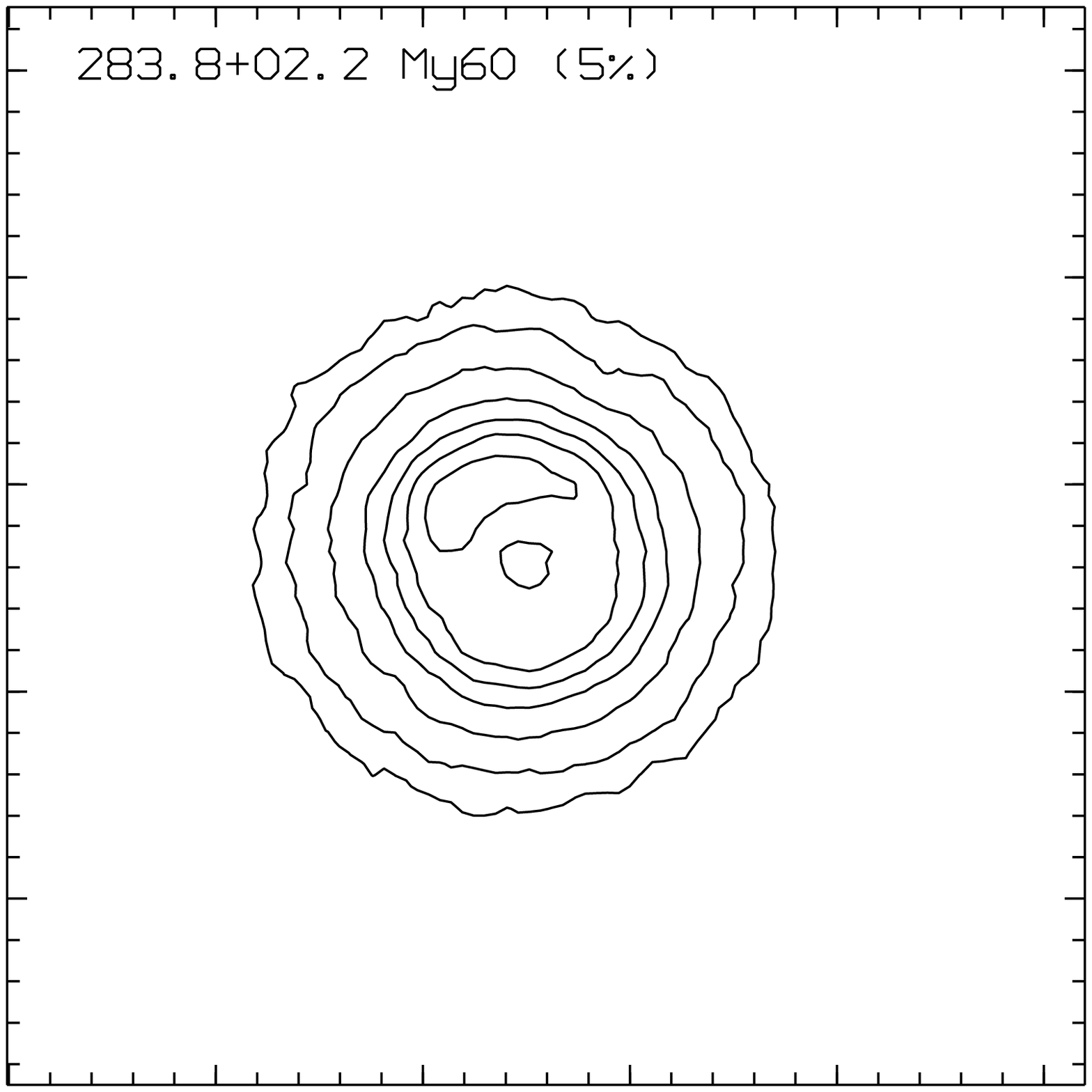}
          \hspace{4.5mm}
          \includegraphics[width=55mm]{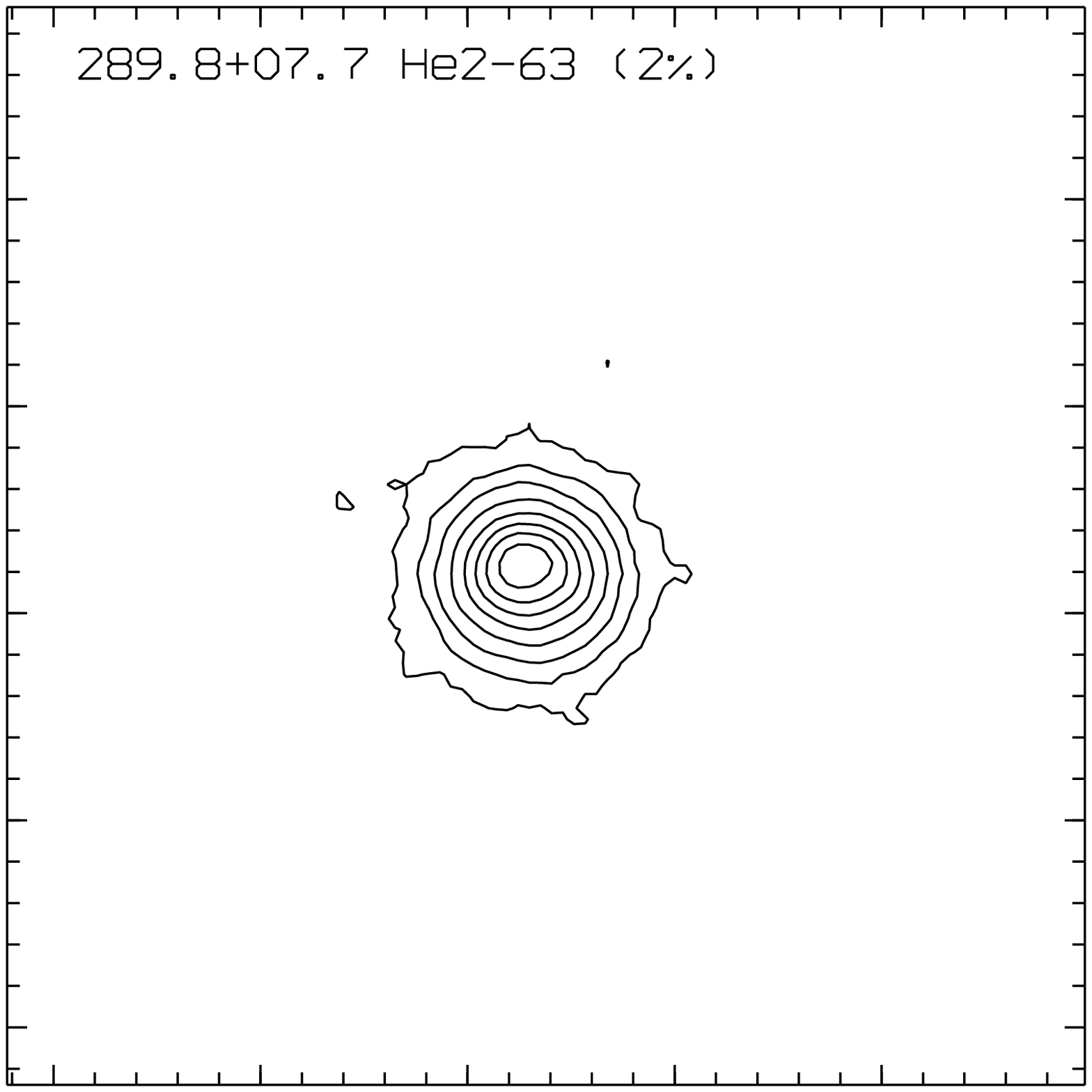}}
    \vspace{3mm}
    \hbox{\includegraphics[width=55mm]{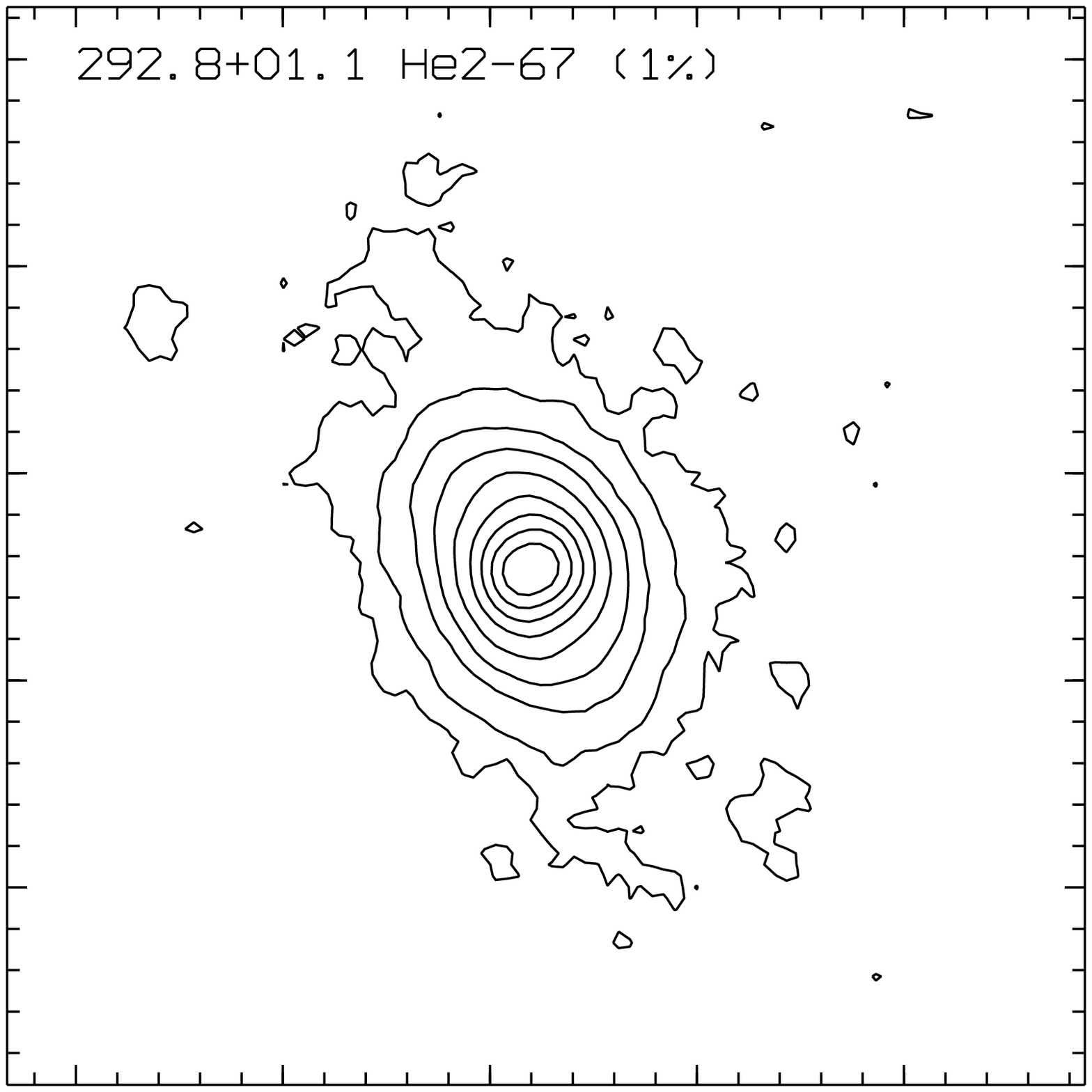}
          \hspace{4.5mm}
          \includegraphics[width=55mm]{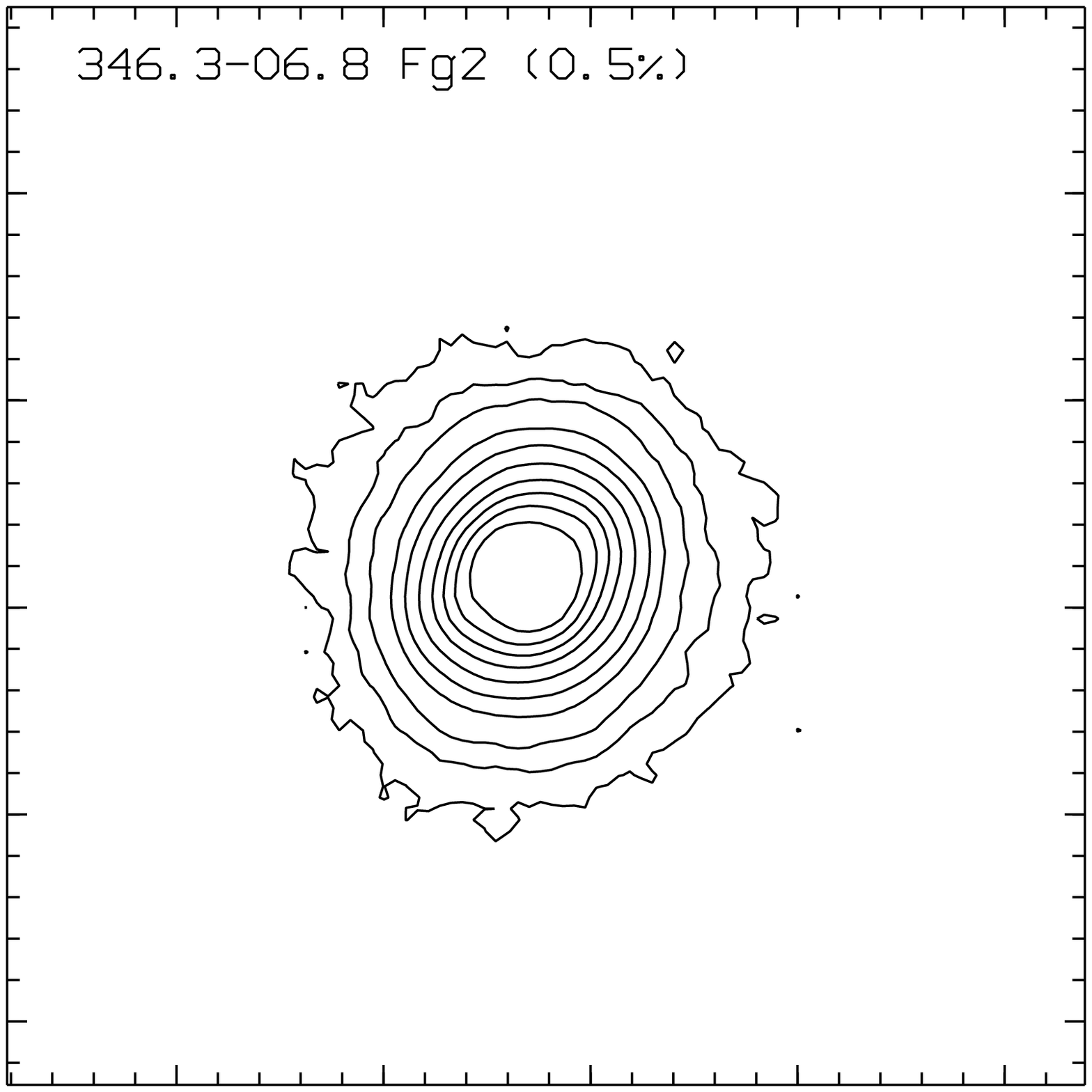}
          \hspace{4.5mm}
          \includegraphics[width=55mm]{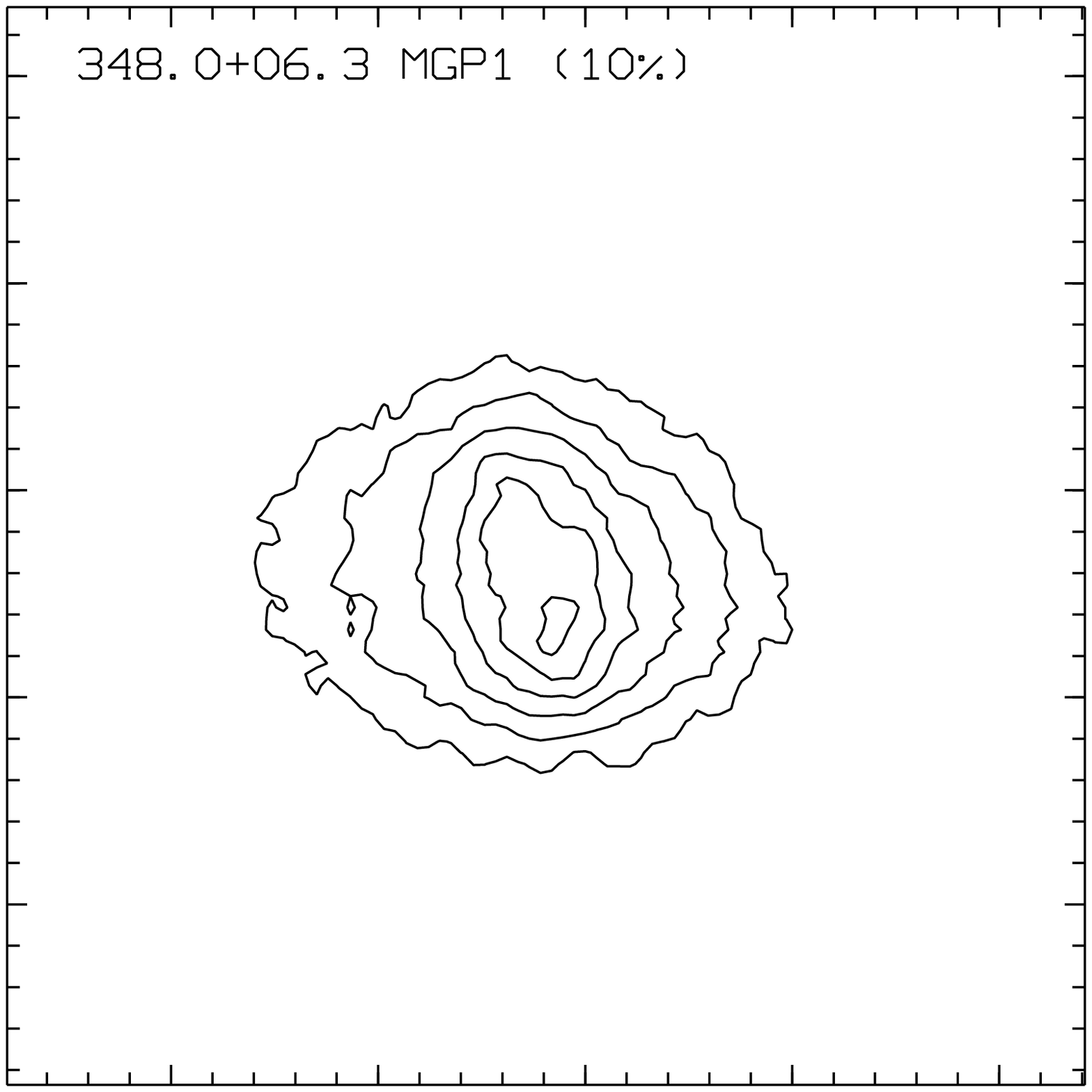}}
    \vspace{3mm}
    \hbox{\includegraphics[width=55mm]{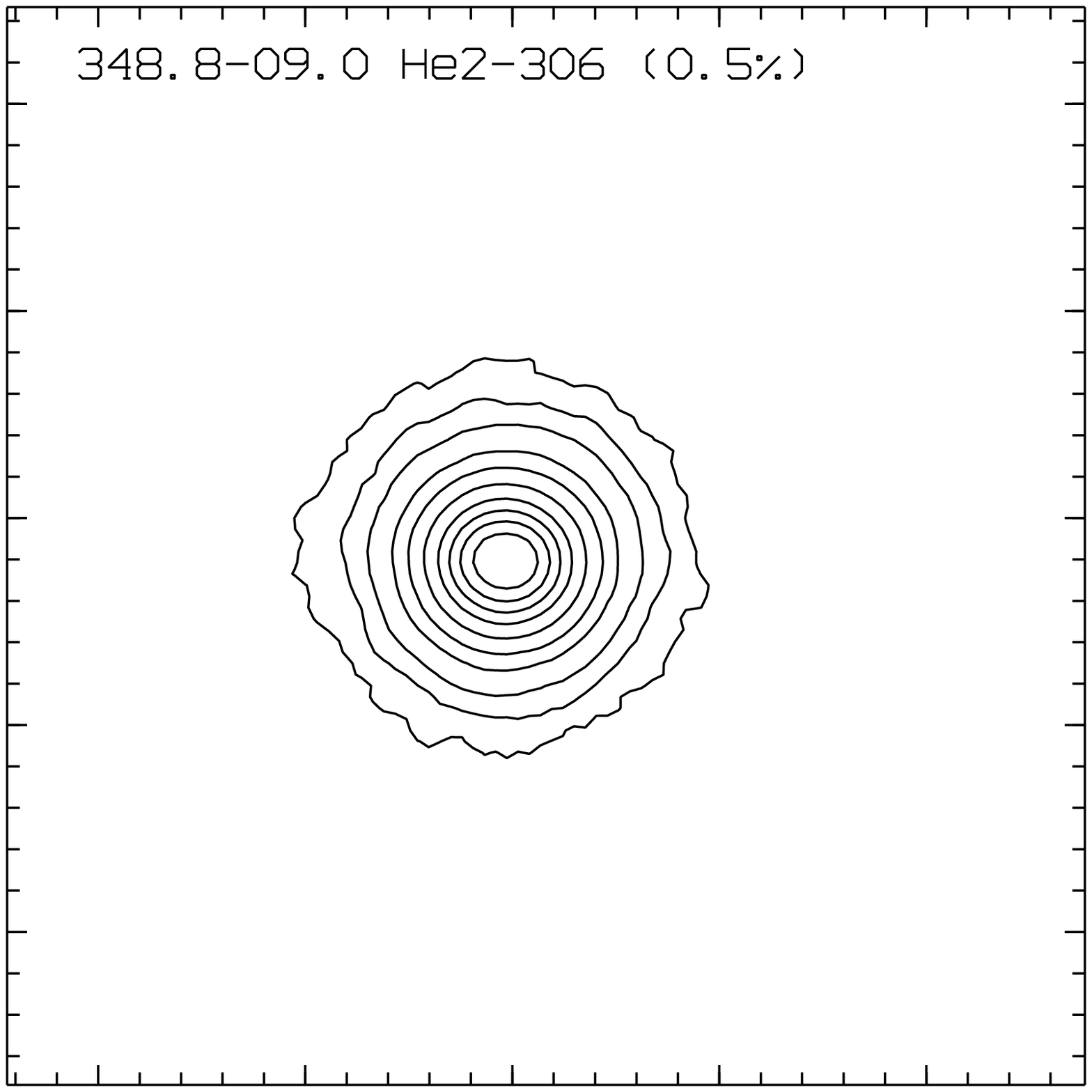}
          \hspace{4.5mm}
          \includegraphics[width=55mm]{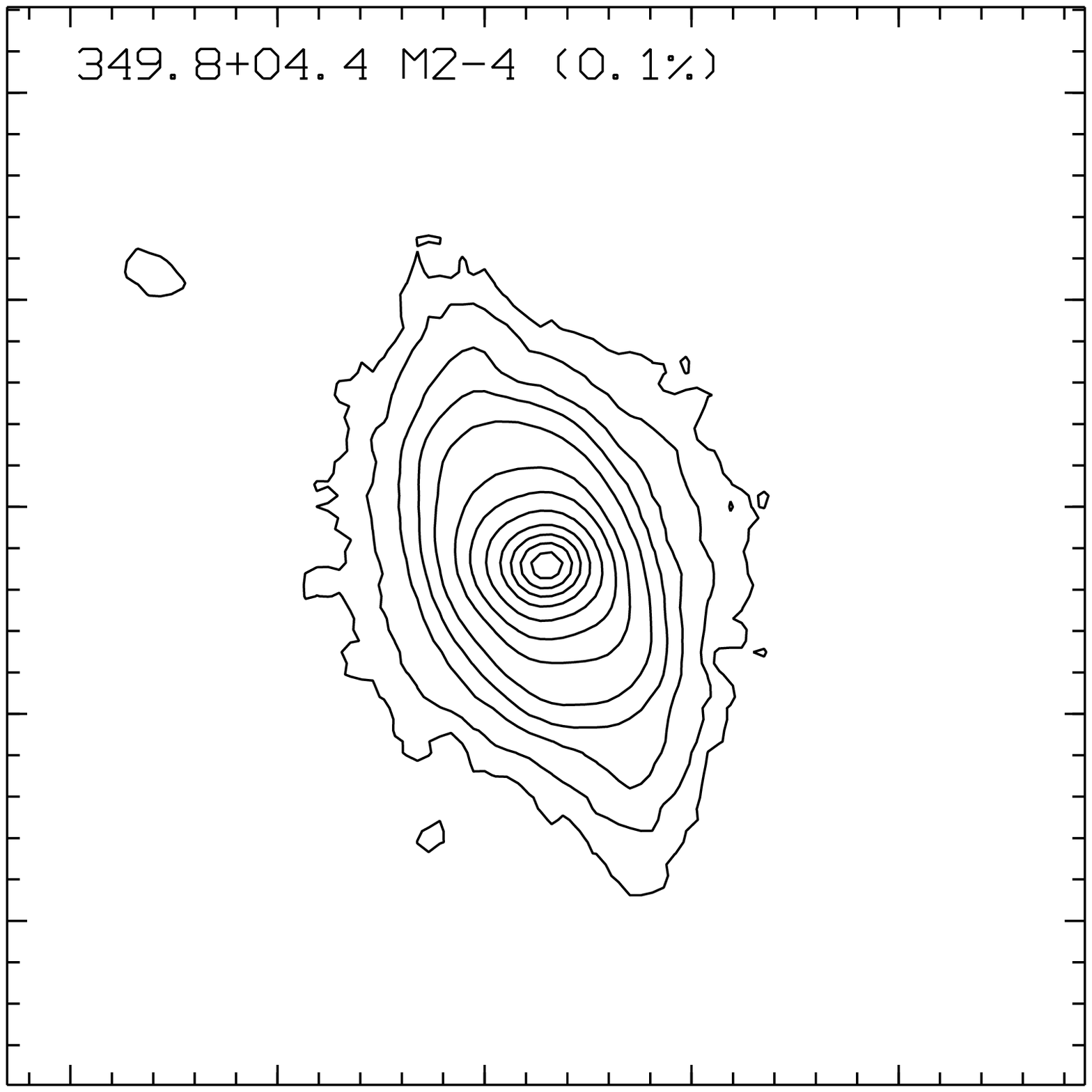}
          \hspace{4.5mm}
          \includegraphics[width=55mm]{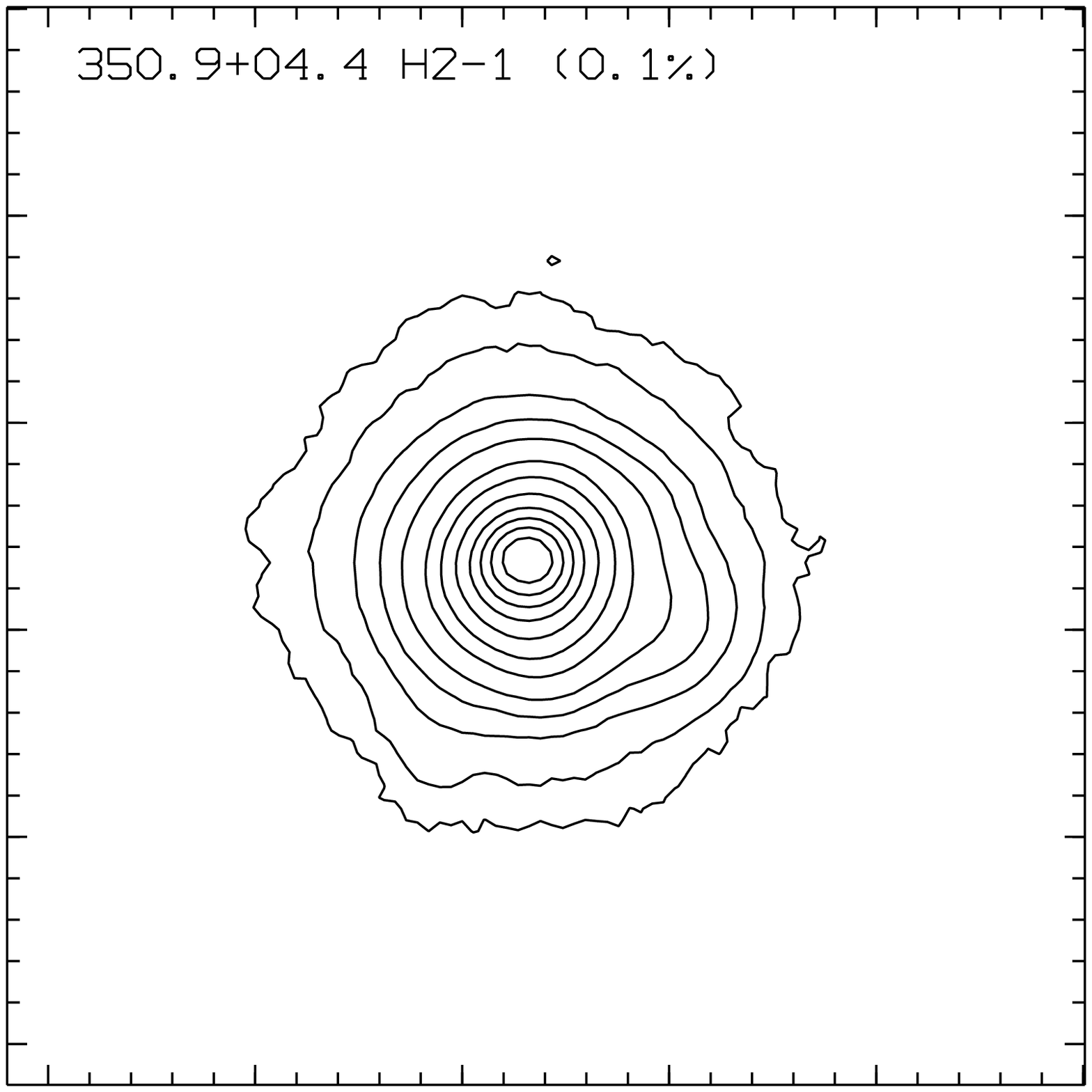}}
    \vspace{3mm}
    \hbox{\includegraphics[width=55mm]{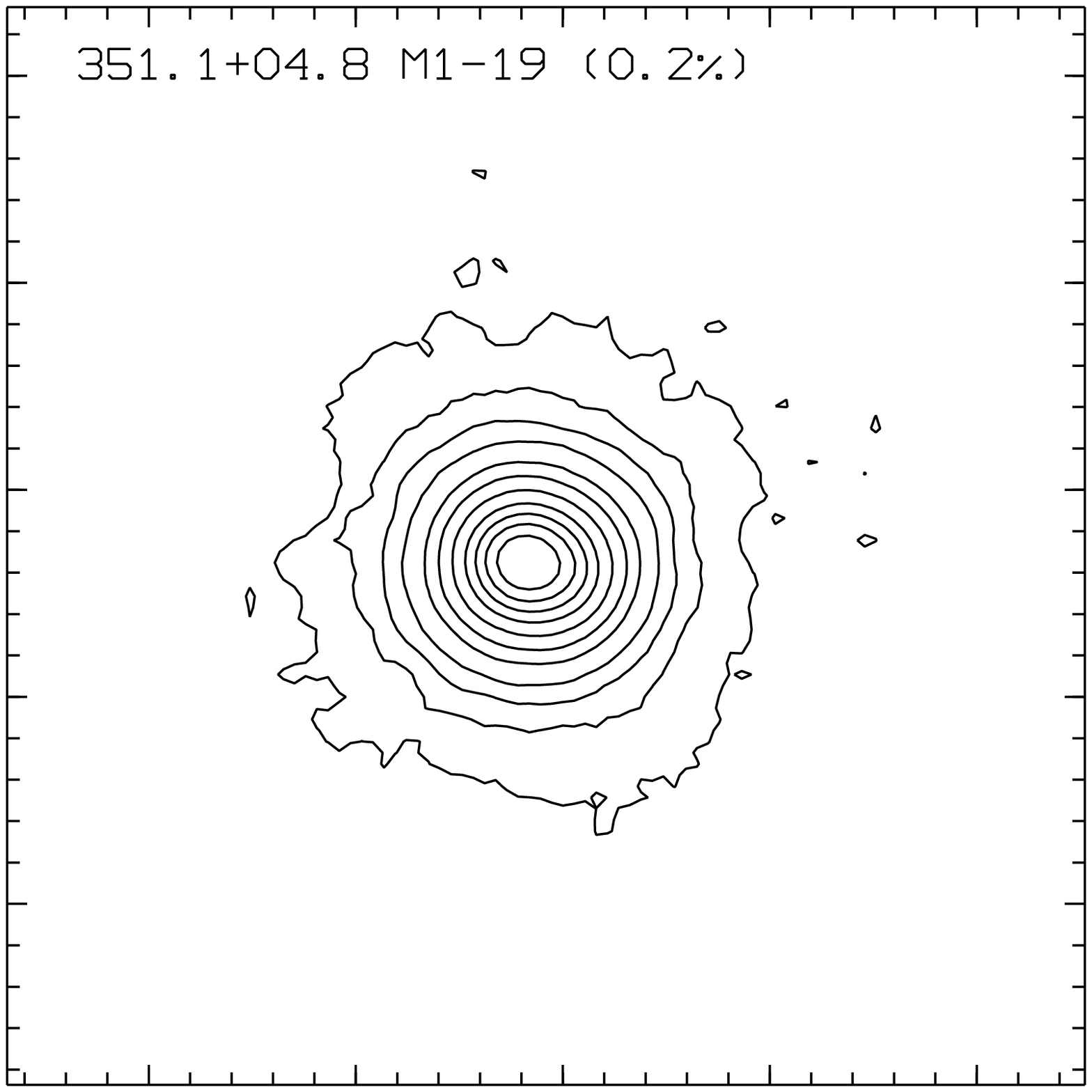}
          \hspace{4.5mm}
          \includegraphics[width=55mm]{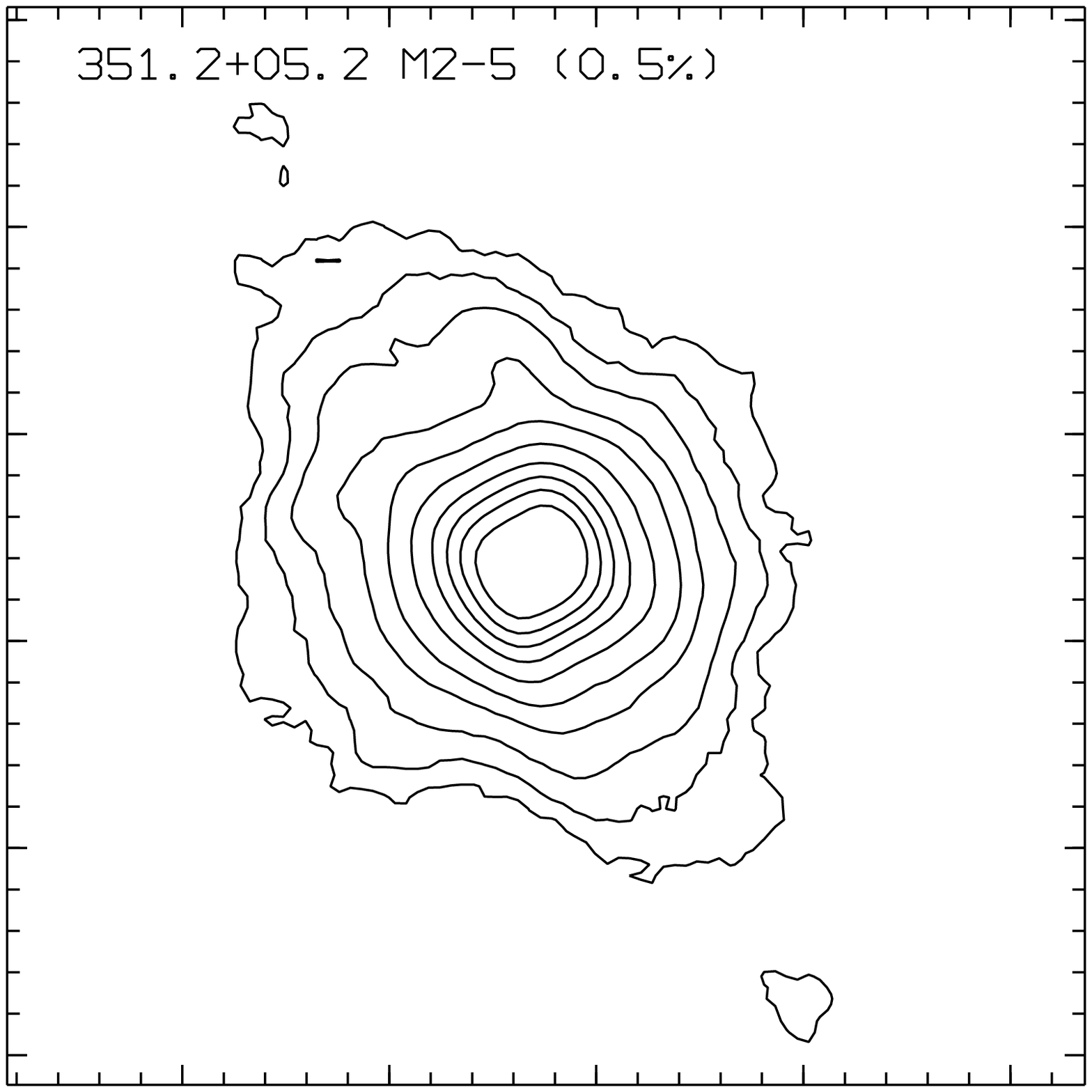}
          \hspace{4.5mm}
          \includegraphics[width=55mm]{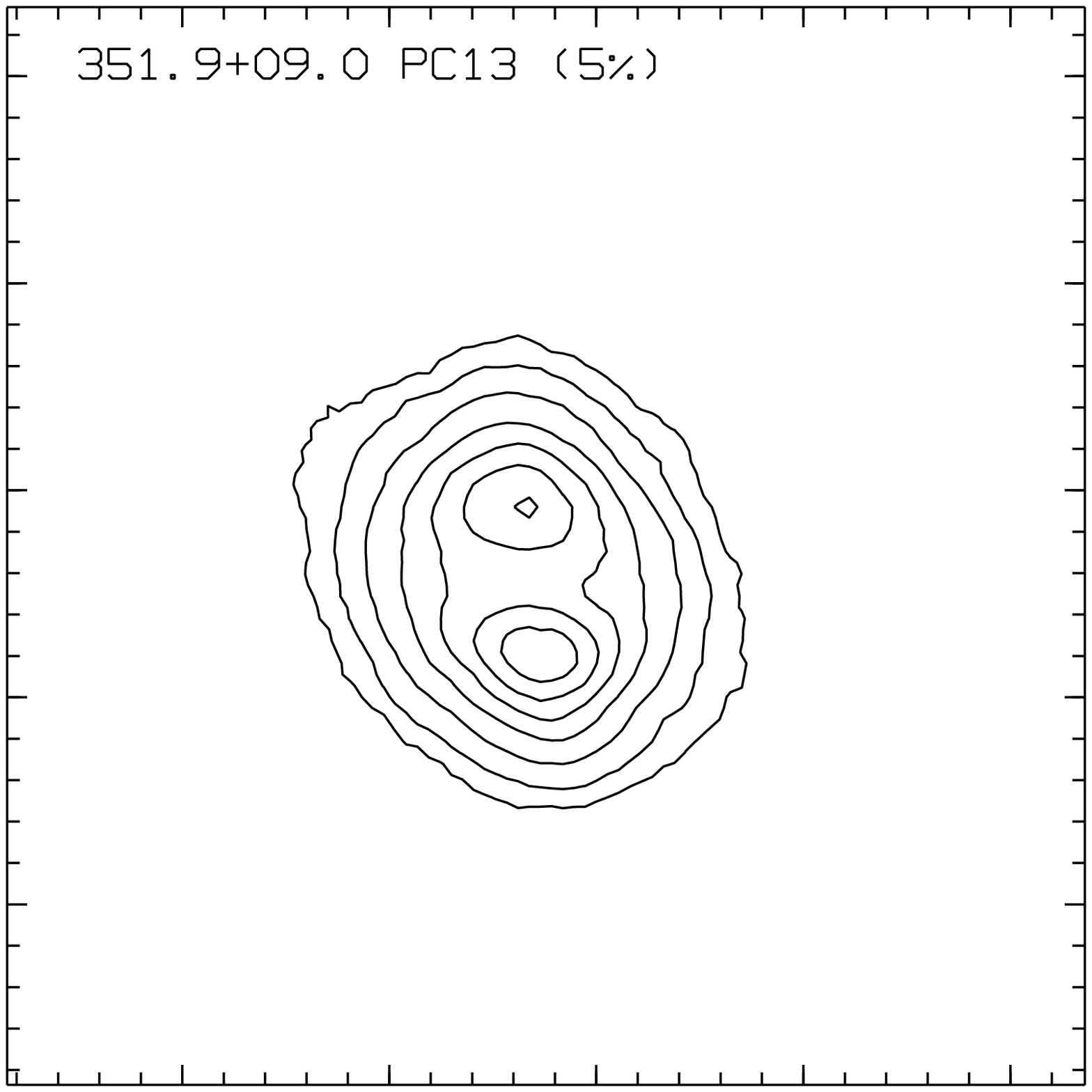}}
    \caption{{\bf(continued)} 
             H$\alpha$ images with contour levels at 80, 65, 50, 35, 20, 10, (5, 2, 1, 0.5, 0.2, 0.1) per cent of the nebula peak. 
             Actual lowest contour level plotted dependent on signal-to-noise ratio, and indicated after PNe name.
             The images have been smoothed with a 3 pixel (0\farcs81) box.}
    \label{plots2}
\end{figure*}

\addtocounter{figure}{-1}

\begin{figure*}
    \hbox{\includegraphics[width=55mm]{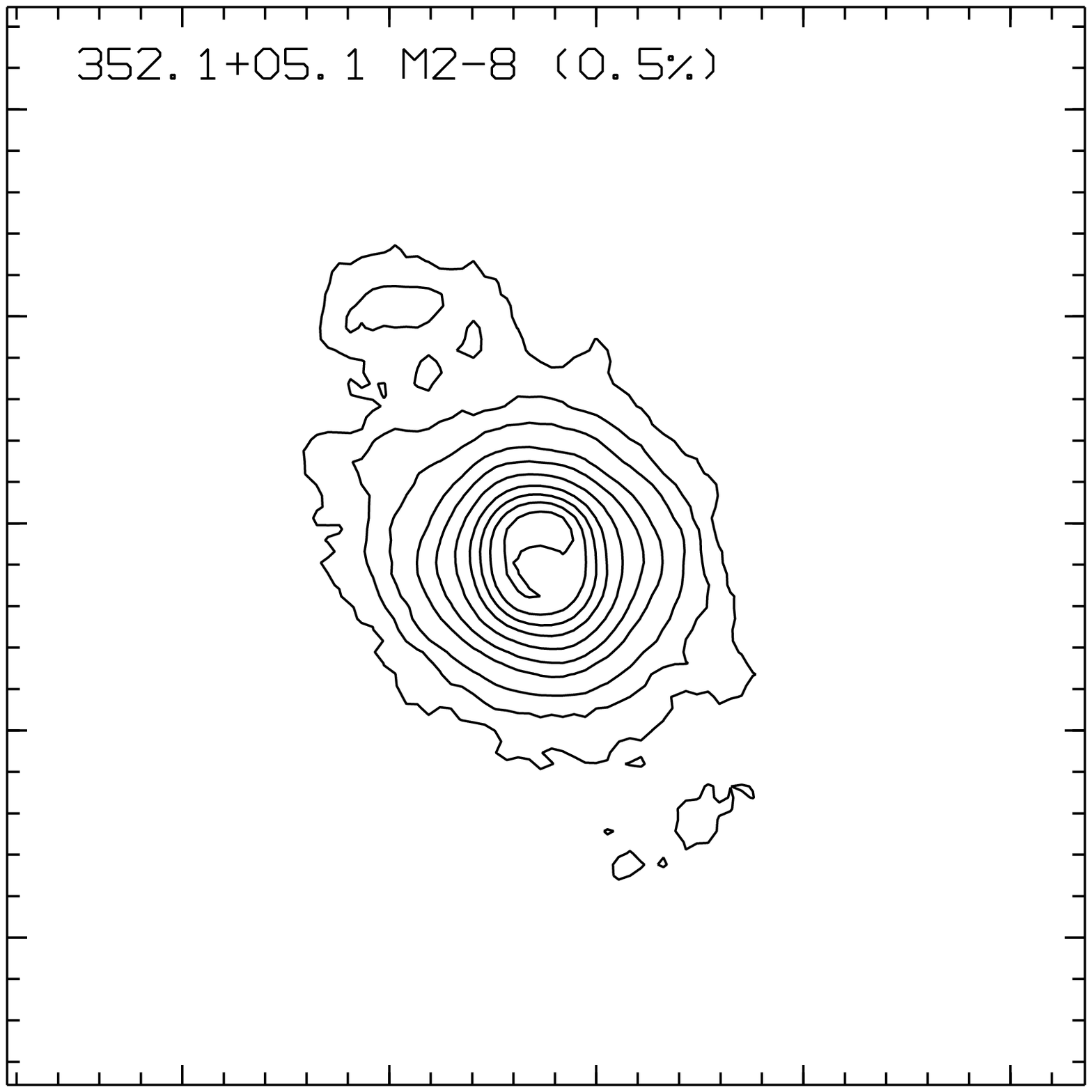}
          \hspace{4.5mm}
          \includegraphics[width=55mm]{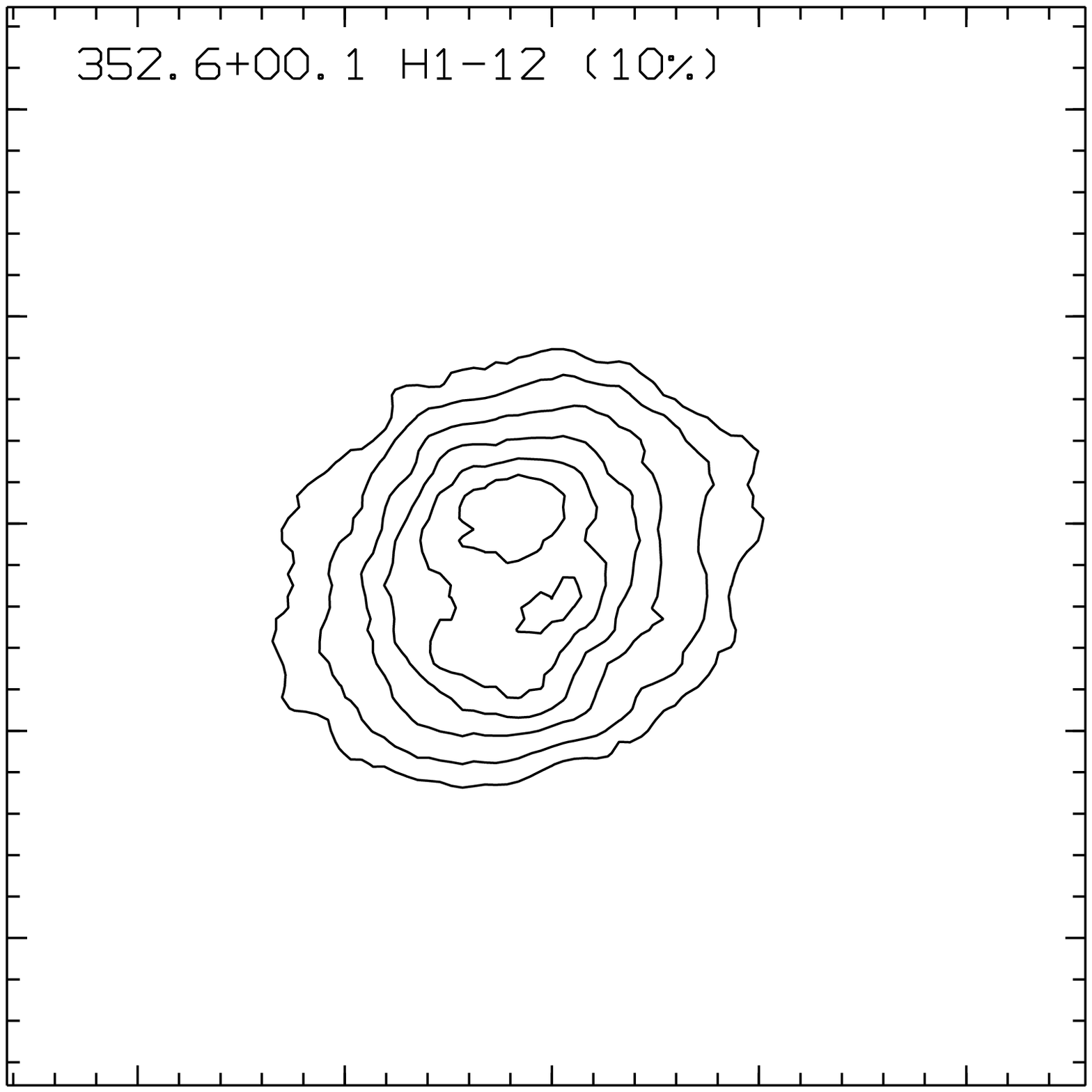}
          \hspace{4.5mm}
          \includegraphics[width=55mm]{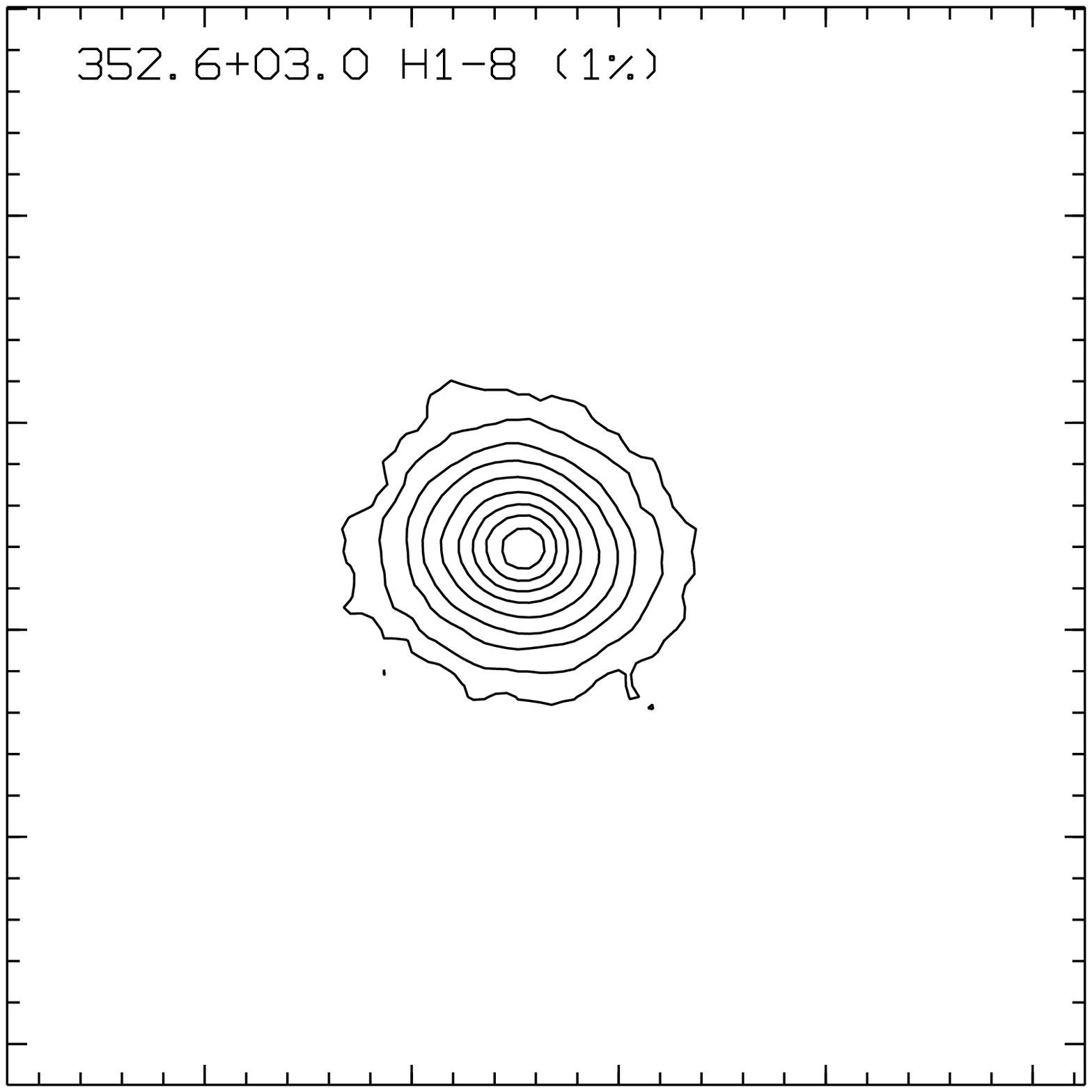}}
    \vspace{3mm}
    \hbox{\includegraphics[width=55mm]{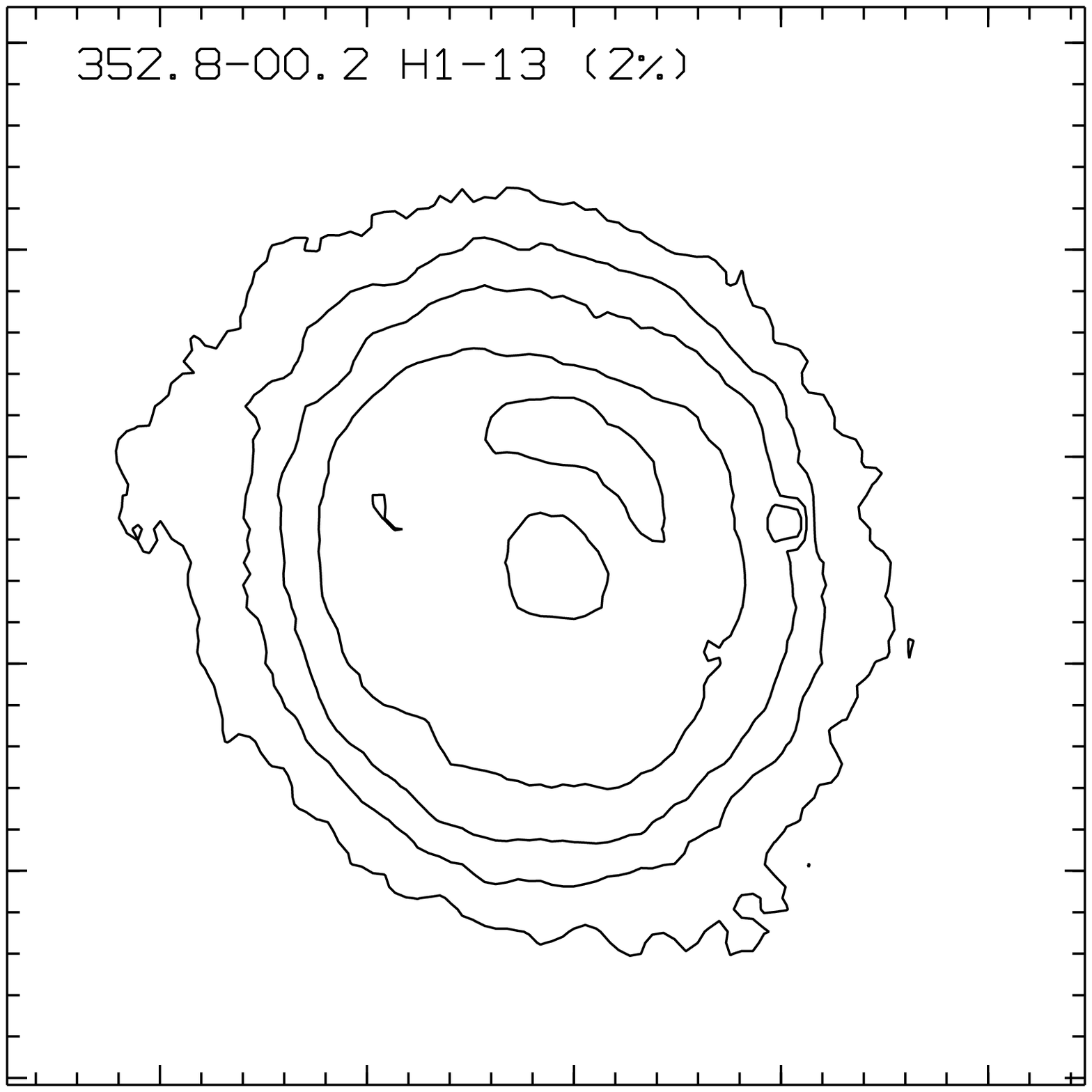}
          \hspace{4.5mm}
          \includegraphics[width=55mm]{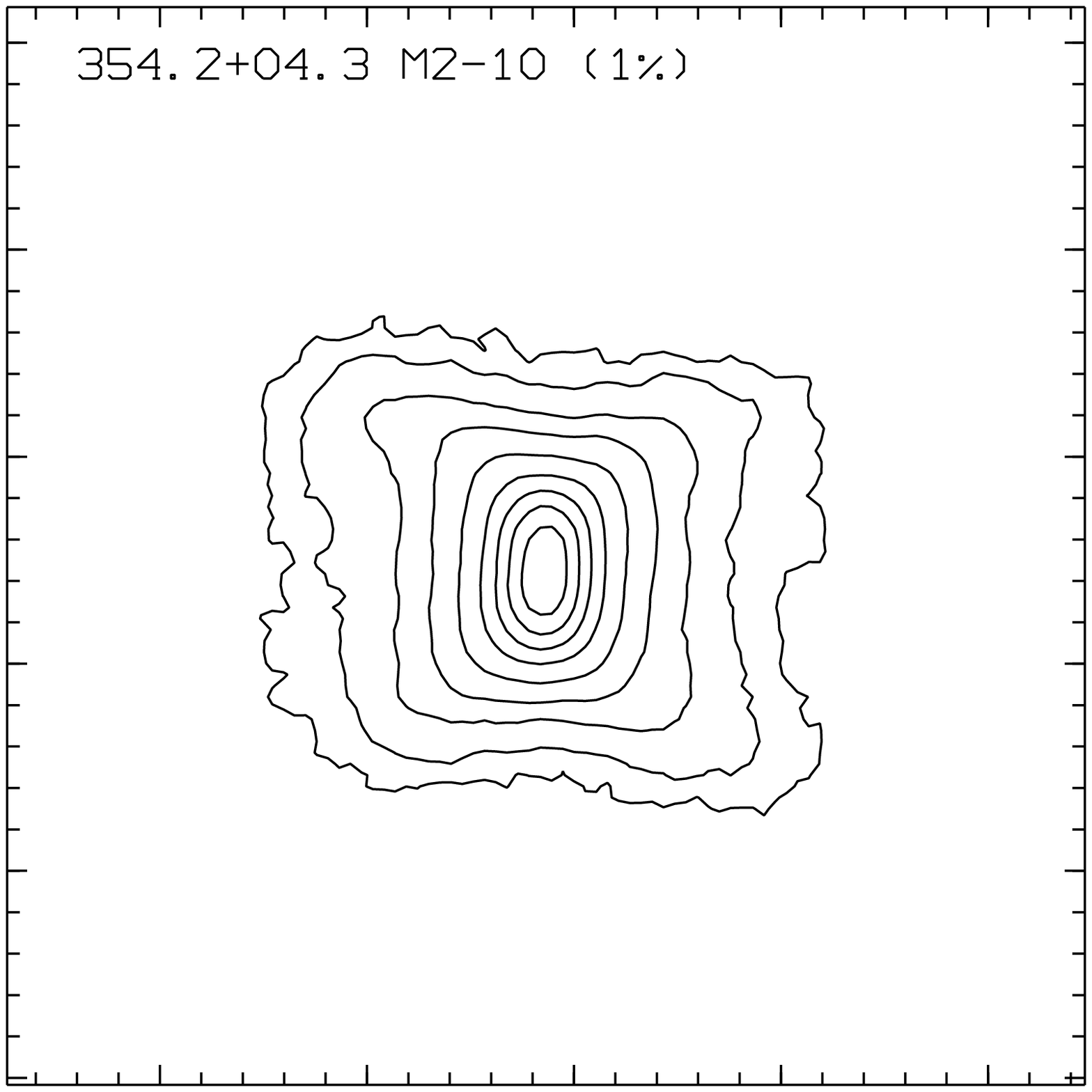}
          \hspace{4.5mm}
          \includegraphics[width=55mm]{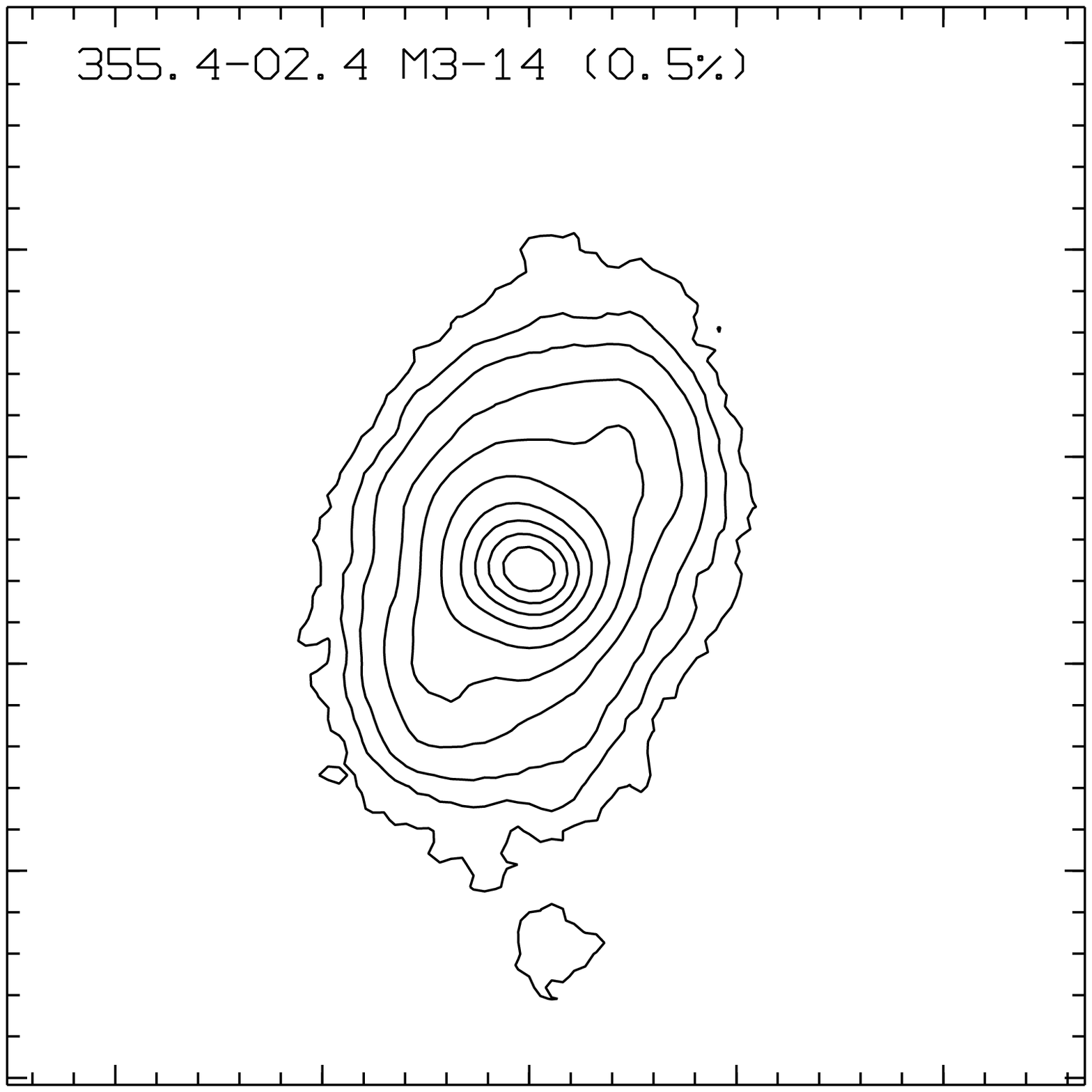}}
    \vspace{3mm}
    \hbox{\includegraphics[width=55mm]{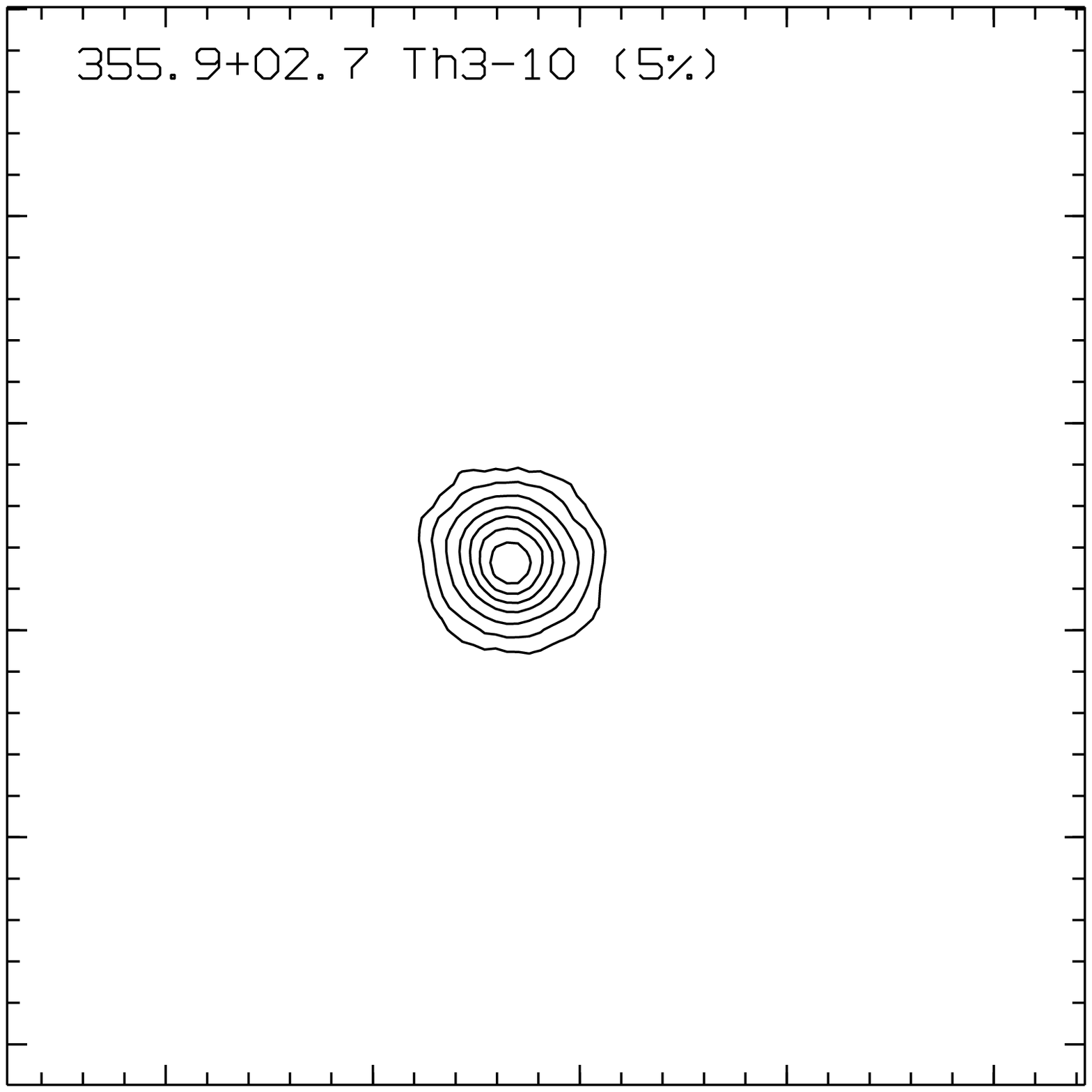}
          \hspace{4.5mm}
          \includegraphics[width=55mm]{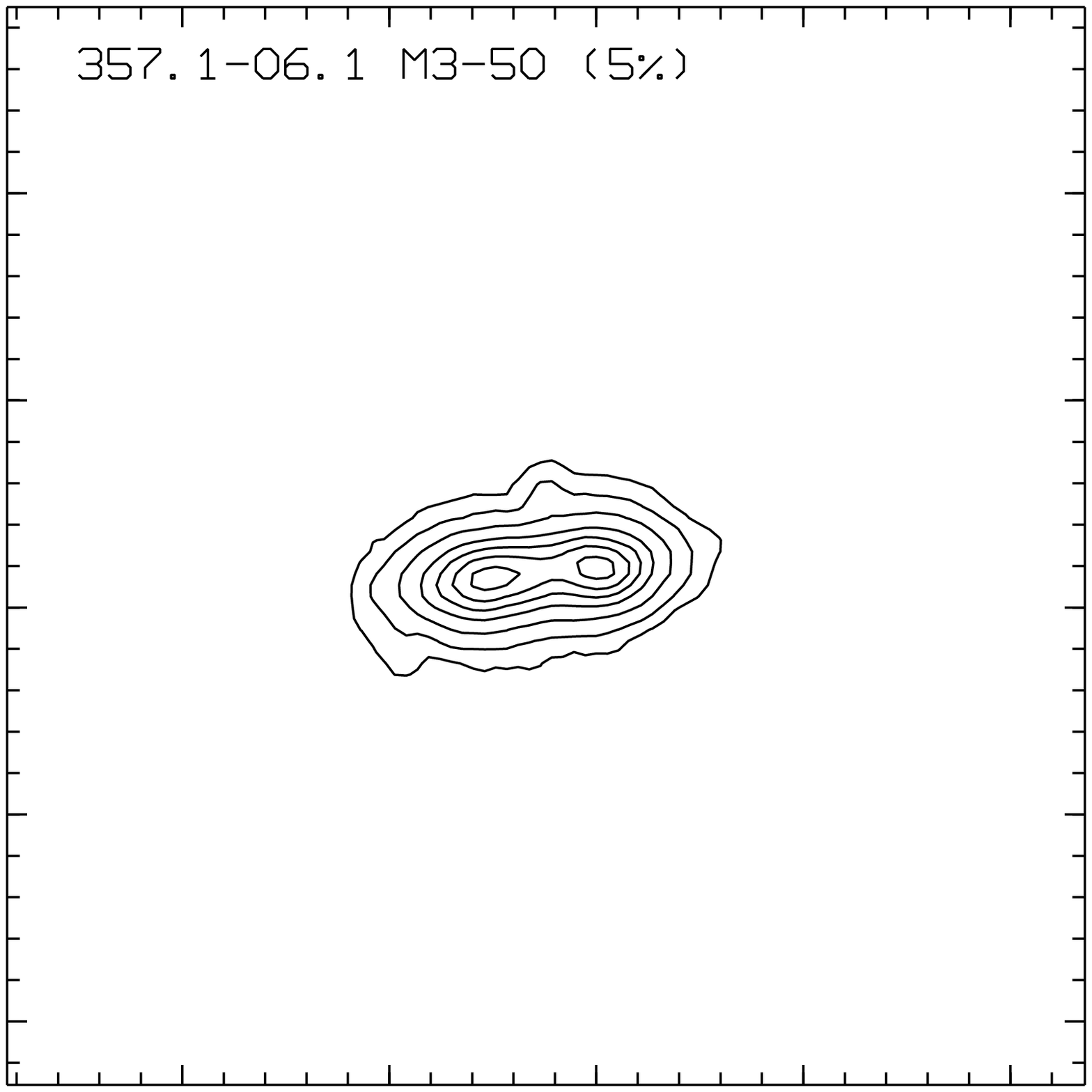}
          \hspace{4.5mm}
          \includegraphics[width=55mm]{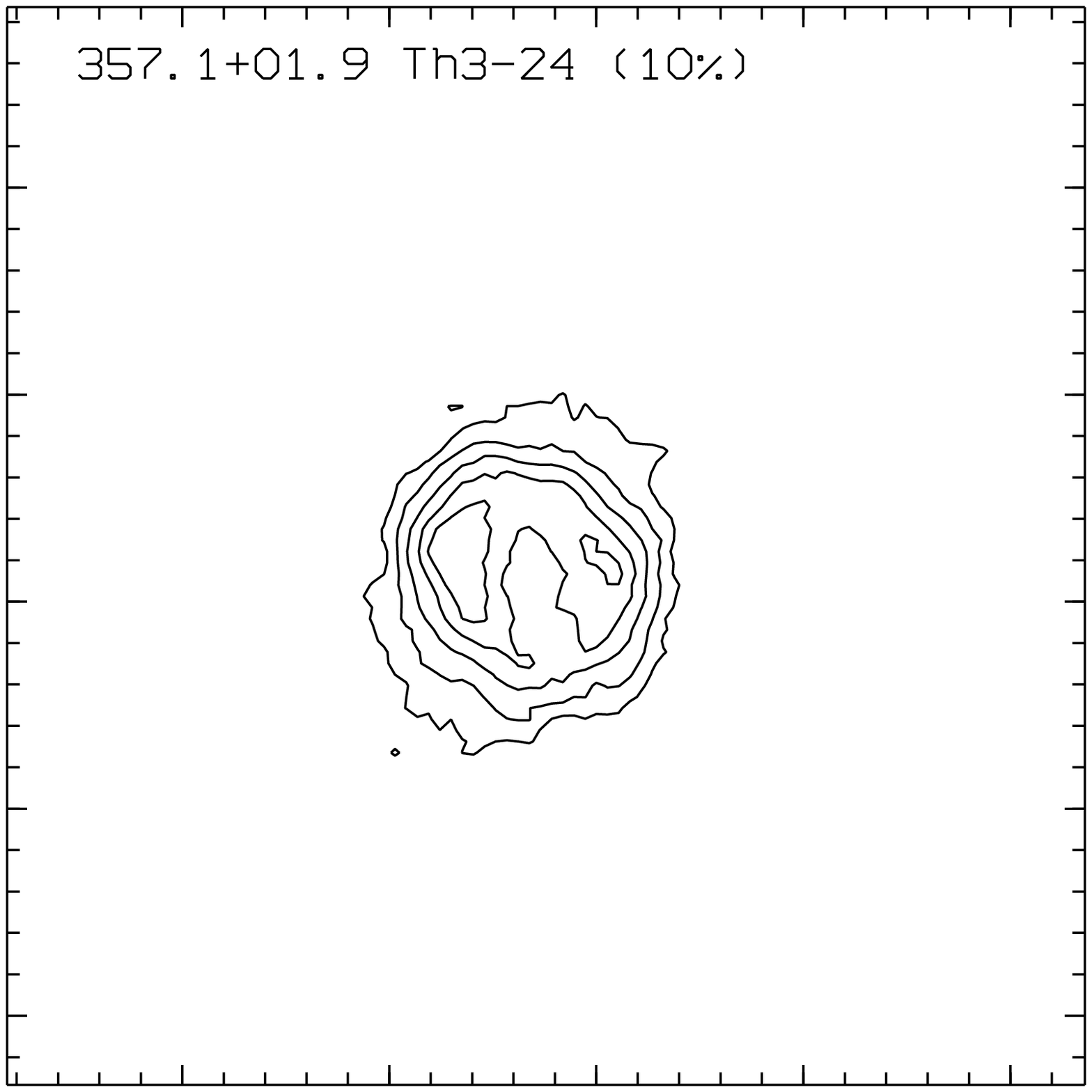}}
    \vspace{3mm}
    \hbox{\includegraphics[width=55mm]{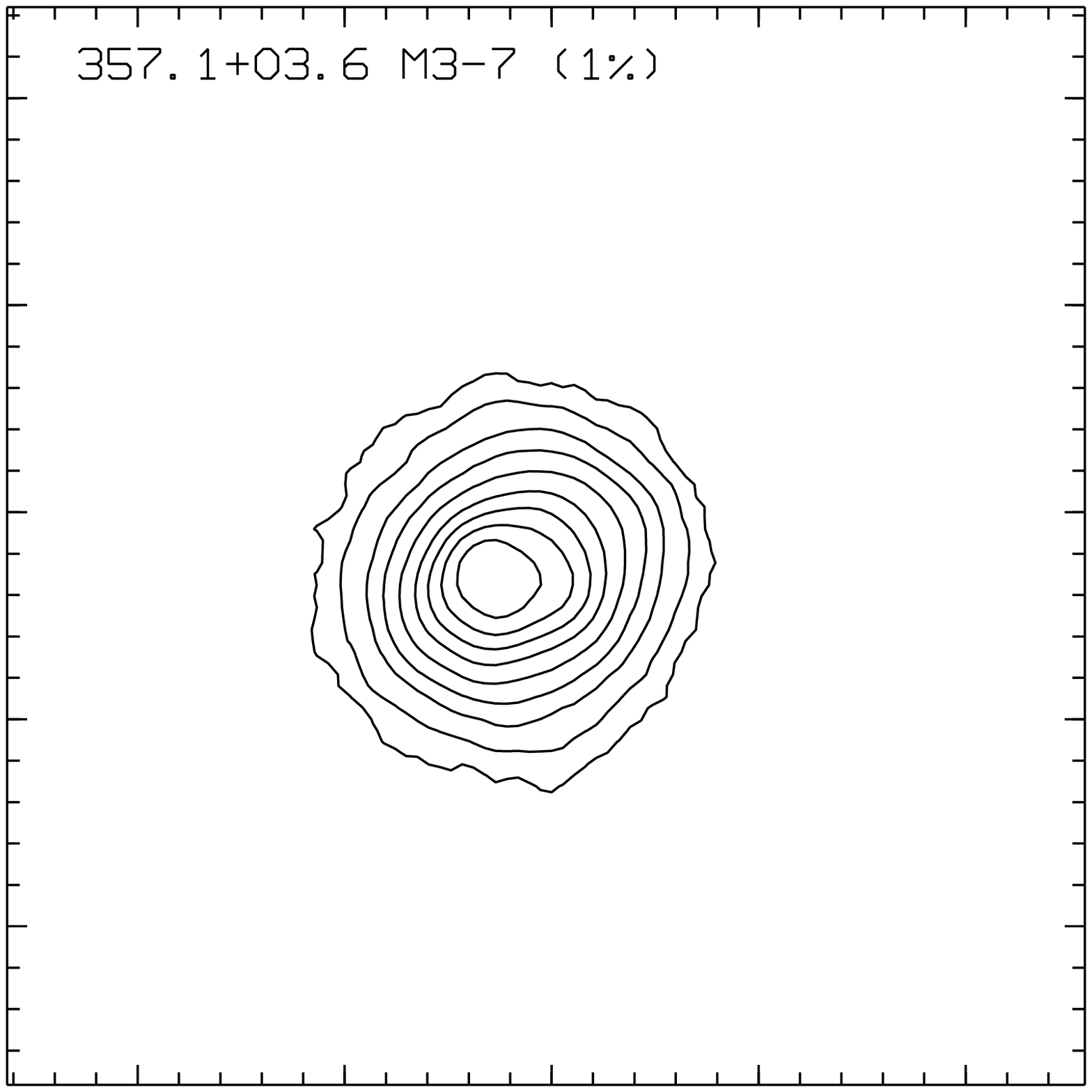}
          \hspace{4.5mm}
          \includegraphics[width=55mm]{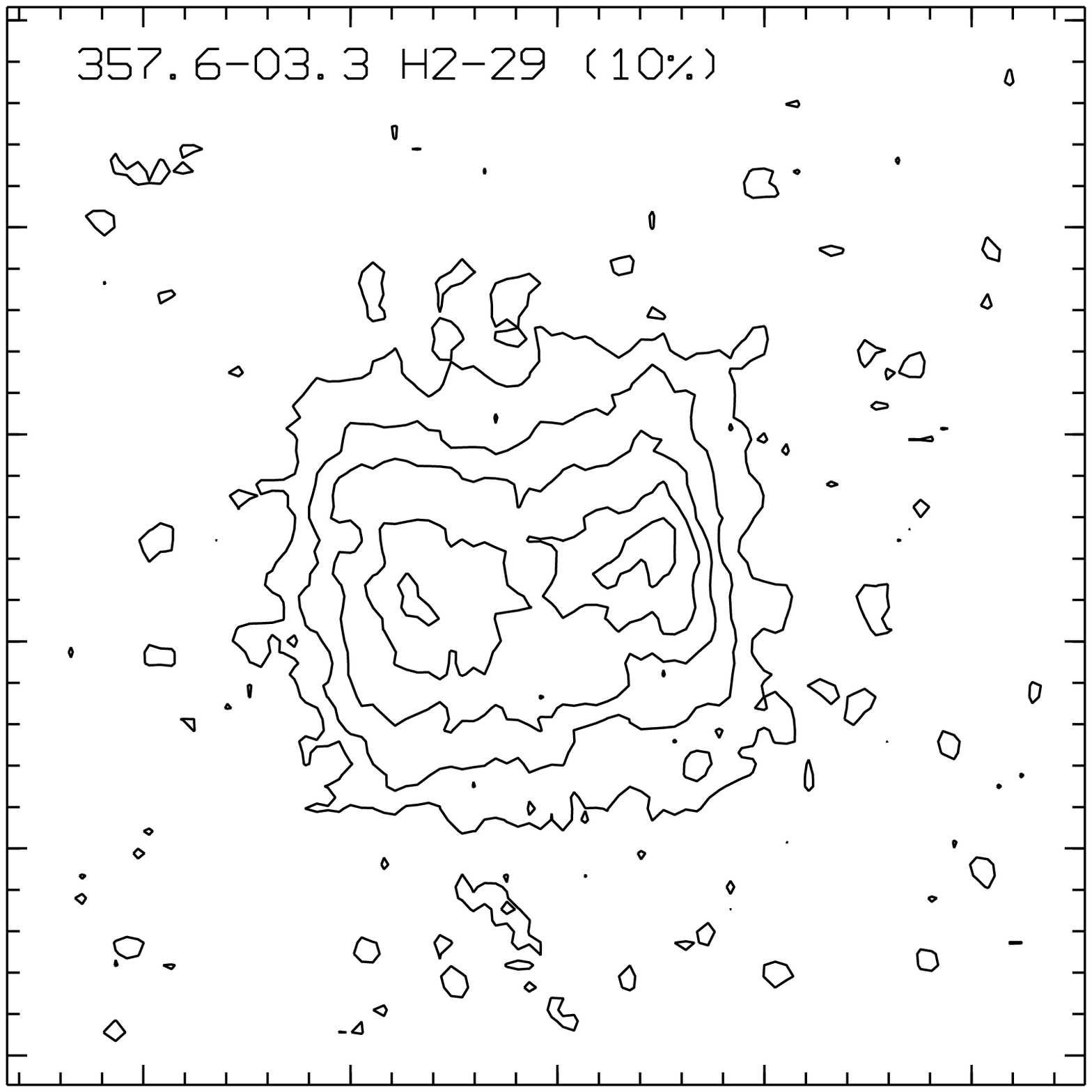}
          \hspace{4.5mm}
          \includegraphics[width=55mm]{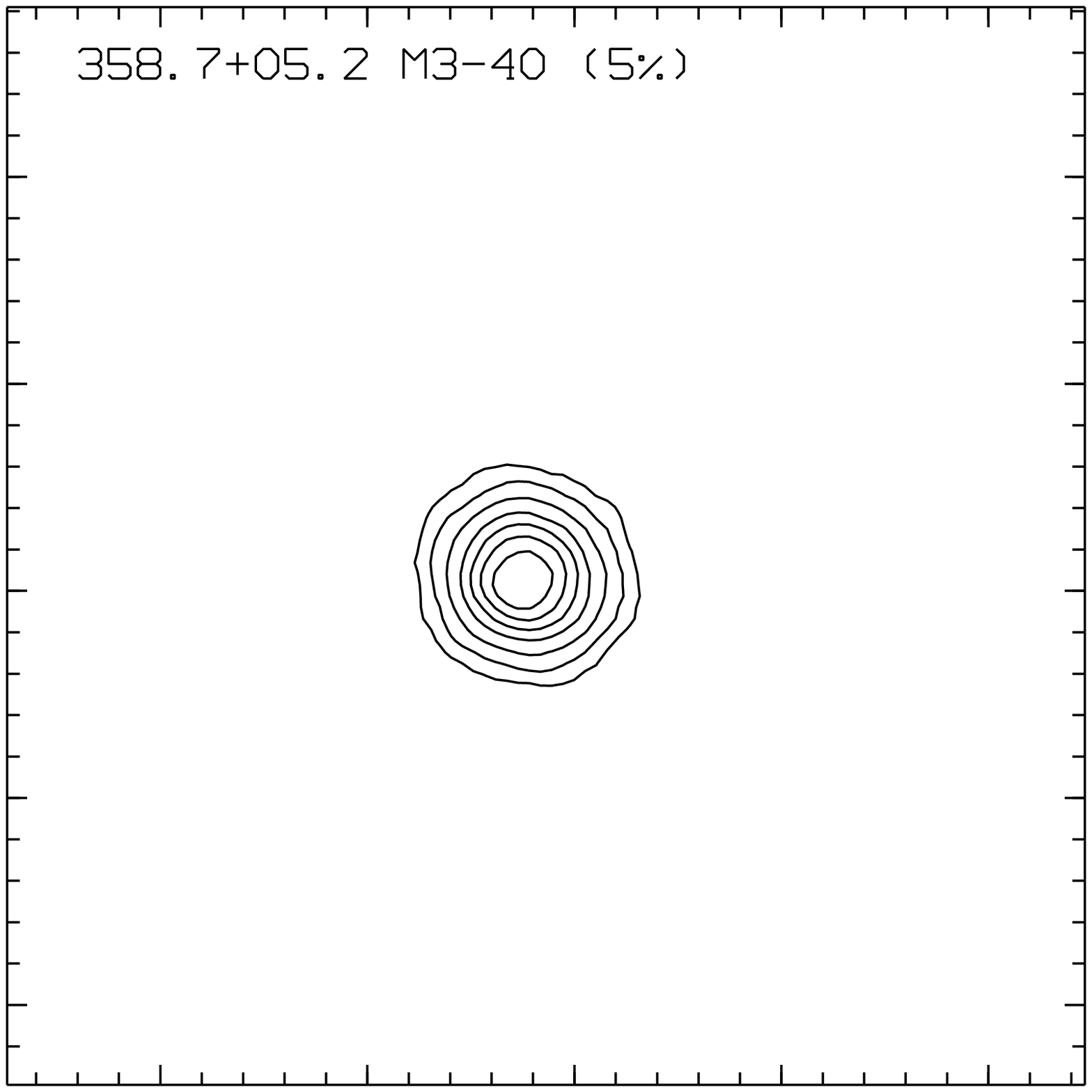}}
    \caption{{\bf(continued)}
             H$\alpha$ images with contour levels at 80, 65, 50, 35, 20, 10, (5, 2, 1, 0.5, 0.2, 0.1) per cent of the nebula peak. 
             Actual lowest contour level plotted dependent on signal-to-noise ratio, and indicated after PNe name.
             The images have been smoothed with a 3 pixel (0\farcs81) box.}
    \label{plots3}
\end{figure*}

\begin{figure}
    \hbox{\hspace{5mm}\includegraphics[width=70mm]{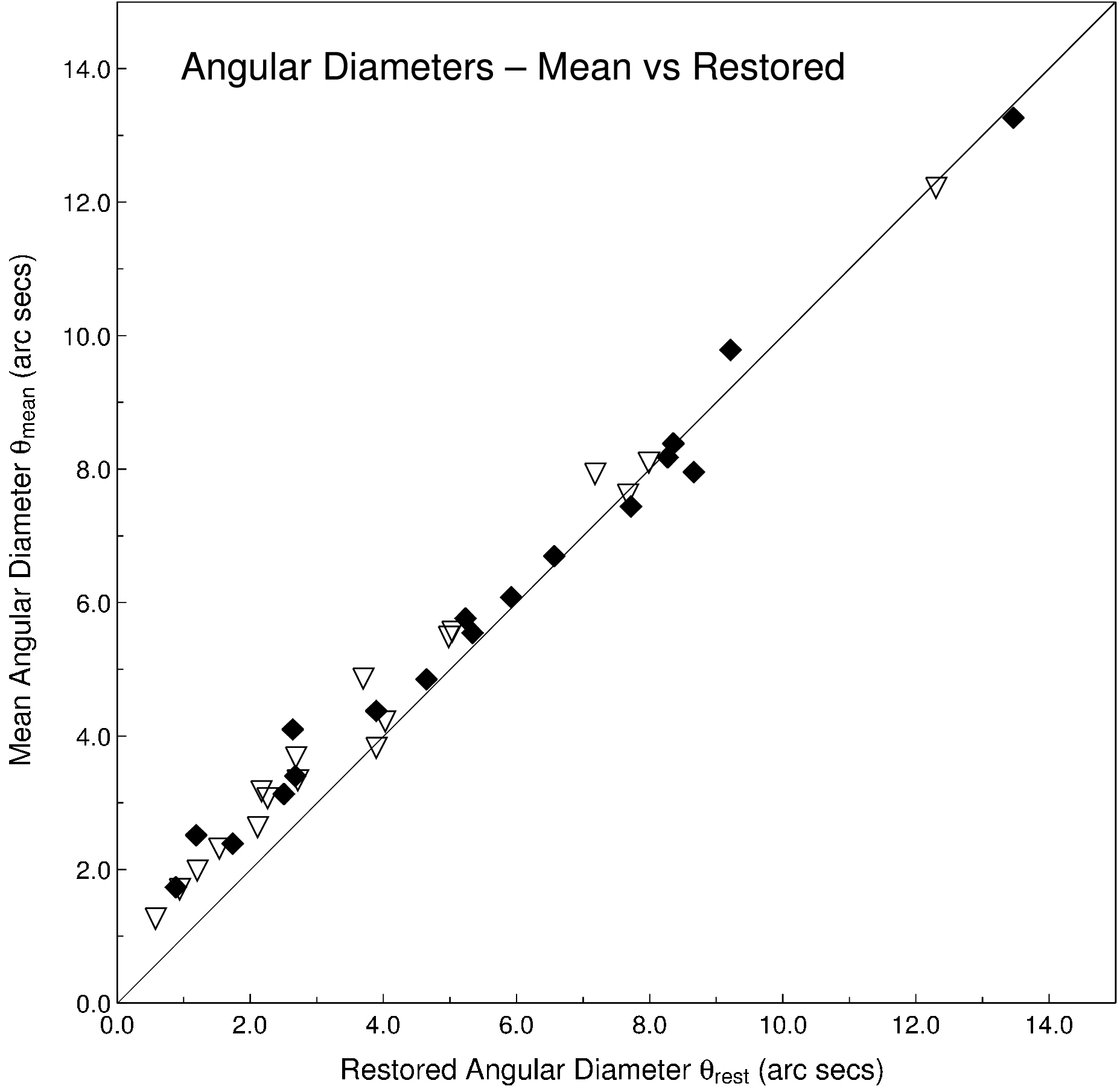}}
    \caption{Comparison of H$\alpha$ angular diameters $\theta_\mathrm{mean}$ and $\theta_\mathrm{rest}$.
             Diamonds denote major axis diameters, triangles minor axis diameters.}
    \label{meanvsrestdias}
\end{figure}

\begin{table*}
\caption[]{Angular diameters of 69 planetary nebulae. Diameters are in arc seconds and are defined in section~\ref{diams}.}
\label{diamsa}
\setlength{\tabcolsep}{4.6pt}
\begin{tabular}{llllccrcccd{2.1}rrl}
\vspace{-10pt} \\
\hline
\vspace{-6pt} \\
PN G & Object & Line & 
\multicolumn{1}{c}{$\theta_\mathrm{PSF}{}^1$} & 
\multicolumn{1}{c}{$\theta_\mathrm{gauss}$}   & 
\multicolumn{1}{c}{$\theta_\mathrm{shell}$}   & 
\multicolumn{1}{c}{$\theta_\mathrm{line}$}    & 
\multicolumn{2}{c}{$\theta_\mathrm{mean}$}    & 
\multicolumn{1}{c}{$\theta_\mathrm{10\%}$}    & 
\multicolumn{1}{c}{$\theta_\mathrm{rad}$}     & 
\multicolumn{1}{c}{$\theta_\mathrm{opt}$}     &
\multicolumn{1}{c}{AR$^2$}                    &
Notes$^3$ \\
\vspace{-6pt} \\
\hline
\vspace{-6pt} \\
000.3$-$04.6 & M2-28      & \ha & $1.57 \pm 0.03$   & 4.10 &  5.3 &  6.4 &  5.8        & $\pm$ 0.5 &  9.0 \x  8.0 &      &  4.8  & 1.09 & fr, ellip   \\
000.3$-$02.8 & M3-47      & \ha & $1.10 \pm 0.09$   & 4.67 &  5.8 &  7.1 &  7.4 \x 5.5 & $\pm$ 0.6 &  9.0 \x  8.0 &      &  St.  & 1.35 & fr, irreg   \\
000.6$-$01.3 & Bl3-15     & \ha & $1.50 \pm 0.03$   & 2.15 &  3.0 &  3.5 &  3.4 \x 3.1 & $\pm$ 0.3 &  6.0 \x  4.5 &      &  3.0  & 1.08 & fr, ellip   \\
000.8$-$01.5 & BlO        & \ha & $1.30 \pm 0.02$   & 1.89 &  2.6 &  3.1 &  2.8        & $\pm$ 0.2 &              &      &  St.  & 1.03 & fr          \\
002.3$-$03.4 & H2-37      & \ha & $1.88 \pm 0.02$   & 2.98 &  4.0 &  4.8 &  5.5 \x 3.2 & $\pm$ 0.4 &  7.0 \x  5.5 &      &  4.2  & 1.54 & fr, ellip   \\
002.8$-$02.2 & Pe2-12     & \ha & $2.00 \pm 0.06$   & 4.70 &  6.1 &  6.6 &  9.8 \x 3.3 & $\pm$ 0.2 & 10.5 \x  5.0 &      &  5.0  & 2.49 & fr, disc    \\
003.0$-$02.6 & KFL4       & \ha & $1.81 \pm 0.04$   & 1.46 &  2.2 &  2.3 &  2.2        & $\pm$ 0.1 &              &      &  3.0  & 1.01 & pr          \\
003.6$+$03.1 & M2-14      & \ha & $1.94 \pm 0.08$   & 1.68 &  2.5 &  2.9 &  3.1 \x 2.2 & $\pm$ 0.2 &              &  2.2 &  St.  & 1.17 & pr          \\
003.7$-$04.6 & M2-30      & \ha & $1.57 \pm 0.06$   & 2.79 &  3.7 &  4.5 &  4.4 \x 3.8 & $\pm$ 0.4 &              &  3.5 &  9.0  & 1.12 & fr          \\
003.8$-$17.1 & Hb8        & \ha & $1.50 \pm 0.09$   & 0.66 &  1.0 &  1.2 &  1.1        & $\pm$ 0.1 &              &  0.8 &  5.0  & 1.04 & br          \\
004.0$-$11.1 & M3-29      & \ha & $1.64 \pm 0.07$   & 5.24 &  6.7 &  8.0 &  7.3        & $\pm$ 0.7 & 10.5 \x  9.5 & 13.0 &  8.2  & 1.06 & fr, ellip   \\
004.7$-$11.8 & He2-418    & \ha & $1.18 \pm 0.05$   & 7.40 &  9.2 & 11.1 & 12.8 \x 7.6 & $\pm$ 1.0 & 14.0 \x  8.5 &      & 13.0  & 1.68 & fr, irreg   \\
004.8$-$22.7 & He2-436    & \ha & $1.38 \pm 0.14$   & 0.32 &  0.5 &  0.6 &  0.6        & $\pm$ 0.0 &              &      & $<10$ & 1.05 & br          \\
004.8$-$05.0 & M3-26      & \ha & $1.54 \pm 0.02$   & 5.93 &  7.5 &  8.0 &  7.7        & $\pm$ 0.3 & 11.0 \x  9.5 &      &  8.6  & 1.03 & fr, disc    \\
005.2$-$18.6 & StWr2-21   & \ha & $1.80 \pm 0.13$   & 1.40 &  2.1 &  2.4 &  2.5 \x 2.0 & $\pm$ 0.2 &              &      & $< 5$ & 1.10 & pr          \\
006.0$-$41.9 & PRMG1      & \ha & $1.38^a$          & 2.65 &  3.5 &  4.2 &  4.4 \x 3.3 & $\pm$ 0.3 &              &      &  8.2  & 1.29 & fr, irreg   \\
006.8$-$19.8 & Wray16-423 & \ha & $1.78 \pm 0.12$   & 1.08 &  1.6 &  1.9 &  1.9 \x 1.6 & $\pm$ 0.1 &              &      &  St.  & 1.05 & pr          \\
008.6$-$07.0 & He2-406    & \ha & $1.56 \pm 0.04$   & 3.72 &  4.8 &  5.8 &  5.8 \x 4.9 & $\pm$ 0.5 &  8.0 \x  7.5 &      &  3.0  & 1.17 & fr, irreg   \\
008.6$-$02.6 & MaC1-11    & \ha & $1.82 \pm 0.04$   & 1.97 &  2.8 &  3.3 &  3.3 \x 2.9 & $\pm$ 0.2 &              &      &  St.  & 1.09 & pr          \\
009.4$-$09.8 & M3-32      & \ha & $1.52 \pm 0.04$   & 4.32 &  5.5 &  6.7 &  6.8 \x 5.4 & $\pm$ 0.6 &  8.5 \x  7.5 &  7.5 &  6.0  & 1.24 & fr, ring    \\
037.5$-$05.1 & A58        & \ox & $1.77 \pm 0.05$   & 0.70 &  1.1 &  1.2 &  1.4 \x 0.9 & $\pm$ 0.0 &              &      &       & 1.06 & br, irreg   \\
283.8$+$02.2 & My60       & \ha & $1.65 \pm 0.02$   & 5.65 &  7.2 &  8.6 &  7.9        & $\pm$ 0.7 & 11.0 \x 10.5 &      &  7.6  & 1.07 & fr          \\
             &            & \ox & $1.97 \pm 0.03$   & 6.43 &  8.2 &  8.8 &  8.5        & $\pm$ 0.3 & 12.5 \x 12.5 &      &       & 1.07 & fr          \\
289.8$+$07.7 & He2-63     & \ha & $1.47 \pm 0.06$   & 1.83 &  2.6 &  3.0 &  3.0 \x 2.6 & $\pm$ 0.2 &              &      &  3.0  & 1.11 & fr          \\
             &            & \ox & $1.55 \pm 0.03$   & 1.87 &  2.6 &  2.8 &  3.1 \x 2.6 & $\pm$ 0.1 &              &      &       & 1.12 & pr          \\
292.8$+$01.1 & He2-67     & \ha & $1.55 \pm 0.01$   & 2.18 &  3.0 &  3.6 &  3.6 \x 3.0 & $\pm$ 0.3 &  6.0 \x  4.5 &      & $< 5$ & 1.16 & fr, ellip   \\
296.4$-$06.9 & He2-71     & \ox & $2.04 \pm 0.20$   & 0.36 &  0.6 &  0.6 &  0.6        & $\pm$ 0.0 &              &      &       & 1.08 & br          \\
305.1$+$01.4 & He2-90     & \ha & $1.39 \pm 0.04$   & 0.48 &  0.8 &  0.9 &  0.8        & $\pm$ 0.1 &              &      & $<10$ & 1.01 & br          \\
             &            & \ox & $1.71 \pm 0.01$   & 0.63 &  1.0 &  1.1 &  1.3 \x 0.8 & $\pm$ 0.0 &              &      &       & 1.06 & br          \\
315.4$+$09.4 & He2-104    & \ha & $1.73 \pm 0.05$   & 0.40 &  0.6 &  0.7 &  0.7        & $\pm$ 0.0 &              &      & $< 5$ & 1.02 & br, outflow \\
321.3$-$16.7 & He2-185    & \ha & $1.73 \pm 0.06$   & 1.53 &  2.2 &  2.6 &  2.8 \x 2.0 & $\pm$ 0.2 &              &      & $< 5$ & 1.17 & pr          \\
322.4$-$00.1 & Pe2-8      & \ha & $1.68 \pm 0.06$   & 0.97 &  1.5 &  1.7 &  1.9 \x 1.3 & $\pm$ 0.1 &              &      &  1.6  & 1.09 & pr          \\
345.0$-$04.9 & Cn1-3      & \ha & $1.48 \pm 0.02$   & 0.74 &  1.2 &  1.3 &  1.2        & $\pm$ 0.1 &              &      & $< 5$ & 1.02 & br          \\
             &            & \ox & $1.75 \pm 0.01$   & 0.71 &  1.1 &  1.2 &  1.2        & $\pm$ 0.0 &              &      &       & 1.02 & br          \\
345.0$+$03.4 & Vd1-4      & \ha & $1.79 \pm 0.03$   & 0.66 &  1.0 &  1.2 &  1.1        & $\pm$ 0.1 &              &      &  5.0  & 1.03 & br          \\
             &            & \ox & $1.62 \pm 0.03$   & 0.61 &  1.0 &  1.0 &  1.0        & $\pm$ 0.0 &              &      &       & 1.03 & br          \\
345.0$+$04.3 & Vd1-2      & \ha & $1.86 \pm 0.01$   & 0.76 &  1.2 &  1.4 &  1.3        & $\pm$ 0.1 &              &      &  St.  & 1.00 & br, diffuse \\
             &            & \ox & $1.70 \pm 0.03$   & 0.79 &  1.2 &  1.3 &  1.3        & $\pm$ 0.0 &              &      &       & 1.70 & br          \\
346.0$+$08.5 & He2-171    & \ha & $1.01 \pm 0.03$   & 0.08 &  0.1 &  0.2 &  0.1        & $\pm$ 0.0 &              &      & $<10$ & 1.04 & br          \\
             &            & \ox & $1.10 \pm 0.07$   & 0.28 &  0.5 &  0.5 &  0.5        & $\pm$ 0.0 &              &      &       & 1.00 & br          \\
346.3$-$06.8 & Fg2        & \ha & $1.17 \pm 0.04$   & 3.20 &  4.1 &  5.0 &  4.9 \x 4.2 & $\pm$ 0.4 &  6.5 \x  5.5 &      & $< 5$ & 1.14 & fr          \\
             &            & \ox & $1.13 \pm 0.03$   & 3.17 &  4.1 &  4.4 &  4.8 \x 4.2 & $\pm$ 0.1 &              &      &       & 1.13 & fr          \\
347.4$+$05.8 & H1-2       & \ha & $1.14 \pm 0.02$   & 0.55 &  0.9 &  1.0 &  1.1 \x 0.8 & $\pm$ 0.1 &              &  0.8 &  St.  & 1.07 & br          \\
             &            & \ox & $1.26 \pm 0.03$   & 0.59 &  0.9 &  1.0 &  1.2 \x 0.8 & $\pm$ 0.0 &              &      &       & 1.07 & br          \\
348.0$+$06.3 & MGP1       & \ha & $1.98 \pm 0.14^c$ & 4.92 &  6.4 &  7.7 &  7.6 \x 6.4 & $\pm$ 0.6 & 12.0 \x  9.5 &  3.8 &       & 1.17 & fr, ellip   \\
348.8$-$09.0 & He2-306    & \ha & $1.40 \pm 0.02$   & 2.04 &  2.8 &  3.3 &  3.1        & $\pm$ 0.3 &              &      &  3.0  & 1.07 & fr          \\
             &            & \ox & $1.57 \pm 0.01$   & 2.01 &  2.8 &  3.0 &  2.9        & $\pm$ 0.1 &              &      &       & 1.06 & pr          \\
349.8$+$04.4 & M2-4       & \ha & $1.12 \pm 0.02$   & 1.40 &  2.0 &  2.3 &  2.1        & $\pm$ 0.2 &              &  2.0 & $< 5$ & 1.09 & fr, ellip   \\
             &            & \ox & $1.21 \pm 0.03$   & 1.32 &  1.9 &  2.0 &  2.0        & $\pm$ 0.1 &              &      &       & 1.07 & pr          \\
350.8$-$02.4 & H1-22      & \ha & $2.13 \pm 0.09$   & 2.17 &  3.1 &  3.7 &  3.4        & $\pm$ 0.3 &              &      &  St.  & 1.01 & pr          \\
350.9$+$04.4 & H2-1       & \ha & $1.38 \pm 0.03$   & 1.79 &  2.5 &  2.9 &  2.7        & $\pm$ 0.2 &              &  2.2 &  5.6  & 1.00 & fr          \\
             &            & \ox & $1.32 \pm 0.04$   & 0.78 &  1.2 &  1.3 &  1.2        & $\pm$ 0.0 &              &      &       & 1.01 & br          \\
351.1$+$04.8 & M1-19      & \ha & $1.32 \pm 0.02$   & 1.99 &  2.7 &  3.2 &  3.2 \x 2.7 & $\pm$ 0.3 &              &  2.6 &  8.0  & 1.13 & fr          \\
351.2$+$05.2 & M2-5       & \ha & $1.33 \pm 0.01$   & 3.41 &  4.4 &  5.3 &  4.9        & $\pm$ 0.4 &  6.5 \x  6.5 &  5.0 &  5.0  & 1.03 & fr, outflow \\
351.3$+$07.6 & H1-4       & \ha & $1.51 \pm 0.02$   & 0.64 &  1.0 &  1.2 &  1.1        & $\pm$ 0.1 &              &      &  St.  & 1.02 & br          \\
351.9$-$01.9 & Wray16-286 & \ha & $1.69 \pm 0.06$   & 1.16 &  1.8 &  2.0 &  2.2 \x 1.6 & $\pm$ 0.1 &              &      &  St.  & 1.12 & pr, ellip   \\
351.9$+$09.0 & PC13       & \ha & $2.07 \pm 0.04$   & 5.15 &  6.7 &  8.0 &  8.6 \x 6.0 & $\pm$ 0.7 & 10.0 \x  8.5 &      &  7.0  & 1.39 & fr, disc    \\
352.0$-$04.6 & H1-30      & \ha & $1.78 \pm 0.05$   & 1.63 &  2.4 &  2.8 &  3.0 \x 2.1 & $\pm$ 0.2 &              &      &  5.4  & 1.20 & pr, ellip   \\
\vspace{-6pt} \\
\hline
\end{tabular}
\vspace{-3pt} \\
\begin{flushleft}
$^1$ Error expressed as the standard deviation of four $\theta_\mathrm{PSF}$ values except: 
     $^a$one value, $^b$two values, $^c$three values. \\
$^2$ Maximum/minimum axis ratio of planetary nebulae from 2D FWHM measurements. \\
$^3$ br, barely resolved, $\theta_\mathrm{line} < \theta_\mathrm{PSF}$; 
     pr, partly resolved, $\theta_\mathrm{PSF}  < \theta_\mathrm{line} < 2 \theta_\mathrm{PSF}$;
     fr, fully resolved,  $\theta_\mathrm{line} \geq 2 \theta_\mathrm{PSF}$. \\
\end{flushleft}
\end{table*}

\addtocounter{table}{-1}

\begin{table*}
\caption[]{{\bf(continued)} 
           Angular diameters of 69 planetary nebulae. Diameters are in arc seconds and are defined in section~\ref{diams}.}
\label{diamsb}
\setlength{\tabcolsep}{5.3pt}
\begin{tabular}{llllcrrcccd{2.1}rrl}
\vspace{-10pt} \\
\hline
\vspace{-6pt} \\
PN G & Object & Line & 
\multicolumn{1}{c}{$\theta_\mathrm{PSF}{}^1$} & 
\multicolumn{1}{c}{$\theta_\mathrm{gauss}$}   & 
\multicolumn{1}{c}{$\theta_\mathrm{shell}$}   & 
\multicolumn{1}{c}{$\theta_\mathrm{line}$}    & 
\multicolumn{2}{c}{$\theta_\mathrm{mean}$}    & 
\multicolumn{1}{c}{$\theta_\mathrm{10\%}$}    & 
\multicolumn{1}{c}{$\theta_\mathrm{rad}$}     & 
\multicolumn{1}{c}{$\theta_\mathrm{opt}$}     &
\multicolumn{1}{c}{AR$^2$}                    &
Notes$^3$ \\
\vspace{-6pt} \\
\hline
\vspace{-6pt} \\
352.1$+$05.1 & M2-8       & \ha & $1.10 \pm 0.01$   & 2.55 &  3.3 &  4.0 &  3.8 \x 3.5 & $\pm$ 0.3 &  5.0 \x  5.0 &  3.7 &  4.2  & 1.10 & fr, outflow \\
             &            & \ox & $1.24 \pm 0.04$   & 2.37 &  3.2 &  3.4 &  3.3        & $\pm$ 0.1 &              &      &       & 1.05 & fr, outflow \\
352.6$+$00.1 & H1-12      & \ha & $1.13 \pm 0.03$   & 5.76 &  7.2 &  8.7 &  7.9        & $\pm$ 0.8 & 12.0 \x 10.0 & 11.0 &  6.8  & 1.00 & fr          \\
352.6$+$03.0 & H1-8       & \ha & $1.36 \pm 0.02$   & 1.92 &  2.6 &  3.1 &  3.1 \x 2.6 & $\pm$ 0.2 &              &      &  3.4  & 1.14 & fr          \\
352.8$-$00.2 & H1-13      & \ha & $1.57 \pm 0.16^c$ & 9.27 & 11.5 & 14.0 & 12.7        & $\pm$ 1.2 & 13.5 \x 12.0 & 14.0 &  9.6  & 1.08 & fr, ring    \\
354.2$+$04.3 & M2-10      & \ha & $1.64 \pm 0.03$   & 2.93 &  3.9 &  4.7 &  5.2 \x 3.4 & $\pm$ 0.4 &  6.5 \x  5.5 &  4.0 &  4.0  & 1.42 & fr, irreg   \\
354.5$+$03.3 & Th3-4      & \ha & $1.84 \pm 0.11$   & 0.97 &  1.5 &  1.7 &  2.0 \x 1.2 & $\pm$ 0.1 &              &      &  St.  & 1.12 & br          \\
354.9$+$03.5 & Th3-6      & \ha & $2.38 \pm 0.08$   & 2.49 &  3.6 &  4.2 &  4.1 \x 3.7 & $\pm$ 0.3 &              &      &  St.  & 1.07 & pr          \\
355.1$-$06.9 & M3-21      & \ha & $1.51 \pm 0.02$   & 1.00 &  1.5 &  1.8 &  1.8 \x 1.5 & $\pm$ 0.1 &              &      & $< 5$ & 1.08 & pr          \\
355.1$+$02.3 & Th3-11     & \ha & $1.42 \pm 0.04$   & 1.21 &  1.8 &  2.1 &  2.1 \x 1.7 & $\pm$ 0.2 &              &      &  St.  & 1.10 & pr          \\
355.4$-$02.4 & M3-14      & \ha & $1.52 \pm 0.02$   & 2.40 &  3.3 &  3.9 &  3.6        & $\pm$ 0.3 &  8.0 \x  5.0 &  2.8 &  7.2  & 1.01 & fr, ellip   \\
355.7$-$03.5 & H1-35      & \ha & $1.40 \pm 0.04$   & 0.90 &  1.4 &  1.6 &  1.5        & $\pm$ 0.1 &              &  1.1 &  2.0  & 1.07 & pr          \\
355.9$+$02.7 & Th3-10     & \ha & $1.24 \pm 0.03$   & 1.58 &  2.2 &  2.6 &  2.4        & $\pm$ 0.2 &              &  2.0 &  St.  & 1.07 & fr          \\
356.1$+$02.7 & Th3-13     & \ha & $1.63 \pm 0.02$   & 0.45 &  0.7 &  0.8 &  1.1 \x 0.2 & $\pm$ 0.1 &              & <  2 &  St.  & 1.07 & br          \\
356.8$+$03.3 & Th3-12     & \ha & $1.75 \pm 0.13$   & 1.22 &  1.8 &  2.1 &  2.3 \x 1.6 & $\pm$ 0.1 &              &  1.2 &  St.  & 1.12 & pr          \\
357.1$-$06.1 & M3-50      & \ha & $1.25 \pm 0.04$   & 3.06 &  4.0 &  4.8 &  6.7 \x 1.7 & $\pm$ 0.4 &  8.5 \x  3.5 &      &  4.0  & 3.03 & fr, disc    \\
357.1$-$04.7 & H1-43      & \ha & $1.71 \pm 0.03$   & 0.91 &  1.4 &  1.6 &  1.7 \x 1.3 & $\pm$ 0.1 &              &  3.0 &  2.0  & 1.08 & br          \\
357.1$+$01.9 & Th3-24     & \ha & $1.24 \pm 0.04$   & 4.61 &  5.8 &  7.0 &  6.4        & $\pm$ 0.6 &  8.0 \x  7.0 &      &  St.  & 1.07 & fr, irreg   \\
357.1$+$03.6 & M3-7       & \ha & $1.39 \pm 0.02$   & 3.15 &  4.1 &  4.9 &  4.9 \x 4.1 & $\pm$ 0.4 &  6.5 \x  5.5 &  4.7 &  5.8  & 1.18 & fr          \\
357.6$-$03.3 & H2-29      & \ha & $1.45 \pm 0.02$   & 7.31 &  9.1 & 11.0 & 12.5 \x 7.7 & $\pm$ 1.0 &              &      &  4.8  & 1.61 & fr, irreg   \\
358.5$+$03.7 & Al2-B      & \ha & $1.56 \pm 0.04$   & 1.62 &  2.3 &  2.7 &  3.0 \x 2.1 & $\pm$ 0.2 &              &      &  St.  & 1.22 & pr, irreg   \\
358.5$-$01.7 & JaSt64     & \ha & $1.19 \pm 0.05$   & 0.98 &  1.5 &  1.7 &  1.6        & $\pm$ 0.1 &              &      &       & 1.04 & pr          \\
358.7$+$05.2 & M3-40      & \ha & $1.49 \pm 0.12^b$ & 1.96 &  2.7 &  3.2 &  3.0        & $\pm$ 0.3 &              &  2.5 &  St.  & 1.05 & fr          \\
359.2$+$04.7 & Th3-14     & \ha & $1.48 \pm 0.05$   & 1.15 &  1.7 &  2.0 &  1.9        & $\pm$ 0.1 &              & <3.3 &       & 1.03 & pr          \\
\vspace{-6pt} \\
\hline
\end{tabular}
\vspace{-3pt} \\
\begin{flushleft}
$^1$ Error expressed as the standard deviation of four $\theta_\mathrm{PSF}$ values except: 
     $^a$one value, $^b$two values, $^c$three values. \\
$^2$ Maximum/minimum axis ratio of planetary nebulae from 2D FWHM measurements. \\
$^3$ br, barely resolved, $\theta_\mathrm{line} < \theta_\mathrm{PSF}$; 
     pr, partly resolved, $\theta_\mathrm{PSF}  < \theta_\mathrm{line} < 2 \theta_\mathrm{PSF}$;
     fr, fully resolved,  $\theta_\mathrm{line} \geq 2 \theta_\mathrm{PSF}$. \\
\end{flushleft}
\end{table*}

\begin{table*}
\caption[]{Flux and extinction values of 70 planetary nebulae. 
           Observed (equation \ref{calphaobs}) and catalogue (equation \ref{calphacat}) 
           extinction are defined in section~\ref{extdets}.}
\label{fluxa}
\setlength{\tabcolsep}{4.2pt}
\begin{tabular}{llld{3.2}d{3.2}d{2.2}d{2.2}d{2.1}d{1.2}d{1.2}d{3.3}d{3.3}l}
\vspace{-10pt} \\
\hline
\vspace{-6pt} \\
PN G  &  Object  &  Line  & 
\multicolumn{2}{c}{Radial Vel. (km\,s$^{-1}$)}  &  
\multicolumn{2}{c}{Flux (log mW\,m$^{-2}$)}     &  
\multicolumn{1}{c}{S$\nu$ 6 cm}                 &  
\multicolumn{2}{c}{\ha\ Extinction}             &  
\multicolumn{1}{c}{$R_\mathrm{V}$}              &
\multicolumn{1}{c}{$\Delta R_\mathrm{V}$}       &
Notes$^5$                                       \\
      &          &        &  
\multicolumn{1}{c}{Observed}                    &  
\multicolumn{1}{c}{Catalogue}                   &  
\multicolumn{1}{c}{Observed}                    &  
\multicolumn{1}{c}{Catalogue}                   &  
\multicolumn{1}{c}{(mJy)$^4$}                   &  
\multicolumn{1}{c}{Observed}                    &  
\multicolumn{1}{c}{Catalogue}         &  &  &  \\
\vspace{-6pt} \\
\hline
\vspace{-6pt} \\
000.3$-$04.6  &  M2-28       & \ha  &  -43.5 &  -29.9 &    -11.70  & -11.73 &   10 &            0.66   & 0.98 &            1.99   &              0.62   & B   \\
000.3$-$02.8  &  M3-47       & \ha  &  -17.9 &  -16.0 &    -12.47  & -12.88 &      &                   & 1.58 &                   &                     &     \\
000.6$-$01.3  &  Bl3-15      & \ha  & -114.8 &        &    -12.59  & -12.92 &      &                   & 2.89 &                   &                     &     \\
000.8$-$01.5  &  BlO         & \ha  &   79.9 &        &    -12.18  & -11.82 &      &                   & 2.04 &                   &                     &     \\
002.3$-$03.4  &  H2-37       & \ha  & -129.8 & -156.8 &    -11.86  & -12.06 &      &                   & 0.69 &                   &                     &     \\
002.8$-$02.2  &  Pe2-12      & \ha  &  108.8 &        &    -11.86  & -12.46 &    2 &            0.13   & 1.06 &            0.57   &              0.41   & lrf \\
003.0$-$02.6  &  KFL4        & \ha  &  -12.9 &   17.0 &    -12.59  & -12.81 &    1 &            0.59   & 1.02 &            1.71   &              0.56   & lrf \\
003.6$+$03.1  &  M2-14       & \ha  &  -25.0 &  -48.2 &    -11.35  & -12.15 &   39 &            0.90   & 1.41 &            1.87   &              0.42   & B   \\
003.7$-$04.6  &  M2-30       & \ha  &  130.0 &  154.9 &    -11.29  & -11.18 &   14 &            0.40   & 0.74 &            1.59   &              0.75   & B   \\
003.8$-$17.1  &  Hb8         & \ha  & -170.1 & -180.5 &    -11.30  & -11.28 &    6 &            0.03   & 0.20 &            0.64   &              2.30   & lrf \\
004.0$-$11.1  &  M3-29       & \ha  &   26.2 &   49.3 &    -11.20  & -11.10 &   12 &            0.25   & 0.12 &           10.81   &             31.50   & B   \\
004.7$-$11.8  &  He2-418     & \ha  &   87.7 &  109.3 &    -12.05  & -11.90 &      &                   & 0.12 &                   &                     &     \\
004.8$-$22.7  &  He2-436     & \ha  &  128.4 &  133.0 &    -11.86  & -11.50 & {\bf3}.{\bf9}  &  0.41   & 0.50 &            2.45   &              1.36   & lrf \\
004.8$-$05.0  &  M3-26       & \ha  &  -24.7 &   -9.8 &    -11.41  & -11.26 &    8 &            0.28   & 0.43 &            1.90   &              1.40   & lrf \\
005.2$-$18.6  &  StWr2-21    & \ha  &  136.3 &  133.0 &    -12.26  & -12.46 &      &                   & 0.32 &                   &                     &     \\
006.0$-$41.9  &  PRMG1       & \ha  &   -8.1 &        &    -12.95  & -12.77 &      &                   & 0.18 &                   &                     &     \\
006.8$-$19.8  &  Wray16-423  & \ha  &  139.2 &  133.0 &    -11.78  & -11.48 & {\bf4}.{\bf7}  &  0.42   & 0.36 &            3.78   &              2.47   & lrf \\
008.6$-$07.0  &  He2-406     & \ha  &   -8.5 &   28.2 &    -12.12  & -12.38 &      &                   & 0.84 &                   &                     &     \\
008.6$-$02.6  &  MaC1-11     & \ha  &  -30.8 &  -89.0 &    -12.33  & -12.25 &      &                   & 1.76 &                   &                     &     \\
009.4$-$09.8  &  M3-32       & \ha  &   40.1 &   58.4 &    -11.39  & -11.22 &   12 &            0.43   & 0.53 &            2.46   &              1.29   & B   \\
009.8$-$07.5  &  GJJC1       & \ox  &        &  -32.0 &    -12.74  &        &      &                   &      &                   &                     &     \\
037.5$-$05.1  &  A58         & \ha  &        &   70.0 &    -13.40  &        &      &                   &      &                   &                     &     \\
              &              & \ox  &        &        &    -12.91  &        &      &                   &      &                   &                     &     \\
283.8$+$02.2  &  My60        & \ha  &        &        & \ge-11.53  & -11.04 &   60 &        \le 1.27   & 0.67 & $\phantom{ 8.33}$ &   $\phantom{ 2.67}$ & npc \\
              &              & \ox  &        &        & \ge-10.68  & -10.64 &      &                   &      &                   &                     &     \\
\vspace{-6pt} \\
\hline
\end{tabular}
\vspace{-3pt} \\
\begin{flushleft}
$^4$ Bold values are for PNe in the Sagittarius dwarf galaxy \citep{dudziak}.
     All other values from the Strasbourg - ESO Catalogue \citep{acker}. \\
$^5$ B, Bulge subset B;
     npc, non-photometric conditions; 
     hbt, high brightness temperature, $T_\mathrm{b} >$ 1000 K; 
     lrf, low radio flux, $S_\nu <$ 10 mJy.
\end{flushleft}
\end{table*}

\addtocounter{table}{-1}

\begin{table*}
\caption[]{{\bf(continued)} 
           Flux and extinction values of 70 planetary nebulae. 
           Observed (equation \ref{calphaobs}) and catalogue (equation \ref{calphacat}) 
           extinction are defined in section~\ref{extdets}.}
\label{fluxb}
\setlength{\tabcolsep}{4.2pt}
\begin{tabular}{llld{3.2}d{3.2}d{2.2}d{2.2}d{2.1}d{1.2}d{1.2}d{3.3}d{3.3}l}
\vspace{-10pt} \\
\hline
\vspace{-6pt} \\
PN G  &  Object  &  Line  & 
\multicolumn{2}{c}{Radial Vel. (km\,s$^{-1}$)}  &  
\multicolumn{2}{c}{Flux (log mW\,m$^{-2}$)}     &  
\multicolumn{1}{c}{S$\nu$ 6 cm}                 &  
\multicolumn{2}{c}{\ha\ Extinction}             &  
\multicolumn{1}{c}{$R_\mathrm{V}$}              &
\multicolumn{1}{c}{$\Delta R_\mathrm{V}$}       &
Notes$^5$                                       \\
      &          &        &  
\multicolumn{1}{c}{Observed}                    &  
\multicolumn{1}{c}{Catalogue}                   &  
\multicolumn{1}{c}{Observed}                    &  
\multicolumn{1}{c}{Catalogue}                   &  
\multicolumn{1}{c}{(mJy)$^4$}                   &  
\multicolumn{1}{c}{Observed}                    &  
\multicolumn{1}{c}{Catalogue}         &  &  &  \\
\vspace{-6pt} \\
\hline
\vspace{-6pt} \\
289.8$+$07.7  &  He2-63      & \ha  &        &  123.2 & \ge-11.92  & -11.91 &   12 &        \le 0.96   & 0.27 & $\phantom{787.7}$ &  $\phantom{142.33}$ & npc \\
              &              & \ox  &        &        & \ge-11.14  & -11.43 &      &                   &      &                   &                     &     \\
292.8$+$01.1  &  He2-67      & \ha  &        &   59.5 & \ge-12.47  & -11.17 &   41 &        \le 1.51   & 0.83 & $\phantom{12.09}$ &   $\phantom{ 4.13}$ & npc \\
296.4$-$06.9  &  He2-71      & \ox  &        &        & \ge-11.71  & -11.69 &      &                   &      &                   &                     & npc \\
305.1$+$01.4  &  He2-90      & \ha  &        &        & \ge-10.71  & -10.52 &   25 & $\phantom{ 0.07}$ & 1.13 & $\phantom{ 0.43}$ &   $\phantom{ 0.37}$ & hbt \\
              &              & \ox  &        &        & \ge-10.73  & -11.14 &      &                   &      &                   &                     &     \\
315.4$+$09.4  &  He2-104     & \ha  &  -70.4 & -143.6 &    -10.83  & -10.81 &  <15 & $\phantom{-0.03}$ & 1.09 & $\phantom{ 0.26}$ &   $\phantom{ 0.37}$ & hbt \\
321.3$-$16.7  &  He2-185     & \ha  &  -12.0 &   -6.0 &    -11.10  & -10.97 &   18 &            0.32   & 0.20 &            6.24   &              7.24   &     \\
322.4$-$00.1  &  Pe2-8       & \ha  &  -23.9 &  -16.3 &    -11.95  & -11.96 &  100 &        \ge 1.91   & 3.16 &            1.78   &              0.18   & hbt \\
345.0$-$04.9  &  Cn1-3       & \ha  &  -68.2 &  -79.8 & \ge-10.89  & -10.58 &      &                   & 0.13 &                   &                     & npc \\
              &              & \ox  &        &        & \ge-10.71  & -10.33 &      &                   &      &                   &                     & npc \\
345.0$+$03.4  &  Vd1-4       & \ha  &   93.4 &   35.1 &    -12.11  & -12.18 &      &                   & 0.84 &                   &                     &     \\
              &              & \ox  &        &        &    -11.73  & -11.80 &      &                   &      &                   &                     &     \\
345.0$+$04.3  &  Vd1-2       & \ha  &        &    3.4 &    -11.57  & -12.30 &      &                   & 1.16 &                   &                     &     \\
              &              & \ox  &        &        &    -13.08  &        &      &                   &      &                   &                     &     \\
346.0$+$08.5  &  He2-171     & \ha  & -100.2 & -101.5 & \ge-11.69  & -11.09 &   10 & $\phantom{ 0.66}$ & 1.54 & $\phantom{ 1.29}$ &   $\phantom{ 0.34}$ & hbt \\
              &              & \ox  &        &        & \ge-11.94  & -11.64 &      &                   &      &                   &                     &     \\
346.3$-$06.8  &  Fg2         & \ha  &   52.5 &   34.5 &    -11.12  & -11.19 &      &                   & 0.37 &                   &                     &     \\
              &              & \ox  &        &        &    -10.47  & -10.59 &      &                   &      &                   &                     &     \\
347.4$+$05.8  &  H1-2        & \ha  & -124.8 & -105.5 &    -11.02  & -10.96 &   62 &        \ge 0.78   & 1.57 &      \ge   1.47   &              0.34   & hbt \\
              &              & \ox  &        &        &    -10.72  & -11.24 &      &                   &      &                   &                     &     \\
348.0$+$06.3  &  MGP1        & \ha  &  -67.9 &  -52.0 &    -12.15  &        & 10.2 &            1.12   &      &                   &                     &     \\
348.8$-$09.0  &  He2-306     & \ha  &  -49.6 &  -40.0 &    -11.14  & -11.23 &      &                   & 0.27 &                   &                     &     \\
              &              & \ox  &        &        &    -11.00  & -11.14 &      &                   &      &                   &                     &     \\
349.8$+$04.4  &  M2-4        & \ha  & -227.0 & -207.2 &    -11.06  & -10.93 &   32 &            0.53   & 1.08 &            1.47   &              0.50   & B   \\
              &              & \ox  &        &        &    -10.97  & -10.94 &      &                   &      &                   &                     &     \\
350.8$-$02.4  &  H1-22       & \ha  & -199.6 & -213.0 &    -11.76  & -11.72 &      &                   & 1.66 &                   &                     &     \\
350.9$+$04.4  &  H2-1        & \ha  &  -35.5 &  -18.8 &    -10.62  & -10.67 &   61 &            0.37   & 0.79 &            1.38   &              0.67   & B   \\
              &              & \ox  &        &        &    -11.57  & -11.63 &      &                   &      &                   &                     &     \\
351.1$+$04.8  &  M1-19       & \ha  &  -68.6 &  -55.2 &    -11.10  & -11.15 &   26 &            0.48   & 0.93 &            1.53   &              0.59   & B   \\
351.2$+$05.2  &  M2-5        & \ha  & -134.6 & -123.4 &    -11.18  & -11.26 &   12 &            0.22   & 0.90 &            0.84   &              0.52   & B   \\
351.3$+$07.6  &  H1-4        & \ha  &   10.1 &    2.7 &    -11.88  & -11.57 &      &                   & 0.52 &                   &                     &     \\
351.9$-$01.9  &  Wray16-286  & \ha  & -143.9 & -152.3 &    -11.79  & -11.82 &      &                   & 1.91 &                   &                     &     \\
351.9$+$09.0  &  PC13        & \ha  &        &  -75.1 &    -11.63  & -11.54 &      &                   & 0.73 &                   &                     &     \\
352.0$-$04.6  &  H1-30       & \ha  &    5.6 &  -12.9 &    -12.01  & -12.12 &      &                   & 1.30 &                   &                     &     \\
352.1$+$05.1  &  M2-8        & \ha  &   18.2 &   25.1 &    -11.31  & -11.72 &   18 &            0.53   & 0.78 &            1.99   &              0.79   & B   \\
              &              & \ox  &        &        &    -11.11  & -11.52 &      &                   &      &                   &                     &     \\
352.6$+$00.1  &  H1-12       & \ha  &        &        &    -11.85  & -12.10 & >719 &            2.67   & 3.88 &            2.02   &              0.16   &     \\
352.6$+$03.0  &  H1-8        & \ha  &   -3.9 & -116.0 &    -11.97  & -11.91 &      &                   & 2.32 &                   &                     &     \\
352.8$-$00.2  &  H1-13       & \ha  &  -11.6 &  -26.3 &    -11.48  & -11.65 & >620 &            2.23   & 2.60 &            2.59   &              0.27   &     \\
354.2$+$04.3  &  M2-10       & \ha  &  -64.2 &  -85.6 &    -11.58  & -11.61 &  9.1 &            0.50   & 1.31 &            1.17   &              0.38   & lrf \\
354.5$+$03.3  &  Th3-4       & \ha  & -122.9 & -165.0 &    -12.26  & -12.03 &      &                   & 2.15 &                   &                     &     \\
354.9$+$03.5  &  Th3-6       & \ha  &  -92.5 &  -73.1 &    -12.28  & -12.40 &      &                   & 1.93 &                   &                     &     \\
355.1$-$06.9  &  M3-21       & \ha  &  -78.6 &  -66.9 &    -10.84  & -10.93 &   30 &            0.28   & 0.08 &          276.33   &             42.22   & B   \\
355.1$+$02.3  &  Th3-11      & \ha  &  -33.1 &        &    -12.81  &        &      &                   &      &                   &                     &     \\
355.4$-$02.4  &  M3-14       & \ha  &  -63.3 &  -82.3 &    -11.47  & -11.33 &   30 &            0.91   & 1.45 &            1.85   &              0.41   & B   \\
355.7$-$03.5  &  H1-35       & \ha  &  135.1 &  123.4 &    -10.92  & -10.58 &   72 &        \ge 0.74   & 0.98 &      \ge   2.24   &              0.66   & hbt \\
355.9$+$02.7  &  Th3-10      & \ha  & -256.5 &        &    -12.70  & -12.96 & 29.5 &            2.14   & 2.55 &            2.51   &              0.27   & B   \\
356.1$+$02.7  &  Th3-13      & \ha  &  -75.3 &  -99.0 &    -12.31  & -12.20 & 14.3 &        \ge 1.43   & 2.69 &      \ge   1.56   &              0.21   & hbt \\
356.8$+$03.3  &  Th3-12      & \ha  &  248.8 &  185.4 &    -12.61  & -12.63 &  3.5 &            1.12   & 1.56 &            2.12   &              0.40   & lrf \\
357.1$-$06.1  &  M3-50       & \ha  &   26.4 &   17.5 &    -12.26  & -12.29 &      &                   & 0.51 &                   &                     &     \\
357.1$-$04.7  &  H1-43       & \ha  &   44.9 &   49.0 &    -11.55  & -11.67 &    6 &            0.29   & 0.93 &            0.99   &              0.52   & lrf \\
357.1$+$01.9  &  Th3-24      & \ha  & -152.4 & -197.0 &    -12.76  & -13.08 &      &                   & 1.57 &                   &                     &     \\
357.1$+$03.6  &  M3-7        & \ha  & -171.7 & -193.6 &    -11.34  & -11.33 &   28 &            0.75   & 1.20 &            1.84   &              0.49   & B   \\
357.6$-$03.3  &  H2-29       & \ha  &  123.4 &  107.0 &    -12.38  & -12.79 &      &                   & 1.30 &                   &                     &     \\
358.5$+$03.7  &  Al2-B       & \ha  & -134.0 & -158.0 &    -12.97  &        &      &                   &      &                   &                     &     \\
358.5$-$01.7  &  JaSt64      & \ha  &   17.6 &        &    -13.13  &        &      &                   &      &                   &                     &     \\
358.7$+$05.2  &  M3-40       & \ha  &   54.0 &   32.8 &    -12.25  & -12.23 & 15.1 &            1.39   & 1.93 &            2.13   &              0.33   & B   \\
359.2$+$04.7  &  Th3-14      & \ha  & -206.7 & -239.2 &    -12.20  & -12.27 &    4 &            0.77   & 1.54 &            1.48   &              0.35   & lrf \\
\vspace{-6pt} \\
\hline
\end{tabular}
\vspace{-3pt} \\
\begin{flushleft}
$^5$ B, Bulge subset B;
     npc, non-photometric conditions; 
     hbt, high brightness temperature, $T_\mathrm{b} >$ 1000 K; 
     lrf, low radio flux, $S_\nu <$ 10 mJy.
\end{flushleft}
\end{table*}

\subsection{Discussion}

By their nature PNe at a given distance are likely to exhibit a range of 
angular diameters depending on where they are in their evolutionary sequence 
of returning shed material from the proginator star to the ISM. 
This is in addition to observed angular size differences for the 
same source, e.g. optical versus radio or specific line emission.

It is apparent from a visual inspection of our images that not all 
PNe have a well-defined outer radius. For such soft boundary PNe a 10 per 
cent radius does not have a direct physical meaning and usually does not 
correlate well with converted deconvolved FWHM diameters. This is not surprising 
as 10 per cent (or lower if signal-to-noise will allow) radius measurements 
`see' into the extended low emission regions surrounding PNe, whereas FWHM 
determinations tend to reveal the compact core. These differing values 
actually give us alternative views of a PNe's morphology and tell us more about 
the object than a single arbitrary radius value.


\section{Flux determinations}\label{fluxes}

\subsection{Method}

The total flux detected is reduced by the transmission of the H$\alpha$ and 
[O\,{\sc iii}] filters. This and other factors must be taken into account 
when determining actual flux values, and we now describe the methods 
used to arrive at the values listed in Table~\ref{fluxa}.

The central position of each object was computed by fitting a Gaussian
to the image distribution and then the magnitude was calculated by
integrating over successively larger diameters of the central area. 
By plotting these radii against magnitude it was possible to determine
the radius at which the flux level matched that of the background
continuum, and therefore the total flux of the object in ADU counts
per 30 second exposure. These profiles were compared with those of
standard stars LTT1020 and LTT624 to determine the aperture correction.

To determine detector flux values in mW\,m$^{-2}$ (erg cm$^{-2}$s$^{-1}$), 
the ratio of ADU count to flux conversion factor was calculated by measuring 
the count in ADUs per ten seconds of LTT1020 and LTT6248, and comparing it 
with tabulated fluxes for these spectrophotometric standards \citep{hamuy}. 
The calibration counts were factored for the shorter exposure time, and the 
aperture correction of $-$5 per cent was applied. Apart from a brief period of 
non-photometric conditions caused by cloud (indicated as `npc' in Table~\ref{fluxa}), 
observations were considered photometric as the computed ADU count to flux ratios 
varied by no more than 2.1 per cent between the two night's observations.

In order to determine actual H$\alpha$ 6565\AA\ flux values, any contribution from [N\,{\sc ii}] 
lines at 6548\AA\ and 6584\AA\ (vacuum), passing through ESO filter \#654, was taken into account 
by assuming the following expression for the total detected flux: 

\begin{equation}
    F_\mathrm{det} = F_{6565} T_{6565} + F_{6584} T_{6584} + F_{6548} T_{6548}
\end{equation}

\noindent
where $F\lambda$ and $T\lambda$ are actual flux values and filter transmission 
coefficients (corrected for the appropriate Doppler shift) respectively. 
Using catalogue [N\,{\sc ii}] to H$\alpha$ flux ratios 
$R_{6584} = F_{6584} / F_{6565}$ and $R_{6548} = F_{6548} / F_{6565}$ 
(which avoid a dependence on absolute values), 
the actual H$\alpha$ 6565\AA\ flux can be calculated using: 

\begin{equation}
    F(\mathrm{H\alpha)_{obs}} = F_\mathrm{det} / (T_{6565} + R_{6584} T_{6584} + R_{6548} T_{6548})
\end{equation}

\noindent
[N\,{\sc ii}]$_{6548}$ flux values were calculated from the [N\,{\sc ii}]$_{6584}$ 
catalogue values \citep{acker} using a factor of 2.97. In the case of 
[O\,{\sc iii}], the flux observed through the filter was factored for the 
appropriate transmission coefficient of the [O\,{\sc iii}] wavelength, 
so as to arrive at actual observed [O\,{\sc iii}] flux values.

The Doppler shifted H$\alpha$ and [O\,{\sc iii}] wavelengths observed at the 
telescope were computed from catalogue radial velocities \citep{durand,acker} 
less a barycentric correction appropriate to the date and time at La Silla, 
or in the absence of a catalogue value, derived from spectra taken during our 
observations. Columns four and five of Table~\ref{fluxa} compare catalogue radial 
velocities with observed values, which we consider accurate to $\pm$25~km~s$^{-1}$.
For three objects it was not possible to determine the radial velocity at all. 
Using the corrected filter response curves (measured at La Silla air pressure), 
the appropriate transmission coefficient was then applied to arrive at actual 
flux values. La Silla-air wavelengths were 6563.3\AA\ for H$\alpha$ and 
5007.3\AA\ for [O\,{\sc iii}], with 6548.5\AA\ and 6583.9\AA\ assumed for 
[N\,{\sc ii}] using the same $\Delta\lambda$ as that of H$\alpha$. 
Although stellar velocities in the Bulge are typically of the order 
of $\pm$100 km s$^{-1}$ \citep{minniti}, or about $\pm 2$\AA, we have 
chosen to calculate precise Doppler wavelength shifts, in order to take 
into account the sensitivity of the H$\alpha$ filter transmission curve 
to small changes in observed wavelength.

\subsection{Results and error factors}

\begin{figure}
    \hbox{\hspace{1mm}\includegraphics[width=79mm]{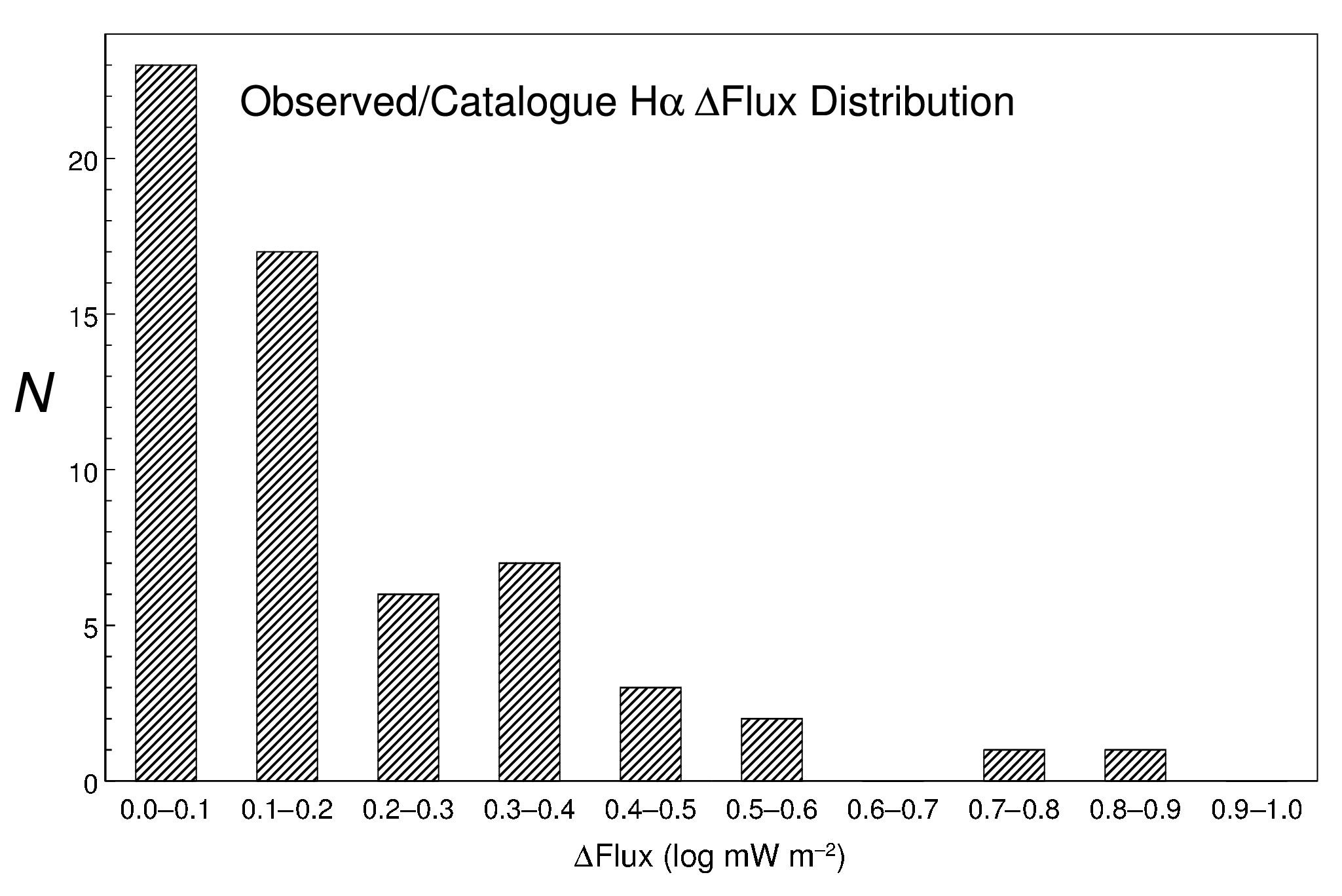}}
    \caption{Distribution of $\Delta F(\mathrm{H}\alpha)$, 
             the absolute difference between observed and 
             catalogued H$\alpha$ flux values (log mW\,m$^{-2}$).}
    \label{dfluxobscat}
\end{figure}

Table~\ref{fluxa} lists the newly determined H$\alpha$ and [O\,{\sc iii}] 
flux densities in column six. For comparison columns seven and eight list 
H$\alpha$ and [O\,{\sc iii}] flux values (calculated from catalogued 
H$\beta$ flux and line intensities) and
the 6~cm radio flux values from the Strasbourg - ESO Catalogue \citep{acker}. 
Radio values in bold are for the two PNe in the Sagittarius dwarf galaxy 
\citep[from][]{dudziak}. Fig.~\ref{dfluxobscat} shows the distribution of 
$\Delta F(\mathrm{H}\alpha)$, the absolute difference between observed
and catalogued H$\alpha$ flux values. 7 out of 60 objects are shown to have
$\Delta F(\mathrm{H}\alpha) > 0.4$ (log mW\,m$^{-2}$). 
A comparison of $\Delta F(\mathrm{H}\alpha)$ with $\theta_\mathrm{line}$ shows no
evidence of a correlation between these flux differences and angular diameter.

\begin{figure}
    \hbox{\hspace{5mm}\includegraphics[width=77mm]{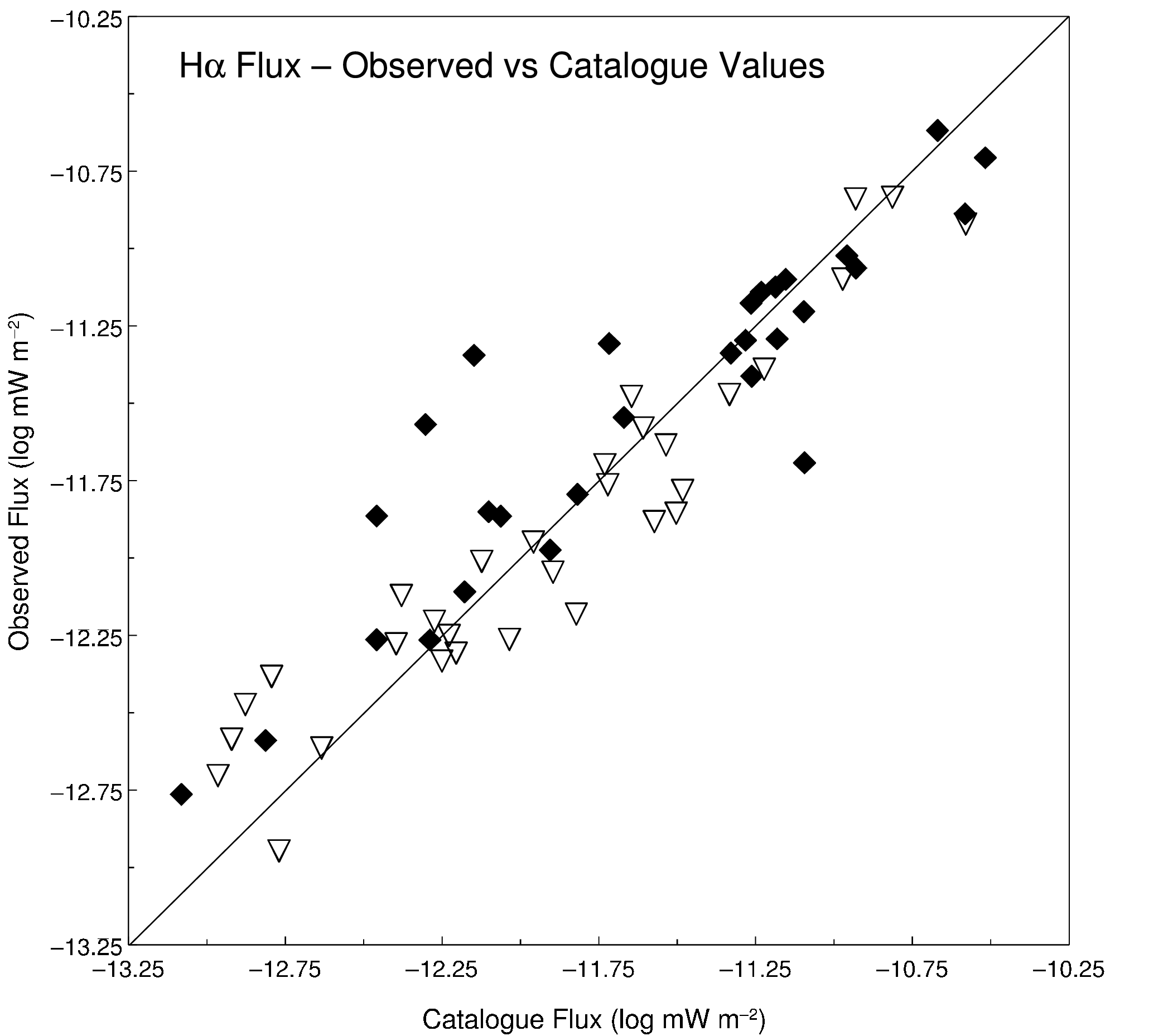}}
    \vspace{3mm}
    \hbox{\hspace{5mm}\includegraphics[width=73mm]{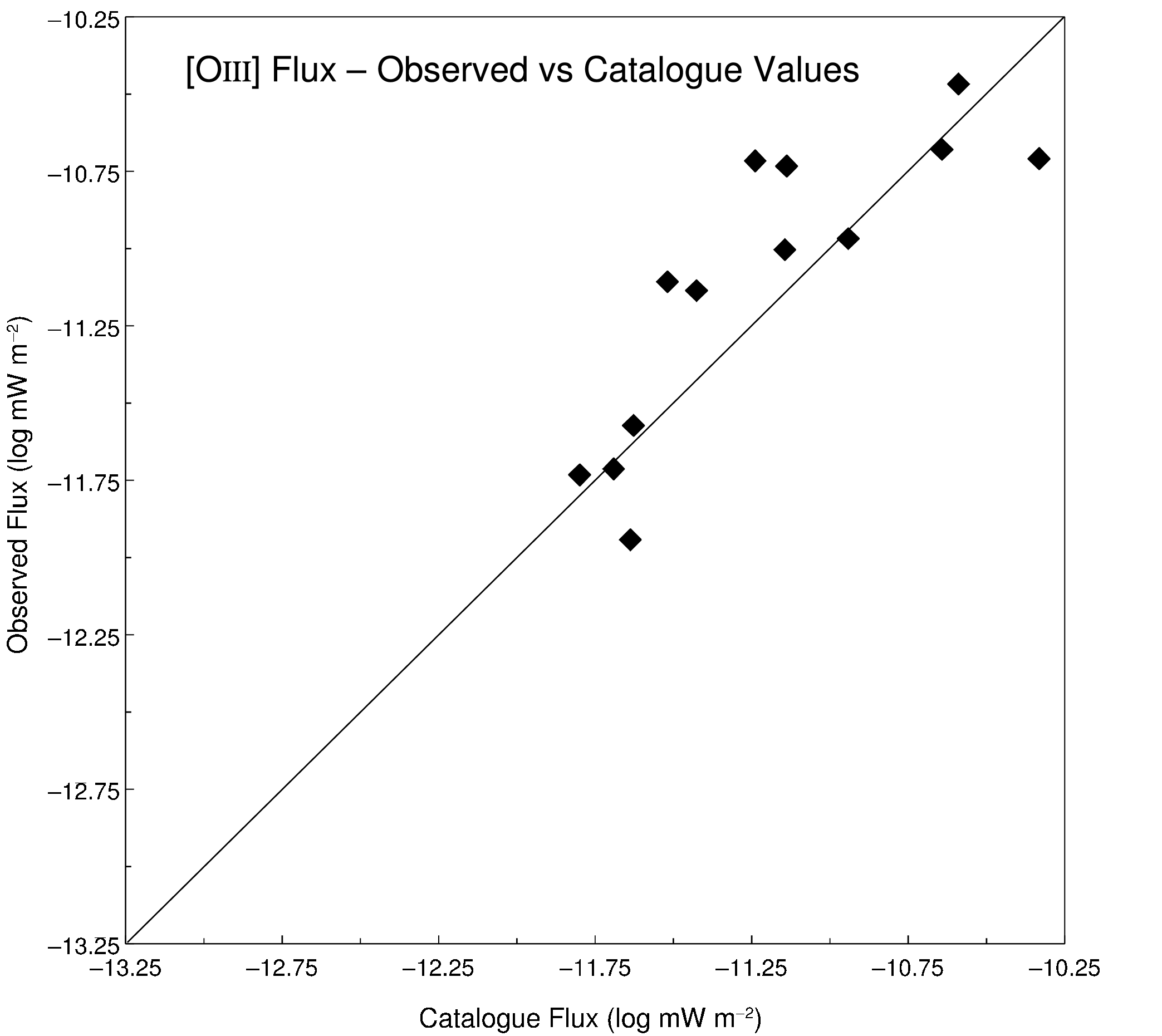}}
    \caption{Comparison of observed and catalogued H$\alpha$ and [O\,{\sc iii}] flux values (log mW\,m$^{-2}$).
             Diamonds denote objects observed on the first night, triangles the second.}
    \label{obscatflux}
\end{figure}

Fig.~\ref{obscatflux} compares observed and catalogued H$\alpha$ and
[O\,{\sc iii}] flux values. Four (non-Bulge) objects (indicated in
Table~\ref{fluxa}) are excluded from Figs.~\ref{dfluxobscat}
\&~\ref{obscatflux} because atmospheric conditions for those exposures
were non-photometric. It can be seen that there is little evidence of 
any systematic differences between observed and catalogued flux values.

Possible sources of error in our flux determinations are: 
poor photometry; 
residual stellar background flux after subtraction of the continuum; 
an off-centred peak in the H$\alpha$ on-band filter response curve; 
estimation of the flux contribution from [N\,{\sc ii}] lines; 
the accuracy of transmission coefficients in the corrected filter response tables 
(particularly for H$\alpha$); 
change in passbands of filters due to their age, input beam and ambient temperature; 
integrating the transmission profile of the filter response curves; 
uncertainties in radial velocities; 
changes in air pressure (airmass); 
and the aperture correction factor. 
Of these, we consider the accuracy of the filter response data the most 
significant and therefore suggest an error factor in the calculated flux 
values (mW\,m$^{-2}$) of $\pm$5 per cent for blue-shifted objects and 
$\pm$10 per cent for red-shifted objects.


\section{Extinction determinations}\label{extdets}

\subsection{Method}

Observed H$\alpha$ flux values are reduced from emitted flux values by 
interstellar extinction. By comparing these observed values with catalogued 
optical and radio fluxes, extinction values can be determined, and we now
describe the methods used to arrive at the values listed in
Table~\ref{fluxa}.

From \citet[][hereafter SP]{pottasch} the radio to H$\beta$ flux ratio is
$S_\nu / F(\mathrm{H}\beta) = 2.51 \times 10^7 T_\mathrm{e}^{0.53}
\nu^{-0.1} Y$ (Jy/mW\,m$^{-2}$) where $T_\mathrm{e}$ is the electron
temperature in K, $\nu$ is the radio frequency in GHz and $Y$ is a
factor incorporating the ionized He/H ratio. Setting $T_\mathrm{e} =
10^4$~K, $\nu = 5$~GHz, $Y = 1.1$ and converting to mJy gives

\begin{equation}
    S_\nu(6\mathrm{cm}) / F(\mathrm{H}\beta)_\mathrm{0} = 3.10 \times 10^{12} \; (\mathrm{mJy/mW\,m}^{-2}).
    \label{fratio}
\end{equation}

\noindent
H$\beta$ extinction ($C_\beta$) can be defined as the ratio of
expected to observed H$\beta$ flux:

\begin{equation}
    C_\beta = \log [ F(\mathrm{H}\beta)_\mathrm{0} / F(\mathrm{H}\beta) ]
    \label{cbeta}
\end{equation}

\noindent
If an electron density of $n_\mathrm{e} = 10^4$ cm$^{-3}$ is assumed,
and $T_\mathrm{e} = 10^4$~K, the Balmer-line intensity ratio,
$F(\mathrm{H}\alpha) / F(\mathrm{H}\beta) = 2.85$.  Combining this
factor with equations (\ref{fratio}) \& (\ref{cbeta}) gives an
expression for observed H$\alpha$ extinction in terms of the observed
H$\alpha$ flux (mW\,m$^{-2}$) and the (catalogue) radio flux (mJy):

\begin{equation}
    C_\alpha = \log [ 2.85 S_\nu(6\mathrm{cm}) / 3.10 \times 10^{12} F(\mathrm{H}\alpha)].
    \label{calphaobs}
\end{equation}

In order to compare observed {\it versus} catalogue extinction values, the
latter can be derived from just the catalogue H$\alpha$ and H$\beta$
flux values. We assume that the $F(\mathrm{H}\alpha) /
F(\mathrm{H}\beta)$ ratio in the catalogue is accurate, even if the
absolute fluxes are in doubt. We start with the full form of
equation~\ref{cbeta}:

\begin{equation}
    C_\beta = \log [ F(\mathrm{H}\beta)_\mathrm{0} / F(\mathrm{H}\beta) ] 
            = A_{4861} E_{B-V} / 2.5
    \label{exbeta}
\end{equation}

\noindent
where $A_\lambda$ is a wavelength dependent extinction coefficient. 
\linebreak It follows that

\begin{equation}
    C_\alpha = \log [ F(\mathrm{H}\alpha)_\mathrm{0} / F(\mathrm{H}\alpha) ] 
             = A_{6561} E_{B-V} / 2.5
    \label{exalpha}
\end{equation}

Combining equations \ref{exbeta} \& \ref{exalpha} with the
Balmer-line intensity ratio of 2.85 gives an expression for
calculating H$\alpha$ extinction values in terms of catalogue
H$\alpha$ and H$\beta$ flux values:

\begin{equation}
    C_{\alpha_\mathrm{opt}} = B \log [ F(\mathrm{H}\alpha)_\mathrm{cat} / 2.85 F(\mathrm{H}\beta)_\mathrm{cat} ],
    \label{calphacat}
\end{equation}

\noindent
where $ B = A_{6561} / ( A_{4861} - A_{6561} ) = 2.28 $ (using data from SP table~V-1). 

An alternate method of calculating the wavelength dependent constant $B$ in equation 
\ref{calphacat} is provided by \citet[][hereafter CCM]{cardelli} in terms of what they 
consider a more fundamental extinction law $A(\lambda) / A(V)$, where $A(\lambda)$ is 
the absolute extinction at $\lambda$ and $A(V)$ is the visual extinction. 
Based on empirical data, CCM provide a mean 
\linebreak $R_\mathrm{V}\,[ = A(V)/E(B-V)]$
extinction law with wavelength dependent coefficients 
($x = \lambda^{-1} \mu \mathrm{m}^{-1}$) that takes the form

\begin{equation}
    \langle A(\lambda) / A(V) \rangle = a(x) + b(x)/R_\mathrm{V},
    \label{rvextlaw}
\end{equation}

\noindent
where $R_\mathrm{V}$ is the ratio of total to selective extinction.
Dividing the expression for $B$ by $A(V)$ and substituting the 
expression $a(x) + b(x)/R_\mathrm{V}$ gives

\begin{equation}
    B = \frac{(a_\alpha + b_\alpha/R_\mathrm{V})}
             {(a_\beta + b_\beta/R_\mathrm{V}) - (a_\alpha + b_\alpha/R_\mathrm{V})},
    \label{brv}
\end{equation}

\noindent
where $a_\alpha = a(1/\lambda_\mathrm{H\alpha})$, $a_\beta =
a(1/\lambda_\mathrm{H\beta})$, etc.  Using $R_\mathrm{V} = 3.1$ and
appropriate values provided by CCM (eqs. 3a \& 3b) for this polynomial 
in $x$ gives $B = 2.36$, which is slightly higher than SP's value of $2.28$. 
However, $B$ is now dependent on the value of $R_\mathrm{V}$ so equation
\ref{calphacat} takes the $R_\mathrm{V}$ dependent form:

\begin{equation}
    C_{\alpha_\mathrm{opt}} = B(R_\mathrm{V}) \log [ F(\mathrm{H}\alpha)_\mathrm{cat} / 2.85 F(\mathrm{H}\beta)_\mathrm{cat}].
    \label{calphacat2}
\end{equation}

\noindent
This enables $R_\mathrm{V}$ to be calculated in terms of the observed
extinction $C_\alpha$ (equation~\ref{calphaobs}) and catalogued
H$\alpha$ and H$\beta$ flux ratios:

\begin{equation}
    R_\mathrm{V} = \frac{C_\alpha (b_\alpha - b_\beta) + D b_\alpha}
                        {C_\alpha (a_\beta - a_\alpha) - D a_\alpha}.
    \label{rvobscat}
\end{equation}

\noindent
where $D = \log [ F(\mathrm{H}\alpha)_\mathrm{cat} / 2.85 F(\mathrm{H}\beta)_\mathrm{cat}]$.

\begin{figure}
    \hbox{\hspace{5mm}\includegraphics[width=72mm]{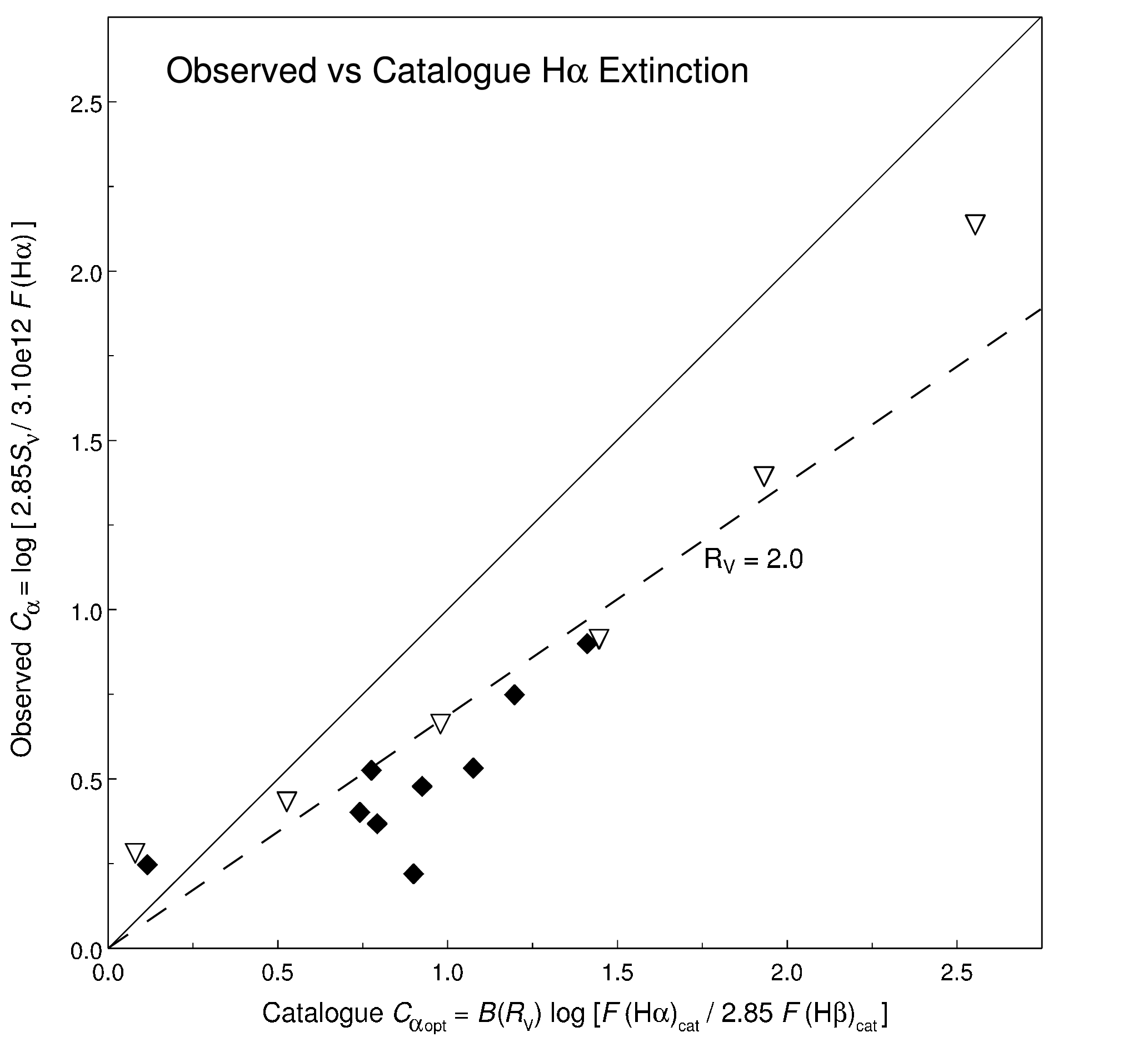}}
    \vspace{3mm}
    \hbox{\hspace{5mm}\includegraphics[width=79mm]{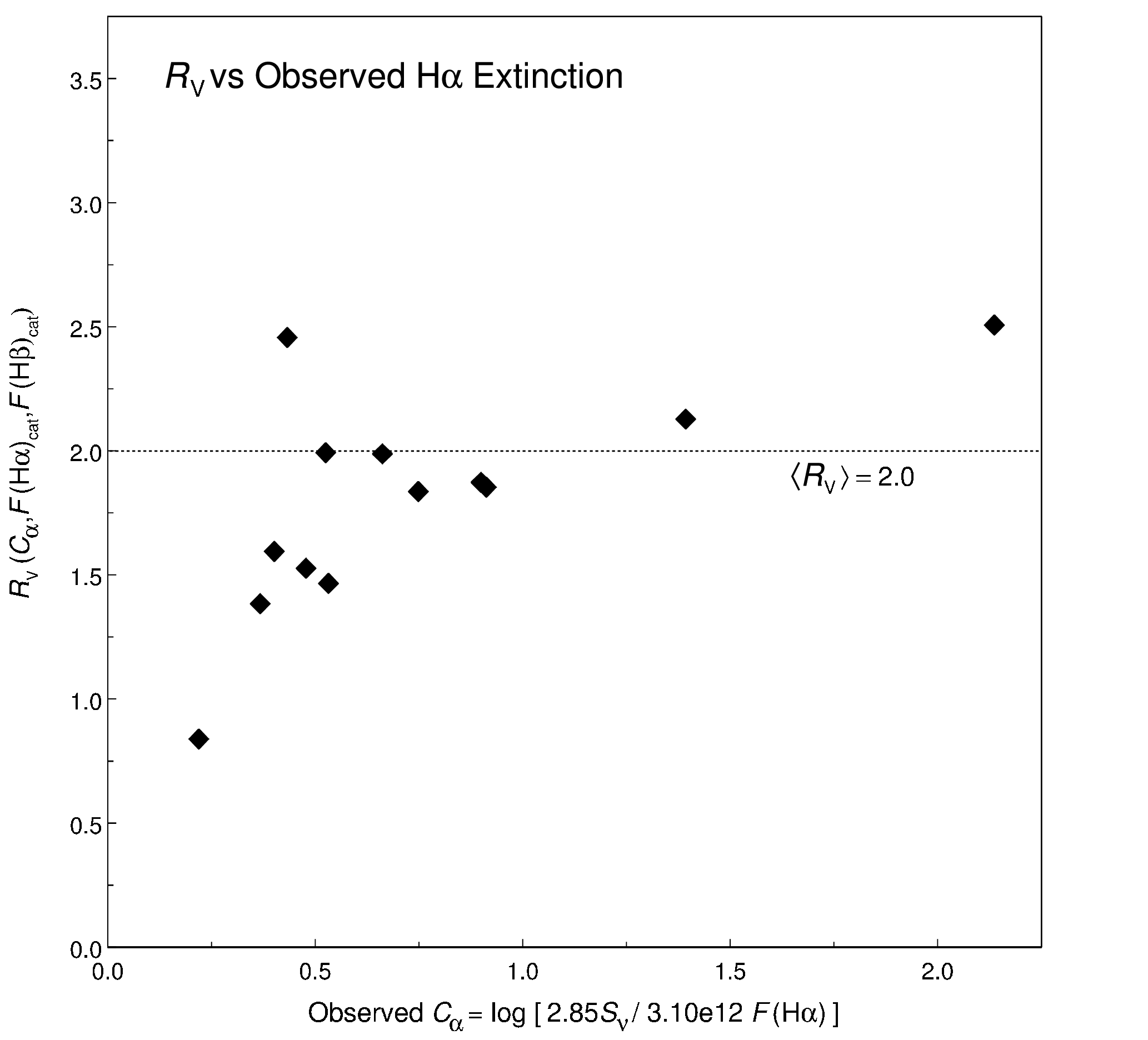}}
    \caption{{\bf(a)} 
             Comparison of observed H$\alpha$ extinction values, 
             $C_\alpha = \log\, [2.85\, S_\nu\,(6\mathrm{cm}) / 
             3.10 \times 10^{12}\, F(\mathrm{H}\alpha)]$ 
             and catalogued  H$\alpha$ extinction values,
             $C_{\alpha_\mathrm{opt}} = B\,(R_\mathrm{V}) 
             \log\, [F(\mathrm{H}\alpha)_\mathrm{cat} / 
             2.85\, F(\mathrm{H}\beta)_\mathrm{cat}]$ 
             where $B\,(3.1) = 2.36$, for Bulge subset B objects. 
             Diamonds denote objects observed on the first night, 
             triangles the second. 
            {\bf(b)} 
             Comparison of $R_\mathrm{V}$ (equation~\ref{rvobscat}) 
             and observed H$\alpha$ extinction values. 
             004.0$-$11.1 (M3-29) and 355.1$-$06.9 (M3-21) are excluded.}
    \label{extinc}
\end{figure}

\subsection{Results}

Our observed (equation \ref{calphaobs}) and catalogue 
(equation \ref{calphacat}) extinction values are listed in columns
nine and ten of Table~\ref{fluxa}.  At high brightness
temperatures, the optical depth of PNe reduces observed radio emission,
so the observed extinction values are likely to be unreliable, as are
those calculated from radio fluxes below 10 mJy (indicated in column
13). Therefore Fig.~\ref{extinc}~(a) only compares observed and
catalogue extinction values for a subset of Bulge objects (hereafter
subset B) with 100 mJy $> S_\nu >$ 10 mJy and $T_\mathrm{b} < 10^3 \mathrm{K}$, 
where $T_\mathrm{b} = 70.62 S_{5\,\mathrm{GHz}} / {\theta_\mathrm{line}}^2$ 
(mJy/\arcsec).  It can be seen that the values determined from
our H$\alpha$ flux and catalogued radio flux tend to be lower than
those calculated with $R_\mathrm{V} = 3.1$ and catalogue H$\alpha$ and
H$\beta$ fluxes. Subset B conforms to the Bulge criteria of \citet{bensby} 
with the exception of 004.0$-$11.1 (M3-29) where $b = -11.1\degr$ and 
350.9$+$04.4 (H2-1) where $S_{5\,\mathrm{GHz}} = 61$ mJy.

Calculated values for $R_\mathrm{V}$ (equation~\ref{rvobscat}) are
listed in column 11 of Table~\ref{fluxa}. 
Fig.~\ref{extinc}~(b) compares $R_\mathrm{V}$ with observed H$\alpha$ 
extinction values for subset B. 
The value of $R_\mathrm{V}$ is seen to be highly sensitive
to differences between $C_\alpha$ and $C_{\alpha_\mathrm{opt}}$ 
with values $>10$ for objects 004.0$-$11.1 (M3-29) and 355.1$-$06.9 (M3-21) 
not included in the figure for the sake of clarity 
(regardless of method, calculated extinction values for these two objects 
are very low with $\Delta F(\mathrm{H}\alpha) = 0.1$, suggesting a possible 
error in the catalogue radio values). Apart from the value for 009.4$-$09.8
(M3-32) and the above two objects, there appears to be a correlation between
increasing values of $R_\mathrm{V}$ and observed H$\alpha$ extinction.

Because of the sensitivity of $R_\mathrm{V}$ to differences between
observed and catalogue extinction values, and the uncertainty in these
values, a reasonable error $\Delta R_\mathrm{V}$ has been calculated
with $C_\alpha \pm 0.1$ using

\begin{equation}
    \Delta R_\mathrm{V} = R_\mathrm{V}(C_\alpha+0.1) - R_\mathrm{V}(C_\alpha-0.1).
    \label{deltarv}
\end{equation}

\noindent
Using $\Delta R_\mathrm{V}$ as the error in $R_\mathrm{V}$, the mean
value $\langle R_\mathrm{V} \rangle$ for subset B is found to be 2.0
(indicated on Fig.~\ref{extinc}).  Values for $\Delta
R_\mathrm{V}$ are listed in column 12 of Table~\ref{fluxa}. 
For subset B a comparison of $R_\mathrm{V}$ with catalogue 6 cm radio values 
shows no evidence of a correlation between low values of $R_\mathrm{V}$ 
and low radio fluxes. In fact, radio flux values would have to increase 
by a factor $\sim 3$ in order for $\langle R_\mathrm{V} \rangle \approx 3.1$.

\begin{figure}
    \hbox{\hspace{2mm}\includegraphics[width=76mm]{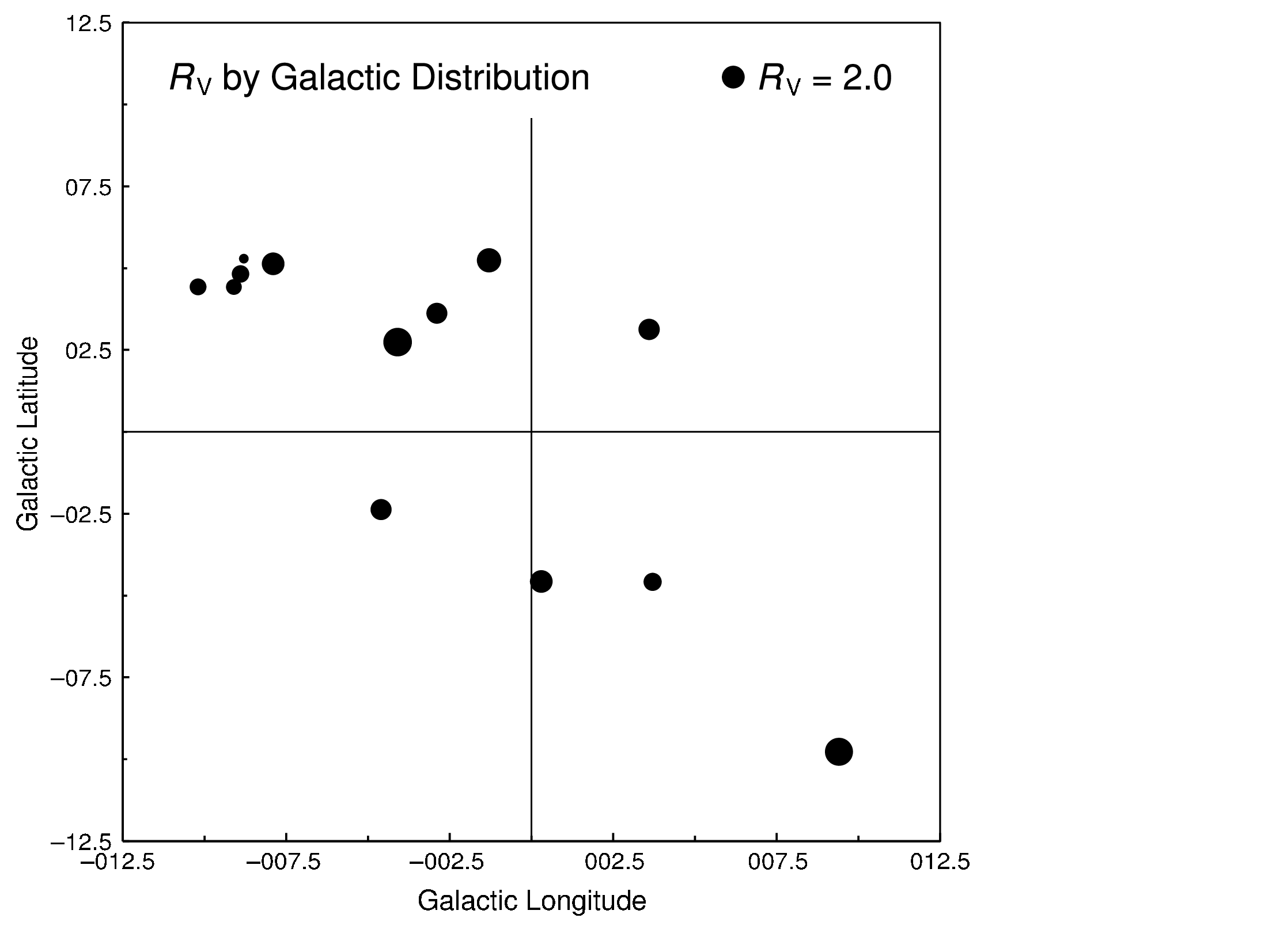}}
    \caption{Comparison of $R_\mathrm{V}$ with Galactic distribution for Bulge subset 
             B objects. Relative $R_\mathrm{V}$ is indicated by dot size with 
             $R_\mathrm{V} = 2.0$ scale indicated at top right.
             004.0$-$11.1 (M3-29) and 355.1$-$06.9 (M3-21) are excluded.}
    \label{rvgaldist}
\end{figure}

Fig.~\ref{rvgaldist} compares $R_\mathrm{V}$ with Galactic
distribution for subset B (again excluding 004.0$-$11.1 and
355.1$-$06.9). For the line of sight towards 351.1$+$04.8, the three
objects close together on the sky (H2-1, M1-19 and M2-5) have, within
their uncertainties, $\langle R_\mathrm{V} \rangle = 1.2$.
It is worth noting that the two Sagittarius dwarf galaxy objects 
004.8$-$22.7 (He2-436) and 006.8$-$19.8 (Wray16-423) have $R_\mathrm{V} = 3.1$ 
within their uncertainties. 

Objects 352.6$+$00.1 (H1-12) and 352.8$-$00.2 (H1-13) have not been included in 
subset B because of their very high radio flux and being in the Galactic plane. 
\citet{caswell8701} identify these positions with H\,{\sc ii} regions with 
$S_{5\,\mathrm{GHz}}$ of 2.1 and 2.8 Jy respectively, giving higher values of 
$R_\mathrm{V}$ (2.4 and 3.5 respectively). 

Values for $R_\mathrm{V\,cat}$ and $\Delta R_\mathrm{V\,cat}$ were also calculated 
using catalogue H$\alpha$ flux values in place of our observed values. 
The mean catalogue value $\langle R_\mathrm{V\,cat} \rangle$ for subset B 
was found to be 2.2 and for the line of sight towards 351.1$+$04.8 it was 1.4. 
A comparison of observed and catalogue $R_\mathrm{V}$ values shows good agreement 
for subset B, apart from values for 009.4$-$09.8 (M3-32), 
352.1$+$05.1 (M2-8) and 003.6$+$03.1 (M2-14). This suggests that in addition to our 
newly observed H$\alpha$ flux values, existing catalogue flux values also predict 
that interstellar extinction toward the Bulge is lower than that corresponding to 
the standard extinction curve $R_\mathrm{V} = 3.1$. 
A comparison of observed and catalogue $S_\nu$/H$\alpha$ extinction values also 
shows good agreement for subset B, apart from values for 352.1$+$05.1 (M2-8)
and 003.6$+$03.1 (M2-14), which have differences between observed and catalogued 
H$\alpha$ flux values, $\Delta F(\mathrm{H}\alpha)$ of 0.4 and 0.8 respectively. 
Excluding the above three objects from subset B has no effect of the calculated values 
of $\langle R_\mathrm{V} \rangle$ and $\langle R_\mathrm{V\,cat} \rangle$. There is 
no evidence for systematic differences due to time of observation or position in 
the sky for these three objects.


\section{Discussion: Extinction}\label{discus}

\subsection{Evidence for anomalous extinction}

In an optical and UV analysis of Bulge PNe, \citet{walton93iau} found that 
$R_\mathrm{V} = 2.29\pm0.50$, concluding that the reddening law is 
steeper towards the Bulge.
\citet[][hereafter STAS]{stasinska} and \citet[][hereafter TASK]{tylenda} 
discuss methods for determining the interstellar extinction of Galactic 
PNe and compare the two most frequently used; the Balmer decrement 
H$\alpha$/H$\beta$ ratio method and the radio/H$\beta$ flux ratio method.
Like all methods, the Balmer decrement method depends on the adopted 
extinction curve (or mean interstellar extinction law).
In addition, it requires good quality optical data acquired by filter photometry, 
or spectrophotometry that encompasses the full extent of the nebula.

In principle, the radio/H$\beta$ flux ratio method is directly capable
of determining the actual value of H$\beta$ extinction, $C_\beta$. 
However, optical thickness of the nebula can lead to an underestimation
of $C_\beta$. In addition, the predicted H$\beta$ flux depends on adopted 
electron temperature and He ionization. Blending of He\,{\sc ii} 
Pickering 8 emission with H$\beta$ flux also adds a (very) small error. 
Interferometric radio data is preferred over single dish
measurements because of the problem of confusion with other sources.
\citet{zijlstra8909} have noted that their VLA fluxes are systematically 
lower than the results from single dish observations, due to possible 
cutting off of emission from fainter, extended regions of PNe. However, 
as noted previously, for subset B these 6 cm radio fluxes would have to be 
underestimated by a factor of $\sim 3$ for $\langle R_\mathrm{V} \rangle 
\approx 3.1$. Allowing that radio fluxes below 10 mJy are often underestimated 
for many faint PN \citep[e.g.][]{potzij94}, strong radio sources still have 
$C_\mathrm{opt}$ higher than $C_\mathrm{rad}$. 


TASK compared the extinction values $C_\mathrm{rad}$ and $C_\mathrm{opt}$ 
for Bulge and Disk PNe (their figs. 6 to 11), and showed that, for all PNe, 
$C_\mathrm{rad}$ becomes systematically smaller than $C_\mathrm{opt}$ for 
increasing extinction. They proposed that one of the possible reasons 
\citep[as mentioned by][]{cahn9209} may lie in the value of $R_\mathrm{V}$ 
adopted in the extinction curve. They estimated that $R_\mathrm{V} = 2.7$ 
would bring $C_\mathrm{opt}$ to a rough agreement with $C_\mathrm{rad}$. 


\citet{koppen9809} analysed the optical extinction of 271 Bulge PNe, and 
found (see their figs. 1 and 2) a mean extinction around $C_\mathrm{opt} = 2$, 
and a systematic decrease of the extinction (from 3.5 to 0.5) with increasing 
Galactic latitude, as well as a genuine scatter about the relation. 
Both are accounted for by a model of small dust clouds randomly distributed 
in an exponential disc. They found an extinction $A_\mathrm{V} = 1.4$ in the 
plane, and = 27 to the centre, in agreement with far IR-studies (COBE results).

The standard extinction law for the diffuse interstellar medium is
based on observations of O and B stars within a few kpc, but these
stars must suffer sufficient extinction to allow measurement of their
reddening. This selection effect favours denser regions of the
interstellar medium in the Solar neighbourhood. For PNe the lines of
sight are longer and could cross a more dilute medium while still
giving measurable extinction.  A lower value of $R_\mathrm{V}$
($<2.5$) could be due to a different size distribution of dust
particles (i.e. smaller) in these more dilute regions.
For the line of sight toward HD 210121, obscured by the high-latitude cloud
DBB 80, \citet{larson9611,larson0004} find $R_\mathrm{V} = 2.1 \pm 0.2$, 
which they attribute to an abundance of small dust grains.

More recently \citet{udalski} presents results of analysis of
photometric data of Bulge fields from the OGLE II microlensing survey
\citep{udalski02}. He shows $R_\mathrm{V}$ in general to be much
smaller than that corresponding to the standard extinction curve
$R_\mathrm{V} = 3.1$, and that it varies considerably along different
lines of sight.  \citet{popowski} has also suggested that interstellar
extinction toward the Bulge might be anomalous.
Using CCM's extinction law model, Udalski calculates values for
$R_\mathrm{V}$ ranging from 1.75 to 2.50 in four small regions of the
Bulge. Assumed mean brightness, metallicity distribution, and large
differences in the mean distance of red clump giants are all dismissed
as unlikely reasons for these highly variable values of $R_\mathrm{V}$.

Our results confirm these previous findings, and we find evidence for
even lower vales of $R_\mathrm{V}$.  We believe our flux and
extinction determinations to be accurate because: our observed
H$\alpha$ fluxes are based on filter photometry; catalogue H$\alpha$
and H$\beta$ flux ratios are normally based on single spectra and are
unlikely to suffer from systematic errors (even if the absolute fluxes
are in doubt); and although catalogue radio fluxes are less certain,
we have used a more reliable subset of Bulge objects ($S_\nu >$ 10 mJy 
and $T_\mathrm{b} < 10^3 \mathrm{K}$) for calculating values for 
$R_\mathrm{V}$ and $\Delta R_\mathrm{V}$. However, radio fluxes should 
be remeasured to obtain the most accurate value for $R_\mathrm{V}$.
Our mean value of $\langle R_\mathrm{V} \rangle = 2.0$ provides further 
evidence that most lines of sight towards the Bulge cross a less dense 
medium containing smaller dust grains. It may be the case that the 
standard extinction law for the diffuse interstellar medium only applies 
to the local region within a few kpc of our Sun.

\subsection{Extinction and the warm ionized medium}

The PNe in our Bulge subset B sample have extinction up to 
$A_\mathrm{V} \approx 5$. The usual conversion between 
extinction and hydrogen column density, e.g.,

\begin{equation}
    N({\rm H}) \approx A_\mathrm{V}\, 2 \times \,10^{22}\,\rm cm^{-2} 
    \label{coldens}
\end{equation}

\noindent
gives column densities up to $10^{23}\,\rm cm^{-2}$. For a line of
sight of 8 kpc, the average density becomes $n_{\rm H} \le 4\,\rm cm^{-3}$.

Components of the interstellar medium are: Molecular clouds; Diffuse
clouds; Warm neutral medium (WNM); Warm ionized medium (WIM); and Hot
ionized medium (HIM).  The warm neutral medium (WNM) has a typical
density of 0.1--10\,cm$^{-3}$, and scale height around 220\,pc. The
warm ionized medium (WIM) has a typical density of 0.3--10\,cm$^{-3}$
in the Galactic plane, with a scale height of 1 kpc \citep{boulares}. 
The HIM has density below 0.01\,cm$^{-3}$ and a scale height of 3\,kpc. 
The WNM, WIM and HIM account for most of the volume in the Galactic disc. 
The WNM is absent near the Sun (within 100\,pc in most directions): 
the local volume is filled with a HIM with embedded cold clouds.

Our extinction values of $A_\mathrm{V} \le 5$\,mag suggest that the lines of
sight do not intercept dense molecular clouds. The density in the HIM
is too low to give appreciable extinction even out to the Bulge.  The
extinction to the Bulge PNe is likely to arise from a mixture of WIM
and WNM.  The line of sight at higher latitudes leaves the WNM within
2--3\,kpc (no Bulge PNe are known within 2 degrees of the plane): the
remainder will be mainly within the WIM.

The relation between $R_\mathrm{V}$ and $C_\alpha$ depicted in Fig.~\ref{extinc}~(b) 
suggests that the lowest $R$ values correspond to the lower density
ISM component. A large fraction of the lines of sight fall in
WIM-dominated regions above/below the disc. The WNM component is more
likely to cause the variable amount of additional extinction.  The
fact that three PNe located within a degree of each other show very
similar -- and very low -- $R_\mathrm{V}$ shows that the low-density 
structures are similar for line of sights within approximately 100\,pc: 
this provides support for the identification of this component with 
the WIM. If correct, the distribution of $R_\mathrm{V}$ in 
Fig.~\ref{extinc}~(b) suggests that the WIM has $R_\mathrm{V} \approx 1.5$, 
and the WNM $R_\mathrm{V} \approx 2.5$. The latter value is close to the 
lowest local values found by CCM.

The anomalous extinction towards the Bulge therefore suggest that the
WIM shows smaller dust grains than found in the WNM or molecular
clouds. Possible explanations include dust-grain evolution, and the
effects of supernovae. The ISM cycles between high-density and low-density
phases. If the dust is slowly destroyed during the low-density phases,
the size of the dust grains may measure the time since the last high
density phase. The large scale height of the WIM suggests its contents 
may cycle much slower.

Supernova-driven shocks can also destroy dust grains. Supernovae are more 
dominant in the inner galaxy \citep{heiles}: The low-density medium in
the inner Galaxy could therefore show systematically smaller dust grains.
This would limit the applicability of the standard Galactic extinction curve 
to  quiescent regions of the Galaxy, with steeper curves found in energetic
environments.

\subsection{Grain sizes}

\begin{figure*}
    \hbox{\hspace{10mm}\includegraphics[width=155mm]{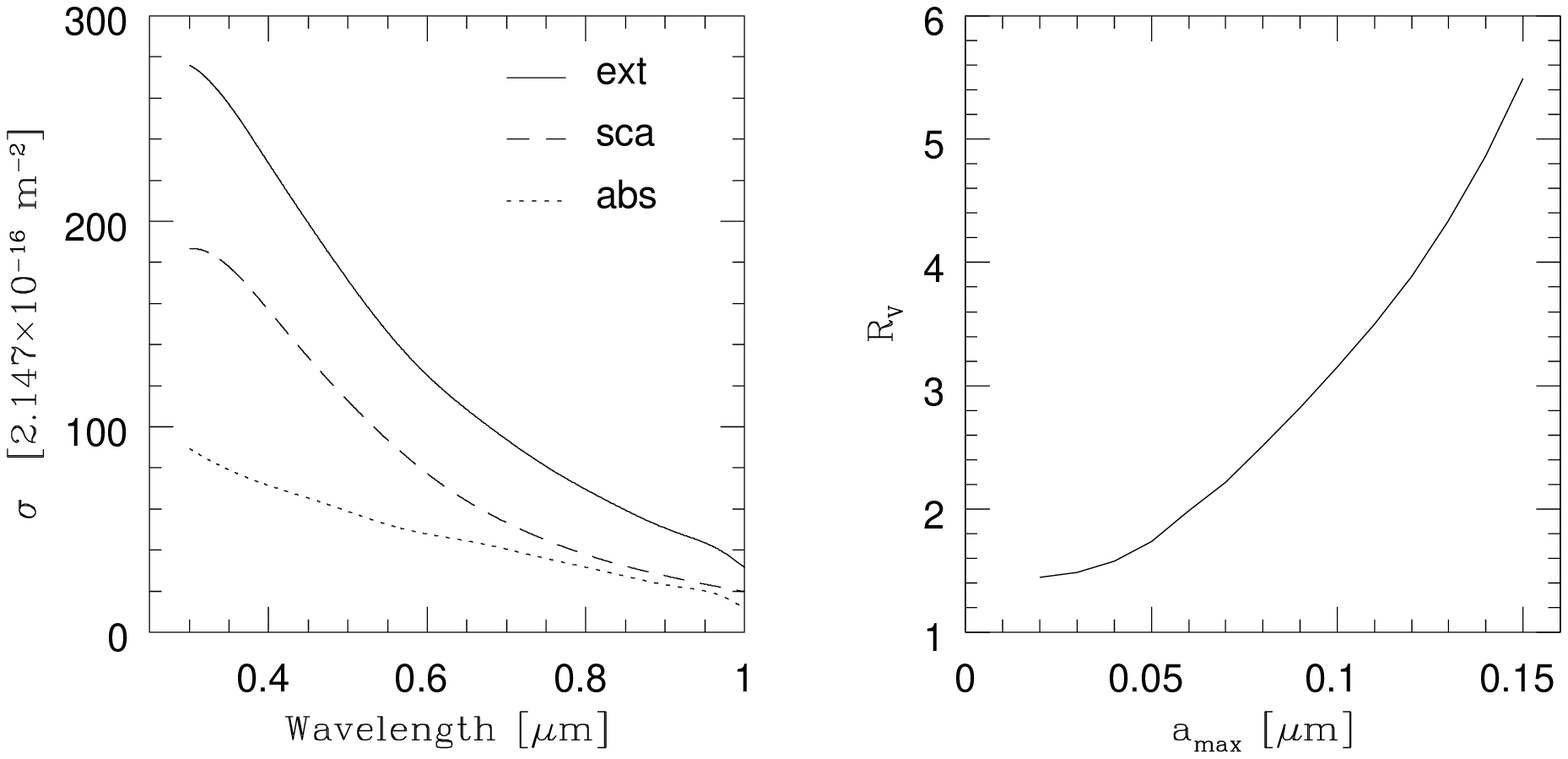}}
    \caption{Left: A plot, between 0.3 and 1.0 microns, of the absorption,
             scattering and extinction cross-sections for a Greenberg model 
             with a turnover radius of 0.08 microns, $q=2$, and core radius 
             equal to 0.042 microns. The cross-sections are measured relative 
             to an area of $2.147 \times 10^{-16}\,\rm m^{-2}$;
             Right: Calculated values for $R_\mathrm{V}$ for a Greenberg dust 
             model, as function of the turnover radius for the size distribution.}
    \label{dust}
\end{figure*}

Although different from the locally derived values of $R_\mathrm{V}$, 
it is possible to show that our low values derived above can be 
reproduced using dust models.

The Greenberg unified dust model \citep{greenberg,li} is capable of
generating a variety of values of $R_\mathrm{V}$, which encompass the
observational value derived in the present work, by varying the
turnover radius, a key parameter of the model.

\citet{li} use a three component model to compute the
entire interstellar extinction law: large composite grains, small
carbonaceous grains and a population of PAH molecules. In the present
work, we consider only the large composite grains, which consist of a
silicate core and a refractory organic mantle. Addition of the smaller
components would tend to increase ultraviolet extinction relative to
that in the visible, and thus drive the model results in the present
work to smaller values of $R_\mathrm{V}$. The size distribution function
used for the composite grains is \citep{li}

\begin{equation}
n(a) = n_{0} \exp \left\{
                     -5 \left(
                             \frac{a - a_{c}}
                                  {a_{m}}
                        \right)^{q}
                  \right\}
\end{equation}

\noindent 
where $n(a)$ is the number density of grains of total radius
$a$. These composite grains have a silicate core of radius $a_{c}$,
and a turnover radius, $a_{m}$. The exponent, $q$, ensures that the
distribution of grains falls rapidly for sizes larger than $a_{m}$. In
the present work we use values of $q=2$, as in \citet{li},
and set $a_{c} = 0.042$\,$\mu$m which was the radius of infinite
cylinders in Li \& Greenberg, but spheres in the present work. The
value of $a_{m}$ was varied to generate a table of theoretical values
of $R_\mathrm{V}$.  The core material used by \citet{greenberg} was
based on values for an amorphous olivine \citep{dorschner}. In
the present work we replace the olivine by amorphous forsterite,
defined by optical constants \citep[in][]{scott}, on the grounds
that the olivine contains iron, whilst interstellar silicate is now
believed to be dominated by magnesium-rich minerals 
\citep[see, for example,][]{kimura}. Optical constants for the organic
refractory mantle were taken from \citet{greenberg}. 

Fig.~\ref{dust} (left panel) shows the optical scattering and
absorption coefficients, and the extinction law, for the adopted
`standard' model, with one particular value of the turnover radius,
$a_{m} = 0.08$\,$\mu$m.  In Fig.~\ref{dust} (right panel), we plot the
value of $R_\mathrm{V}$ as a function of the turnover radius, $a_{m}$. The
theoretical values shown encompass all values of $R_\mathrm{V}$ from the
smallest-but-one of the observational figures obtained in the present
work, through the rest of the present samples and the more traditional
values of $\sim 3.1$ expected for dust in diffuse clouds, to values in
excess of $5$, expected for dense, cold, clouds. We note that the key
parameter is the amount of grain material found in large grains (with
radii larger than $\sim 0.1$\,$\mu$m).

We also note that, for any value of $a_{m}$, a smaller value of 
$R_\mathrm{V}$ may be obtained by adding the populations of small 
carbonaceous grains and PAHs, whilst a larger value is computable by 
overlaying the composite grains with layers of ices, as is likely in 
cold clouds. For a comparison of ice-coated and bare grains, see \citet{gray}.


\section{Conclusions}\label{concl}

\subsection{Angular diameters}

We have calculated angular diameters for 69 PNe by deconvolving the
point spread function of field stars from Gaussian FWHM profiles. A
correction factor, $\gamma$, was applied to determine the true diameter. 
$\gamma$ was approximated by a simple analytic function and diameters 
were calculated using two models: a constant emissivity shell 0.8,
$\theta_\mathrm{shell}$, and photoionization line emission, 
$\theta_\mathrm{line}$. We present the mean value derived from 
the two models, $\theta_\mathrm{mean}$, as our best estimate. 
For some objects the low intensity contour plots revealed 
an elliptical structure that was not always apparent from the FWHM
measurements. Axis averaged mean angular diameters of fully resolved 
objects ranged from $2.8 \pm 0.1$ to $12.7 \pm 1.2$ arc seconds. 
18 H$\alpha$ images were restored with the \citeauthor{rich72}-\citeauthor{lucy7406} algorithm 
and restored diameters tended to be around one arc second smaller below 4 arc seconds.

\subsection{Fluxes}

The total flux detected through the filters was calculated for 70 PNe. 
Doppler shifted H$\alpha$, [O\,{\sc iii}] (and [N\,{\sc ii}]) wavelengths 
observed at the telescope were computed from radial velocities, and filter 
transmission coefficients applied to arrive at actual flux densities. 
For H$\alpha$ and [O\,{\sc iii}] there is little evidence of any 
systematic differences between observed and catalogued flux values. 
The accuracy of filter response data was considered the most significant 
source of error and estimated at between $\pm$5 to $\pm$10 per cent.

\subsection{Extinction}

Observed H$\alpha$ extinction in the direction of the Bulge was
determined using the radio to H$\beta$ flux ratio, the expected to
observed H$\beta$ flux ratio and the Balmer-line intensity ratio,
giving an expression in terms of observed H$\alpha$ flux and catalogue
radio flux. Catalogue H$\alpha$ extinction values were also derived
from catalogued H$\alpha$ and H$\beta$ flux values in terms of the
fundamental extinction law $A(\lambda) / A(V)$ by means of an
$R_\mathrm{V}$ dependent function. 
We compared observed and catalogue extinction values for a subset of 
Bulge objects and found that values determined from observed H$\alpha$ 
flux and catalogued radio flux tended to be lower than those calculated 
with $R_\mathrm{V} = 3.1$ and catalogue H$\alpha$ and H$\beta$ fluxes. 
A method for determining $R_\mathrm{V}$ was derived in terms of observed 
extinction and catalogued H$\beta$ and H$\alpha$ flux values. 
We found $R_\mathrm{V}$ to be highly sensitive to differences between
observed and catalogue extinction values. There also appears to be a
correlation between increasing values of both $R_\mathrm{V}$ and
observed H$\alpha$ extinction, which is in agreement with the findings 
of STAS discussed above. Estimating a reasonable error in $R_\mathrm{V}$, 
we find observed $\langle R_\mathrm{V} \rangle = 2.0$, which fits within 
the range calculated by Udalski and agrees with his conclusion that 
toward the Bulge interstellar extinction is steeper than $R_\mathrm{V} = 3.1$.
Using catalogue H$\alpha$ and $S_\nu$ flux values we find good agreement, 
with $\langle R_\mathrm{V\,cat} \rangle = 2.2$. 
For one small region in our sample we find $\langle R_\mathrm{V} \rangle$ 
as low as 1.2, which suggests that $R_\mathrm{V}$ does indeed vary 
considerably along different lines of sight. 
We have shown that our values of $R_\mathrm{V}$ can be reproduced 
using dust models with a turnover radius of 0.08 microns. A lower value 
of $R_\mathrm{V}$ may also be obtained by adding the populations of small 
carbonaceous grains and PAHs.


\section*{Acknowledgements}

We are grateful to ESO for making telescope time available on the NTT.
PMER acknowledges receipt of a PPARC studentship. 
Astrophysics at UMIST is supported by PPARC. 
DM is supported by FONDAP Center for Astrophysics 15010003. 


\bsp            

\label{lastpage}


\begin{thebibliography}{}

    \bibitem[\protect\citeauthoryear{Acker et al.}{1992}]{acker}
        Acker~A., Marcout~J., Ochsenbein~F., Stenholm~B., Tylenda~R., 1992, 
        \emph{Strasbourg - ESO catalogue of galactic planetary nebulae}, 
        ESO, Garching
    \bibitem[\protect\citeauthoryear{Balick}{1987}]{balick}
        Balick~B., 1987,
        \aj, 94, 671
    \bibitem[\protect\citeauthoryear{Bedding \& Zijlstra}{1994}]{bedding}
        Bedding~T.~R., Zijlstra~A.~A., 1994,
        \aap, 283, 955 (BZ)
    \bibitem[\protect\citeauthoryear{Bensby \& Lundstr{\" o}m}{2001}]{bensby}
        Bensby~T., Lundstr{\" o}m, I., 2001, 
        \aap, 374, 599 
    \bibitem[\protect\citeauthoryear{Boulares \& Cox}{1990}]{boulares}
        Boulares~A., Cox~D.~P., 1990,
        \apj, 365, 544
    \bibitem[\protect\citeauthoryear{Cahn, Kaler \& Stanghellini}{1992}]{cahn9209}
        Cahn~J.~H., Kaler~J.~B., Stanghellini~L., 1992, 
        \aaps, 94, 399
    \bibitem[\protect\citeauthoryear{Cardelli, Clayton \& Mathis}{1989}]{cardelli}
        Cardelli~J.~A., Clayton~G.~C., Mathis~J.~S., 1989, 
        \apj, 345, 245 (CCM)
    \bibitem[\protect\citeauthoryear{Caswell \& Haynes}{1987}]{caswell8701}
        Caswell~J.~L., Haynes~R.~F., 1987, 
        \aap, 171, 261
    \bibitem[\protect\citeauthoryear{Dorschner et al.}{1995}]{dorschner}
        Dorschner~J., Begemann~B., Henning~T., Jaeger~C.,  Mutschke~H., 1995,  
       \aap, 300, 503
    \bibitem[\protect\citeauthoryear{Dudziak et al.}{2000}]{dudziak}
        Dudziak~G., P{\' e}quignot~D., Zijlstra~A.~A., Walsh~J.~R., 2000, 
        \aap, 363, 717 
    \bibitem[\protect\citeauthoryear{Durand et al.}{1998}]{durand}
        Durand~S., Acker~A., Zijlstra~A.~A., 1998,
        \aaps, 132, 13 
    \bibitem[\protect\citeauthoryear{Gray}{2001}]{gray}
        Gray~M.~D., 2001, 
        \mnras, 324, 57
    \bibitem[\protect\citeauthoryear{Greenburg \& Li}{1996}]{greenberg}
        Greenberg~J.~M., Li~A., 1996, 
        \aap, 309, 258
    \bibitem[\protect\citeauthoryear{Hamuy et al.}{1994}]{hamuy}
        Hamuy~M., Suntzeff~N.~B., Heathcote~S.~R., Walker~A.~R., Gigoux~P., Phillips~M.~M., 1994, 
        \pasp, 106, 566 
    \bibitem[\protect\citeauthoryear{Heiles}{1987}]{heiles}
        Heiles~C., 1987, 
        \apj,  315, 555     
    \bibitem[\protect\citeauthoryear{Kimura et al.}{2003}]{kimura}
        Kimura~H., Mann~I., Jessberger~E.~K., 2003, 
        \apj, 583, 314
    \bibitem[\protect\citeauthoryear{Koppen \& Vergely}{1998}]{koppen9809}
        Koppen~J., Vergely~J.-L., 1998, 
        \mnras, 299, 567 
    \bibitem[\protect\citeauthoryear{Larson et al.}{1996}]{larson9611}
        Larson~K.~A., Whittet~D.~C.~B., Hough~J.~H., 1996, 
        \apj, 472, 755
    \bibitem[\protect\citeauthoryear{Larson et al.}{2000}]{larson0004}
        Larson~K.~A., Wolff~M.~J., Roberge~W.~G., Whittet~D.~C.~B., He~L., 2000,
        \apj, 532, 1021
    \bibitem[\protect\citeauthoryear{Li \& Greenberg}{1997}]{li}
        Li~A., Greenberg~J.~M., 1997, 
        \aap, 323, 566
    \bibitem[\protect\citeauthoryear{Lucy}{1974}]{lucy7406}
        Lucy~L.~B., 1974,
        \aj, 79, 745
    \bibitem[\protect\citeauthoryear{Minniti}{1996}]{minniti}
        Minniti~D., 1996, 
        \apj, 459, 175
    \bibitem[\protect\citeauthoryear{Popowski}{2000}]{popowski}
        Popowski~P., 2000, 
        \apjl, 528, L9 
    \bibitem[\protect\citeauthoryear{Pottasch}{1984}]{pottasch}
        Pottasch~S.~R., 1984,
        \emph{Planetary Nebulae}, 
        Reidel, Dordrecht, p. 41, 92, 95 (SP)
    \bibitem[\protect\citeauthoryear{Pottasch \& Zijlstra}{1994}]{potzij94}
        Pottasch~S.~R.~\& Zijlstra~A.~A., 1994,
        \aap, 289, 261 
    \bibitem[\protect\citeauthoryear{Reid}{1993}]{reid}
        Reid~M., 1993, 
        \araa, 31, 345
    \bibitem[\protect\citeauthoryear{Richardson}{1972}]{rich72}
        Richardson~W.~H., 1972, 
        Jou. Opt. Soc. Am., 62, 55
    \bibitem[\protect\citeauthoryear{Scott \& Duley}{1996}]{scott}
        Scott~A., Duley~W.~W., 1996, 
        \apjs, 105, 401
    \bibitem[\protect\citeauthoryear{Stasinska et al.}{1992}]{stasinska} 
        Stasinska~G., Tylenda~R., Acker~A., Stenholm~B., 1992, 
        \aap, 266, 486 (STAS)
    \bibitem[\protect\citeauthoryear{Tylenda et al.}{1992}]{tylenda} 
        Tylenda~R., Acker~A., Stenholm~B., Koeppen~J., 1992, 
        \aaps, 95, 337 (TASK)
    \bibitem[\protect\citeauthoryear{Tylenda et al.}{2003}]{tylenda0307} 
        Tylenda~R., Si{\' o}dmiak~N., G{\' o}rny~S.~K., Corradi~R.~L.~M., Schwarz~H.~E., 2003,
        \aap, 405, 627
    \bibitem[\protect\citeauthoryear{Udalski}{2003}]{udalski}
        Udalski~A., 2003, 
        \apj, 590, 284 
    \bibitem[\protect\citeauthoryear{Udalski et al.}{2002}]{udalski02}
        Udalski~A., Szymanski~M., Kubiak~M., Pietrzynski~G., 
        Soszynski~I., Wozniak~P., Zebrun~K., Szewczyk~O., Wyrzykowsk~L., 2002, 
        Acta Astron., 52, 217 
    \bibitem[\protect\citeauthoryear{van Hoof}{2000}]{vanhoof}
        van~Hoof~P.~A.~M., 2000,
        \mnras, 314, 99 (VH)
    \bibitem[\protect\citeauthoryear{Walton, Barlow \& Clegg}{1993}]{walton93iau}
        Walton~N.~A., Barlow~ M.~J., Clegg~R.~E.~S., 1993, 
        IAU Symp.~153: Galactic Bulges, 153, 337 
    \bibitem[\protect\citeauthoryear{Zijlstra, Pottasch \& Bignell}{1989}]{zijlstra8909}
        Zijlstra~A.~A., Pottasch~S.~R., Bignell~C., 1989,
        \aaps, 79, 329
\end{thebibliography}
\end{document}